\documentclass[11pt]{article}

\usepackage[utf8]{inputenc}
\usepackage[T1]{fontenc}
\usepackage[letterpaper,margin=1in]{geometry}

\usepackage{amsmath,amssymb,amsthm,amsfonts,mathtools,mathrsfs}
\usepackage{graphicx}
\usepackage{setspace}
\usepackage{verbatim}
\usepackage{indentfirst}
\usepackage{caption,subcaption}
\usepackage{threeparttable}
\usepackage{booktabs}
\usepackage{multirow}
\usepackage{longtable}
\usepackage[dvipsnames]{xcolor}
\usepackage[linesnumbered,ruled,vlined]{algorithm2e}
\usepackage{xspace,makecell,tikz}
\usepackage{siunitx}
\usepackage{float}
\usepackage[sort,numbers]{natbib}
\usepackage{hyperref}
\usepackage[capitalize]{cleveref}
\usepackage{aliascnt}
\usepackage{bbm}
\usepackage{enumerate}
\usepackage[shortlabels]{enumitem}
\usepackage{tocloft}
\usepackage{comment}
\usepackage[toc,page,title]{appendix}
\usepackage{subfiles}
\numberwithin{equation}{section}
\hypersetup{
linktocpage=true,
colorlinks=true,
linkcolor=magenta,
filecolor=magenta,
urlcolor=cyan,
citecolor=blue,
}

\allowdisplaybreaks
\numberwithin{equation}{section}

\newcommand{\sharedtheorem}[2]{%
  \newaliascnt{#1}{theorem}%
  \newtheorem{#1}[#1]{#2}%
  \aliascntresetthe{#1}%
  \counterwithin*{#1}{section}%
}
\newtheorem{theorem}{Theorem}[section]
\sharedtheorem{lemma}{Lemma}
\sharedtheorem{proposition}{Proposition}
\sharedtheorem{corollary}{Corollary}
\sharedtheorem{claim}{Claim}
\theoremstyle{definition}
\sharedtheorem{definition}{Definition}
\sharedtheorem{remark}{Remark}
\sharedtheorem{fact}{Fact}

\newcommand{\E}{\mathbb E}
\newcommand{\Prb}{\mathbb P}
\newcommand{\Z}{\mathbb Z}
\newcommand{\TV}{\operatorname{TV}}
\newcommand{\KL}{D_{\mathrm{KL}}}
\newcommand{\Var}{\operatorname{Var}}
\newcommand{\ind}{\mathbf 1}
\newcommand{\norm}[1]{\lVert #1\rVert}
\newcommand{\floor}[1]{\left\lfloor #1\right\rfloor}
\newcommand{\ceil}[1]{\left\lceil #1\right\rceil}
\newcommand{\err}{\operatorname{err}}
\newcommand{\Rk}{R}

\crefname{equation}{Eq.}{Eqs.}
\Crefname{equation}{Eq.}{Eqs.}
\Crefname{theorem}{Theorem}{Theorems}
\Crefname{lemma}{Lemma}{Lemmas}
\Crefname{proposition}{Proposition}{Propositions}
\Crefname{corollary}{Corollary}{Corollaries}
\Crefname{definition}{Definition}{Definitions}
\Crefname{remark}{Remark}{Remarks}
\Crefname{fact}{Fact}{Facts}
\Crefname{claim}{Claim}{Claims}

\crefname{appendix}{Appendix}{Appendices}
\Crefname{appendix}{Appendix}{Appendices}

\makeatletter
\let\standardappendix\appendix
\renewcommand{\appendix}{%
  \standardappendix
  \def\sectionname{\appendixname}%
  \def\@seccntformat##1{\@ifundefined{##1name}{}{\csname ##1name\endcsname\ }%
    \csname the##1\endcsname\quad}%
  \crefalias{section}{appendix}%
}
\makeatother

\DeclarePairedDelimiter{\bk}{(}{)}
\DeclarePairedDelimiter{\Bk}{[}{]}

\newcommand{\defn}[1]{\textbf{#1}}

\newcommand{\varbar}{\overline{\Var}}

\newcommand{\misep}{\mathrel{;}}
\newcommand{\cond}{\mathrel{}\middle|\mathrel{}}

\newcommand{\MI}[2]{I\bk*{#1\misep #2}}
\newcommand{\MIc}[3]{I\bk*{#1\misep #2\cond #3}}

\newcommand{\Ec}[2]{\E\Bk*{#1\cond #2}}
\newcommand{\Varc}[2]{\Var\bk*{#1\cond #2}}
\newcommand{\varbarc}[2]{\varbar\bk*{#1\cond #2}}

\newcommand{\indepsym}{\perp}
\newcommand{\CIc}[3]{#1\mathrel{\indepsym}#2\mid #3}

\newcommand{\poly}{\text{poly }}

\title{Tight Bounds for Rank Estimation in Turnstile Streams}
\author{Yang S. Hu\thanks{Carnegie Mellon University. \href{mailto:yanghu2@andrew.cmu.edu}{yanghu2@andrew.cmu.edu}. Partially supported by NSF grant CNS-2504471, by NSF CAREER award CCF-2542165, and by a Jane Street Research Grant.}\and Cheng Jiang\thanks{Department of Electrical Engineering and Computer Science, Massachusetts Institute of Technology, \href{mailto:chengj36@mit.edu}{chengj36@mit.edu}.}\and Yinchen Liu\thanks{Institute for Interdisciplinary Information Sciences, Tsinghua University. \href{mailto:liuyinch23@mails.tsinghua.edu.cn}{liuyinch23@mails.tsinghua.edu.cn}.}\and Huacheng Yu\thanks{Department of Computer Science, Princeton University. \href{mailto:yuhch123@gmail.com}{yuhch123@gmail.com}. Supported in part by an NSF CAREER award CCF-2339942.}}
\date{}

\begin{document}

\setcounter{page}{0}
\maketitle
\thispagestyle{empty}

\begin{abstract}
We study rank estimation in turnstile streams. A frequency vector $f$ over
the universe $[U]$ receives insertions and deletions. Given a query $x\in [U]$,
the goal is to estimate the rank $\sum_{u<x}f_u$ within additive error
$\varepsilon\norm f_1$ with constant probability. The standard approach is the
Dyadic CountSketch by Luo, Wang, Yi and Cormode [The VLDB Journal '16], which decomposes the universe into dyadic
intervals and maintains each level with a CountSketch, using $\widetilde{O}(\varepsilon^{-1}\log^{3/2}U\log n)$ bits of space, where $n$ is the final $\ell_1$ norm of $f$. Recently, Bul{\'a}nek,
Kumar, Meka, Nelson, and Sarl{\'o}s [Preprint '26] placed this approach in a
matrix-factorization framework. They improved the space of Dyadic CountSketch to
$O(\varepsilon^{-1}\log^{3/2}U)$ words, and proved a lower bound of
$\Omega(\varepsilon^{-1}\log^2(\varepsilon U))$ bits for all streaming
algorithms when $U\le n\le U^{O(1)}$ and $\varepsilon=U^{-1+\Omega(1)}$. This leaves a gap of $\sqrt{\log U}$.

In this work, we fully determine the space cost of rank estimation up to an absolute constant. Our main result is a space lower bound for all streaming algorithms: for
$0<\varepsilon\le1/8$ and $n\ge\log^2U$, the space is at least
\[
  \Omega\left(\min\left\{
    \varepsilon^{-1}L^{3/2}\log\left(2+\frac{n\varepsilon}{\sqrt L}\right),\
    \varepsilon^{-2}\log U\log n,\
    \log\binom{U+n}{n}
  \right\}\right)\quad\text{bits},
\]
where $L=\log(2+U\varepsilon)$. When $U\le n\le U^{O(1)}$ and
$\varepsilon=U^{-1+\Omega(1)}$, this is simplified to
\begin{align*}
    \Omega(\min\{\varepsilon^{-1}\log^{5/2}U,\ \varepsilon^{-2}\log^2U\})
\end{align*}bits. Our proof identifies a surprising hard distribution, and proves lower bound over it using novel information-theoretic techniques. We believe that our techniques are of independent interest.

The above bound matches Dyadic CountSketch when $n\ge U$ and $U^{-1+\Omega(1)}\le \varepsilon\le O(1/\sqrt{\log U})$. Surprisingly, the lower bound is also tight in the remaining regimes. For $\varepsilon\ge(\log U)^{-1/2}$, there exists an
algorithm that uses $O(\varepsilon^{-2}\log U\log n)$ bits using $\ell_1$ samplers. For $\varepsilon\le U^{-1+o(1)}$, we refine the
Dyadic CountSketch so that its space in bits matches the first term of the
bound.
\end{abstract}

\newpage

\section{Introduction}

We study \emph{rank estimation in turnstile streams}. A stream over the ordered
universe $[U]=\{0,\ldots,U-1\}$ maintains a frequency vector, initially zero,
through updates $(u,\Delta)$ that add an integer $\Delta$ to coordinate $u$.
The final frequency vector $f\in\Z_{\ge0}^U$ is promised to be nonnegative,
and $n$ is an upper bound on its final total frequency $\norm f_1$.
A rank query at $x$ asks for $R_f(x)=\sum_{u<x}f_u$. Given an accuracy
parameter $\varepsilon>0$, the goal is to estimate $R_f(x)$ within additive
error $\varepsilon\norm f_1$ with constant success probability for each fixed query.

Rank estimation provides a way to approximate the empirical distribution of a
dataset. Understanding this distribution reveals its typical values, variability,
and tail behavior. For example, the distribution of response times
describes both the experience of a typical user and the frequency of unusually
slow requests. 
When data arrive as a stream, storing all items may be prohibitively
expensive, motivating compact summaries that approximate their empirical
distribution. Rank and quantile estimation are essentially equivalent to
approximating the empirical cumulative distribution function
(CDF)~\citep{KarninLL16}. Normalized ranks (i.e., the fraction of data items smaller than a given query $x$) give CDF values, while quantiles
describe its inverse. Such summaries support tasks including estimating the
selectivity of database queries and partitioning data into ranges of approximately
equal size~\citep{GreenwaldK01}.

Rank and quantile estimation have been studied in two models. In the
\emph{comparison-based model}, items belong to a totally ordered universe and
can only be accessed through comparisons. In the \emph{fixed-universe model},
items belong to a known integer
universe, and algorithms may exploit their representations. Both models have
been extensively studied in insertion-only streams, including comparison-based
summaries such as Greenwald--Khanna and KLL~\citep{GreenwaldK01,KarninLL16},
and fixed-universe summaries such as $q$-digest and the sketch of Gupta,
Singhal, and Wu~\citep{ShrivastavaBAS04,GuptaSW24}.

Many applications, however, require summarizing a dataset that changes through
both insertions and
deletions. Database records may be removed or modified, and network monitoring
may concern the distribution of currently active sessions: a session enters the
dataset when it begins and leaves when it ends. In these settings, the summary
must reflect the distribution of the current dataset as observations are added
and removed~\citep{GilbertKMS05}.

Arbitrary deletions pose a fundamental obstacle in the comparison-based model.
Imagine that we insert $m$ distinct items, then delete all but one. Any item could be the
survivor, and a discarded item cannot be reconstructed using comparisons alone.
A quantile summary must therefore retain $\Omega(m)$ items to recover the
survivor with constant success probability~\citep[cf.][]{LuoWY16}.\footnote{For rank
queries, consider query values immediately below and above the survivor. If the
survivor has been discarded, these queries are indistinguishable through
comparisons with the retained items, although their ranks are $0$ and $1$.
For $\varepsilon<1/2$, success probability $2/3$ likewise requires retaining
$\Omega(m)$ items.}
This motivates studying turnstile rank estimation in the fixed-universe model.

Consider the common regime of $n=U^{\Theta(1)}$.
A standard approach to handling turnstile updates is to dyadically decompose the universe and maintain a Count-Min or CountSketch~\citep{CormodeM05,LuoWY16} for each interval (see more in \Cref{sec:dyadic-hard-distribution}). 
Sketches based on the \emph{Dyadic CountSketch} (DCS for short) take $O(\varepsilon^{-1}\log^{3/2}U)$ words, i.e., $O(\varepsilon^{-1}\log^{5/2}U)$ bits, of space~\citep{Nelson26}.
It was also known that any streaming algorithm must take
$\Omega(\varepsilon^{-1}\log^2(\varepsilon U))$ bits when $U\le n\le U^{O(1)}$ and $\varepsilon=U^{-1+\Omega(1)}$~\citep{Nelson26}.\footnote{These bounds hold for constant success probability and sufficiently
small $\varepsilon$. The lower bound in Theorem~5.1 of~\citep{Nelson26} is stated and proved
for approximate quantile (select) queries; the surrounding discussion also
refers to rank queries.} This leaves a gap of
$\sqrt{\log U}$ in this regime.

When $\varepsilon$ is close to a constant, note that another approach is to draw $O(\varepsilon^{-2})$ independent samples,
each selecting item $u\in[U]$ with probability
$f_u/\norm f_1$. Multiplying the fraction of samples below the query by
$\norm f_1$ gives a rank estimate. A streaming algorithm that draws one such
sample is called an \emph{$\ell_1$ sampler}. Running
$O(\varepsilon^{-2})$ independent copies of the perfect $\ell_1$ sampler
(see~\citep{JayaramW21,SwartworthWZ25}) yields an
$O(\varepsilon^{-2}\log^2U)$-bit algorithm.\footnote{This bound assumes free access to randomness.}
This is a better space bound when $\varepsilon\geq \Omega\left((\log U)^{-1/2}\right)$.

When $\varepsilon$ is tiny, the error bound essentially forces exact queries.
In this case, one can use a sketch that recovers the exact frequency vector with space $O(\log\binom{U+n}{n})$ bits.

Our main result shows that the smaller of these three bounds is optimal up to constant factors for almost the full range of parameters.

\begin{theorem}[Main result]
\label{thm:intro-main}
Let $U,n,\varepsilon$ be parameters such that $0<\varepsilon\le1/8$ and $n\ge\log^2 U$.
Set $L=\log(2+U\varepsilon)$. Rank sketches for turnstile streams
of final $\ell_1$ norm at most $n$, with success probability at least $2/3$
for each fixed query, have optimal space complexity
\[
\mathbf{RQ}(\varepsilon,n,U):=\Theta\!\left(\min\left\{
    \varepsilon^{-1}L^{3/2}
      \log\!\left(2+\frac{n\varepsilon}{\sqrt L}\right),\;
    \varepsilon^{-2}\log U\,\log n,\;\log\binom{U+n}{n}
  \right\}\right)
  \qquad\text{bits}.
\]
\end{theorem}

The lower bound holds for rank queries, while our upper bounds support both
rank and select queries. A \emph{select query} specifies a target rank
$r\in[0,\norm f_1]$ and asks for an item $\widehat x\in[U]$ satisfying
\[
  R_f(\widehat x+1)\ge r-\varepsilon\norm f_1,
  \qquad
  R_f(\widehat x)\le r+\varepsilon\norm f_1.
\]
Note that by reducing from rank queries using a binary search, one can obtain a weaker lower bound for select queries with a $\log(1/\varepsilon)$-factor loss.

The three terms in the theorem correspond, respectively, to the Dyadic CountSketch, sampling, and
exact recovery of the frequency vector. In the most common parameter regime,
$n=U^{\Theta(1)}$ and $\varepsilon^{-1}=U^{o(1)}$, the bound simplifies to
\[
  \Theta\!\left(
    \min\left\{
      \varepsilon^{-1}\log^{5/2}U,\,
      \varepsilon^{-2}\log^2U
    \right\}
  \right)
  \qquad\text{bits}.
\]
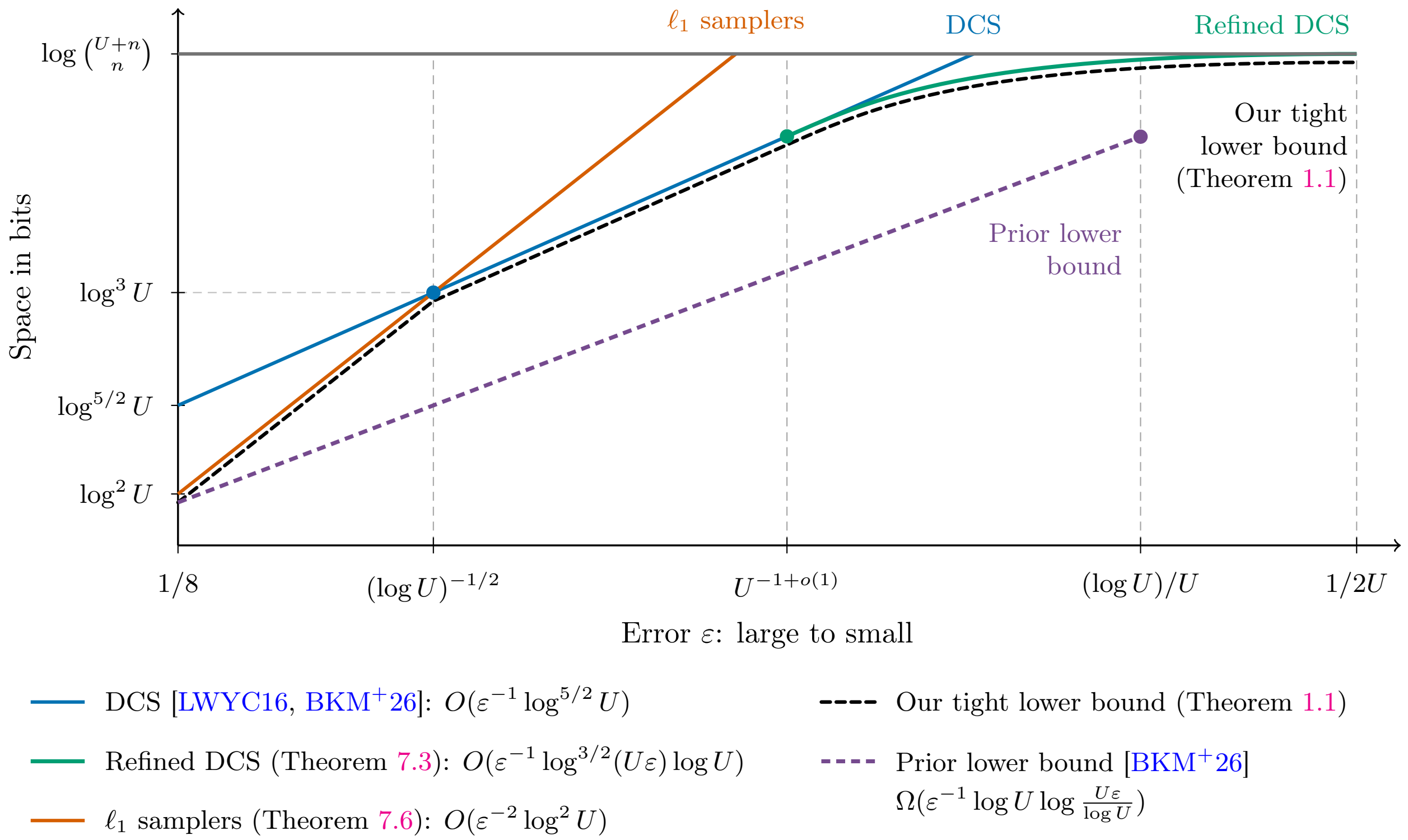
\begin{figure}[!t]
  \centering
  \begingroup
  \definecolor{bounddcs}{RGB}{0,114,178}
  \definecolor{boundsampling}{RGB}{213,94,0}
  \definecolor{boundprior}{RGB}{117,74,142}
  \definecolor{boundrefined}{RGB}{0,158,115}
  \begin{tikzpicture}[
      x=1.16cm, y=1cm, font=\small,
      dcs/.style={draw=bounddcs, line width=1.15pt, solid},
      sampling/.style={draw=boundsampling, line width=1.15pt, solid},
      prior/.style={draw=boundprior, line width=1.35pt, densely dashed},
      refined/.style={draw=boundrefined, line width=1.35pt, solid},
      exact/.style={draw=black!55, line width=1.15pt, solid},
      ours/.style={draw=black, line width=1.2pt, densely dashed, line cap=round,
        line join=round},
      ticklabel/.style={font=\footnotesize},
      curvelabel/.style={font=\footnotesize, inner sep=2pt},
      legendtext/.style={font=\footnotesize, anchor=base west, inner sep=0pt}
    ]
    \def\crossoverx{2.6}
    \pgfmathsetmacro{\dcsbase}{0.45-0.10/1.35+\crossoverx*(3.65-0.45)/9.8}
    \pgfmathsetmacro{\dcsslope}{(4.3-\dcsbase)/8.1}
    \pgfmathsetmacro{\crossoverheight}{\dcsbase+\crossoverx*\dcsslope}
    \pgfmathsetmacro{\samplerend}{(4.3-0.45)*\crossoverx/(\crossoverheight-0.45)}
    \begin{scope}[y=1.35cm]
    \foreach \x in {\crossoverx,6.2,9.8,12} {
      \draw[black!35, densely dashed] (\x,0) -- (\x,4.3);
    }
    \draw[black!25, densely dashed] (0,\crossoverheight) -- (\crossoverx,\crossoverheight);
    \draw[->, line width=0.65pt] (0,0) -- (12.45,0);
    \draw[->, line width=0.65pt] (0,0) -- (0,4.7);
    \node[rotate=90, font=\small] at (-1.58,2.35)
      {Space in bits};

    \foreach \y/\lab in {
        0.45/{\log^2U},
        \dcsbase/{\log^{5/2}U},
        \crossoverheight/{\log^3U},
        4.3/{\log\binom{U+n}{n}}
      } {
      \draw (-0.07,\y) -- (0.07,\y);
      \node[ticklabel, anchor=east] at (-0.13,\y) {$\lab$};
    }
    \foreach \x/\lab in {
        0/{1/8},
        \crossoverx/{(\log U)^{-1/2}},
        6.2/{U^{-1+o(1)}},
        9.8/{(\log U)/U},
        12/{1/2U}
      } {
      \draw (\x,-0.07) -- (\x,0.07);
      \node[ticklabel, anchor=north] at (\x,-0.13) {$\lab$};
    }

    \def\samplinghead{-- (\crossoverx,\crossoverheight)}
    \def\dcsmiddle{-- (6.2,{\dcsbase+6.2*\dcsslope})}
    \def\refinedtail{
      .. controls (7.15,{\dcsbase+7.15*\dcsslope}) and (8.1,4.3) .. (12,4.3)}
    \begin{scope}[yshift=-0.10cm]
      \draw[ours] (0,0.45) \samplinghead \dcsmiddle \refinedtail;
      \node[curvelabel, anchor=north east, align=right, fill=white]
        at (11.98,4.0)
        {Our tight\\lower bound\\(\Cref{thm:intro-main})};
    \end{scope}
    \begin{scope}[overlay]
      \clip (-0.05,0) rectangle (12.05,4.3);
      \draw[dcs] (0,\dcsbase) -- (8.1,4.3);
      \draw[sampling] (0,0.45) -- (\samplerend,4.3);
    \end{scope}
    \draw[refined] (6.2,{\dcsbase+6.2*\dcsslope}) \refinedtail;
    \draw[exact] (0,4.3) -- (12,4.3);
    \fill[bounddcs] (\crossoverx,\crossoverheight) circle[radius=2.4pt];
    \fill[boundrefined] (6.2,{\dcsbase+6.2*\dcsslope}) circle[radius=2.4pt];
    \node[curvelabel, text=boundsampling, anchor=south] at (\samplerend,4.42)
      {$\ell_1$ samplers};
    \node[curvelabel, text=bounddcs, anchor=south] at (8.1,4.42)
      {DCS};
    \node[curvelabel, text=boundrefined, anchor=south east, align=right]
      at (12,4.42)
      {Refined DCS};
    \begin{scope}[yshift=-0.10cm]
      \draw[prior] (0,0.45) -- (9.8,3.65);
      \fill[boundprior] (9.8,3.65) circle[radius=2.4pt];
      \node[curvelabel, text=boundprior, anchor=north east, align=right]
        at (9.68,2.95)
        {Prior lower\\bound};
    \end{scope}
    \end{scope}

    \node[font=\small] at (6,-1.06)
      {Error $\varepsilon$: large to small};

    \draw[dcs] (-1.5,-1.85) -- (-0.95,-1.85);
    \node[legendtext] at (-0.75,-1.95)
      {DCS~\citep{LuoWY16,Nelson26}:
       $O(\varepsilon^{-1}\log^{5/2}U)$};
    \draw[ours] (6.55,-1.85) -- (7.1,-1.85);
    \node[legendtext] at (7.3,-1.95)
      {Our tight lower bound (\Cref{thm:intro-main})};

    \draw[refined] (-1.5,-2.55) -- (-0.95,-2.55);
    \node[legendtext] at (-0.75,-2.65)
      {Refined DCS (\Cref{thm:small-error-upper}):
       $O(\varepsilon^{-1}\log ^{3/2}(U\varepsilon)\log U)$};
    \draw[prior] (6.55,-2.55) -- (7.1,-2.55);
    \node[legendtext, anchor=north west, align=left] at (7.3,-2.39)
      {Prior lower bound~\citep{Nelson26}\\[2pt]
       $\Omega(\varepsilon^{-1}\log U
         \log\frac{U\varepsilon}{\log U})$};

    \draw[sampling] (-1.5,-3.25) -- (-0.95,-3.25);
    \node[legendtext] at (-0.75,-3.35)
      {$\ell_1$ samplers (\Cref{thm:priority-bits}):
       $O(\varepsilon^{-2}\log^2U)$};

  \end{tikzpicture}
  \endgroup
  \caption{Space versus decreasing error for $U^2<n=U^{\Theta(1)}$,
    where $\varepsilon$ ranges from $1/8$ to $1/2U$. Constant factors
    are suppressed; the small gap
    between matching upper and lower bounds is for readability.}
  \label{fig:space-bounds}
\end{figure}
Across all parameter regimes, we give implementations with polylogarithmic
update and query time. Their space bounds, including the conditions under which
they match the optimal bounds in bits, are detailed in \Cref{sec:upper-bounds}.

To prove the space lower bound, we first prove a lower bound for linear sketches, which store linear
measurements of the frequency vector. The key property of linear sketches that we use is the
\emph{history independence}: for fixed randomness, the memory state depends only on
the final frequency vector. This lets us analyze how much information about each item the
sketch must retain to answer rank queries.
To extend the result to general turnstile streaming algorithms, we build on the
recent work of Jiang, Liu, and Yu~\citep{JiangLY26}, which simulates algorithms
on polynomial-length streams using linear sketches. We adapt this simulation
to rank queries supplied after the stream, obtaining the same lower bound in
bits.

\subsection{Related Work}
\label{sec:intro-related-work}

Dynamic quantile estimation maintains summaries under both insertions and
deletions. Gilbert, Kotidis, Muthukrishnan, and Strauss~\citep{GilbertKMS05}
use random subset sums to maintain a multiresolution representation of the
data distribution. A common strategy is to decompose a prefix into
$O(\log U)$ dyadic intervals and estimate their total frequencies using
sketches that support positive and negative updates. Cormode and
Muthukrishnan~\citep{CormodeM05} implement this strategy using Count-Min
sketches. Luo, Wang, Yi, and Cormode~\citep{LuoWY16} analyze the Dyadic CountSketch and give an extensive experimental comparison of quantile summaries.

More recently, Bul{\'a}nek, Kumar, Meka, Nelson, and Sarl{\'o}s~\citep{Nelson26}
place this approach in a general matrix-factorization framework: factor the
prefix-sum matrix as $Q=AB$, sketch $Bf$, and combine the estimates using $A$.
For constant success probability, this framework improves the prior Dyadic CountSketch space
bound for rank queries by a factor of $\log^{3/2}(\varepsilon^{-1}\log U)$,
removing the extra dependence on $\log(1/\varepsilon)$ and $\log\log U$.
Their analysis of correlations between CountSketch errors allows a single
CountSketch to serve all levels of the dyadic decomposition and yields the rank
upper bound discussed above. They also prove a quantile-query lower bound and
show that the dyadic factorization is nearly optimal within their framework.
Our results determine the optimal space over all linear sketches.

Insertion-only quantile summaries have been studied in both comparison-based
and finite-universe models. The Greenwald--Khanna summary~\citep{GreenwaldK01}
and KLL sketch~\citep{KarninLL16} are deterministic and randomized
comparison-based constructions, respectively. For a fixed query and constant
failure probability, KLL uses $O(\varepsilon^{-1})$ words.
Cormode and Vesel{\'y}~\citep{CormodeV20} prove
that the Greenwald--Khanna space bound is optimal for deterministic
comparison-based summaries. For integer universes, the $q$-digest~\citep{ShrivastavaBAS04}
and the more recent sketch of Gupta, Singhal, and Wu~\citep{GuptaSW24}
exploit the representation of item values to obtain different space tradeoffs.
These distinctions between computational models are important when comparing
quantile lower bounds.

Relative-error quantile estimation asks for error proportional to the query
rank, providing finer accuracy in the tails of the distribution. This setting
has been studied by Cormode, Karnin, Liberty, Thaler, and Vesel{\'y}~\citep{CormodeKLTV23} and, more recently,
Gribelyuk, Sawettamalya, Wu, and Yu~\citep{GribelyukSWY25}, whose elastic compactors give
near-optimal space bounds for insertion-only streams. Domes and
Vesel{\'y}~\citep{DomesV27} further improve the logarithmic factors in this
space bound and obtain full mergeability using adaptive compactors.

The broader $\ell_p$-sampling problem draws coordinate $u$ with probability
proportional to $|f_u|^p$. Approximate samplers were developed
in~\citep{MonemizadehW10,AndoniKO11,JowhariST11}, followed by the perfect
samplers of Jayaram and Woodruff~\citep{JayaramW21} and the faster
implementations of Swartworth, Woodruff, and Zhou~\citep{SwartworthWZ25}.
The case $p=1$ samples items in proportion to their frequencies, while $p=2$
emphasizes coordinates with larger frequencies. Our upper bounds also use
ideas from priority sampling, which estimates arbitrary subset sums from a
weighted sample~\citep{DuffieldLT07}.

The relationship between turnstile streaming and linear sketching was
established by Li, Nguyen, and Woodruff~\citep{LiNW14} and further developed
by Ai, Hu, Li, and Woodruff~\citep{AHLW16}, under assumptions allowing very long streams.
Kallaugher and Price~\citep{KallaugherP20} show that restrictions on stream
length or intermediate coordinate magnitudes can separate the two models.
The recent polynomial-length simulation of Jiang, Liu, and Yu~\citep{JiangLY26}
provides the basis for transferring our sketching lower bounds to streaming.

\section{Technical Overview}
\label{sec:overview}

\subsection{Sketching Lower Bound for the Key Regime}

In the following, we restrict our attention to lower bounds on \emph{linear sketches} for the turnstile rank estimation problem. This suffices because all turnstile streaming algorithms can be simulated by linear sketches with only a small space blowup (see \cite{JiangLY26}).

As the optimal tradeoff curve consists of multiple pieces, it is difficult to directly prove a tight lower bound that holds for all regimes. Instead, we begin by studying the \emph{key regime}: The case where $\varepsilon=1/\sqrt{\log U}$, and the $\ell_1$ norm $n=\Theta(\log U)$. In the key regime, we aim to prove a space lower bound of $\Omega(\log^2 U)$ bits.

\paragraph{Justification for the key regime.} In the key regime, we let $\varepsilon=\Theta\bk*{1/\sqrt{\log U}}$ because it is the \emph{unique} point where two algorithms agree on the space cost (both costs equal $O(\log^2 U\log n)$ bits), and we let $n=\Theta(\log U)$ because it is the smallest $n$ for which we can prove a lower bound of $\Omega(\log^2U)$ bits. For even smaller $n$, rank estimation can be solved in $n\cdot \log U=o(\log^2 U)$ bits of space.

Another (surprising) reason for studying the key regime is that it is essentially the \emph{only} case for which we need to prove a lower bound. Later, we will show that our lower bound for the key regime can be rather easily generalized to all the regimes that we care about.

\subsubsection{A Non-Uniform Hard Distribution that Exploits the Dyadic Structure}
\label{sec:dyadic-hard-distribution}

We prove the lower bound by finding a distribution over input streams (each consisting of $d = \Theta(\log U)$ insertion updates $V_1,\dots,V_{d}\in [U]$ and one query location $X\in [U]$ in the end), such that any linear sketch $\mathcal{A}$ that answers the query within an additive error of $\varepsilon \cdot d$ with probability $0.99$ over the distribution must use $\Omega(\log^2 U)$ bits of space. In this section, we discuss the observations that lead to such a hard input distribution.

\paragraph{The structure of the Dyadic CountSketch.} We first briefly describe the Dyadic CountSketch algorithm~\citep{LuoWY16,Nelson26}. Understanding how it works is very important for designing the hard distribution.

Assume for simplicity that $U$ is a power of $2$. The DCS algorithm builds a dyadic structure (i.e., a perfect binary tree) of $\log U+1$ levels over the universe $[U]$, such that level $j$ ($0\le j\le \log U$) consists of $2^j$ disjoint intervals, each of length $U/2^j$ (these are called \defn{dyadic intervals}).

For each $0\le j\le \log U$, the algorithm then maintains a count sketch, which is responsible for estimating the total frequency of each dyadic interval at level $j$. That is, given an index $a\in [2^j]$, the count sketch at level $j$ outputs an estimate for the total frequency of the dyadic interval $[a\cdot U/2^j,(a+1)\cdot U/2^j)$. In the key regime, to control the error of the estimates, each count sketch uses $\tilde{O}(\log U)$ bits of space (we omit the calculation), so the total space is $\tilde{O}(\log^2 U)$ bits.

To answer a rank query at location $X$, the algorithm decomposes $[0,X)$ into at most $\log U$ disjoint dyadic intervals, uses the count sketches to estimate the total frequency of each interval, and outputs the sum of the estimates.

\paragraph{The uniform distribution is easy for (a variation of) the Dyadic CountSketch.} The most natural idea for a hard distribution is simply to sample a uniformly random stream of updates $V_1,\dots,V_{\log U}\in [U]$, followed by a uniformly random query location $X\in [U]$. This distribution, however, allows for a very space-efficient algorithm, which is a simple variation of DCS. 

Consider how the interval $[0,X)$ is decomposed into dyadic intervals when answering a query in the Dyadic CountSketch: Start at leaf $X$ of the dyadic structure and move towards the root. If some ancestor of leaf $X$ is a right child, then the sibling of this ancestor is added to the list of dyadic intervals. Observe that, if there exists a level $j$ dyadic interval in the decomposition, then \emph{any key in this interval is at distance at most $U/2^{j-1}$ from $X$}.

On the other hand, in the uniform distribution, with constant probability, \emph{any update key $V_i$ is at distance at least $\Omega(U/\log U)$ from $X$}. This means that, when a count sketch at level $j\ge \Omega(\log\log U)$ is called during query, it is likely that the level $j$ dyadic interval in question does not contain any update key, so its total frequency is $0$.

As a result, we can improve the space usage of the Dyadic CountSketch by only storing the first $O(\log\log U)$ levels, and simply outputting $0$ for all the lower levels. This reduces the total space to merely $\tilde{O}(\log U)$ bits.

\paragraph{The actual hard distribution.} Inspired by the above counterexample, we now propose a hard distribution over which the Dyadic CountSketch must store the count sketches for a large number of levels. We construct the hard distribution based on a simple observation: \emph{The count sketch at level $j$ is needed when there exists a key at distance $\approx 2^{\log U-j}$ from $X$}.

\begin{figure}[t]
    \centering
    \begin{tikzpicture}[
        updatepath/.style={draw=black, line width=0.55pt},
        querypath/.style={draw=black, line width=1.5pt},
        update/.style={circle, draw=black, fill=black, inner sep=0.85pt},
        query/.style={circle, draw=black, fill=black, inner sep=0.95pt},
        meet/.style={circle, draw=black, fill=white, line width=1.05pt, inner sep=2.05pt},
        leaflabel/.style={font=\scriptsize, text=black, anchor=north},
        meetlabel/.style={font=\scriptsize\bfseries, text=black},
        levelmark/.style={font=\scriptsize, text=black, anchor=east}
    ]
        \begin{scope}
            \node[font=\bfseries] at (3.58,6.15) {Uniform Updates};
            \node[font=\small] at (3.58,5.80) {};
            \coordinate (u-0-0) at (3.575000,5.050000);
            \coordinate (u-1-0) at (2.062500,4.580000);
            \coordinate (u-1-1) at (5.087500,4.580000);
            \coordinate (u-2-0) at (1.306250,4.110000);
            \coordinate (u-2-1) at (2.818750,4.110000);
            \coordinate (u-2-2) at (4.331250,4.110000);
            \coordinate (u-2-3) at (5.843750,4.110000);
            \coordinate (u-3-0) at (0.928125,3.640000);
            \coordinate (u-3-1) at (1.684375,3.640000);
            \coordinate (u-3-2) at (2.440625,3.640000);
            \coordinate (u-3-3) at (3.196875,3.640000);
            \coordinate (u-3-4) at (3.953125,3.640000);
            \coordinate (u-3-5) at (4.709375,3.640000);
            \coordinate (u-3-6) at (5.465625,3.640000);
            \coordinate (u-3-7) at (6.221875,3.640000);
            \coordinate (u-4-1) at (1.117188,3.170000);
            \coordinate (u-4-3) at (1.873438,3.170000);
            \coordinate (u-4-5) at (2.629688,3.170000);
            \coordinate (u-4-7) at (3.385937,3.170000);
            \coordinate (u-4-9) at (4.142188,3.170000);
            \coordinate (u-4-10) at (4.520313,3.170000);
            \coordinate (u-4-11) at (4.898438,3.170000);
            \coordinate (u-4-13) at (5.654687,3.170000);
            \coordinate (u-4-15) at (6.410937,3.170000);
            \coordinate (u-5-2) at (1.022656,2.700000);
            \coordinate (u-5-6) at (1.778906,2.700000);
            \coordinate (u-5-10) at (2.535156,2.700000);
            \coordinate (u-5-14) at (3.291406,2.700000);
            \coordinate (u-5-18) at (4.047656,2.700000);
            \coordinate (u-5-21) at (4.614844,2.700000);
            \coordinate (u-5-22) at (4.803906,2.700000);
            \coordinate (u-5-26) at (5.560156,2.700000);
            \coordinate (u-5-30) at (6.316406,2.700000);
            \coordinate (u-6-5) at (1.069922,2.230000);
            \coordinate (u-6-13) at (1.826172,2.230000);
            \coordinate (u-6-21) at (2.582422,2.230000);
            \coordinate (u-6-29) at (3.338672,2.230000);
            \coordinate (u-6-37) at (4.094922,2.230000);
            \coordinate (u-6-42) at (4.567578,2.230000);
            \coordinate (u-6-45) at (4.851172,2.230000);
            \coordinate (u-6-53) at (5.607422,2.230000);
            \coordinate (u-6-61) at (6.363672,2.230000);
            \coordinate (u-7-11) at (1.093555,1.760000);
            \coordinate (u-7-27) at (1.849805,1.760000);
            \coordinate (u-7-43) at (2.606055,1.760000);
            \coordinate (u-7-59) at (3.362305,1.760000);
            \coordinate (u-7-75) at (4.118555,1.760000);
            \coordinate (u-7-85) at (4.591211,1.760000);
            \coordinate (u-7-91) at (4.874805,1.760000);
            \coordinate (u-7-107) at (5.631055,1.760000);
            \coordinate (u-7-123) at (6.387305,1.760000);
            \coordinate (u-8-23) at (1.105371,1.290000);
            \coordinate (u-8-55) at (1.861621,1.290000);
            \coordinate (u-8-87) at (2.617871,1.290000);
            \coordinate (u-8-119) at (3.374121,1.290000);
            \coordinate (u-8-151) at (4.130371,1.290000);
            \coordinate (u-8-170) at (4.579395,1.290000);
            \coordinate (u-8-183) at (4.886621,1.290000);
            \coordinate (u-8-215) at (5.642871,1.290000);
            \coordinate (u-8-247) at (6.399121,1.290000);
            \coordinate (u-4-0) at (0.739063,3.170000);
            \coordinate (u-4-2) at (1.495313,3.170000);
            \coordinate (u-4-4) at (2.251562,3.170000);
            \coordinate (u-4-6) at (3.007812,3.170000);
            \coordinate (u-4-8) at (3.764062,3.170000);
            \coordinate (u-4-12) at (5.276562,3.170000);
            \coordinate (u-4-14) at (6.032812,3.170000);
            \coordinate (u-5-0) at (0.644531,2.700000);
            \coordinate (u-5-1) at (0.833594,2.700000);
            \coordinate (u-5-3) at (1.211719,2.700000);
            \coordinate (u-5-4) at (1.400781,2.700000);
            \coordinate (u-5-5) at (1.589844,2.700000);
            \coordinate (u-5-7) at (1.967969,2.700000);
            \coordinate (u-5-8) at (2.157031,2.700000);
            \coordinate (u-5-9) at (2.346094,2.700000);
            \coordinate (u-5-11) at (2.724219,2.700000);
            \coordinate (u-5-12) at (2.913281,2.700000);
            \coordinate (u-5-13) at (3.102344,2.700000);
            \coordinate (u-5-15) at (3.480469,2.700000);
            \coordinate (u-5-16) at (3.669531,2.700000);
            \coordinate (u-5-17) at (3.858594,2.700000);
            \coordinate (u-5-19) at (4.236719,2.700000);
            \coordinate (u-5-20) at (4.425781,2.700000);
            \coordinate (u-5-23) at (4.992969,2.700000);
            \coordinate (u-5-24) at (5.182031,2.700000);
            \coordinate (u-5-25) at (5.371094,2.700000);
            \coordinate (u-5-27) at (5.749219,2.700000);
            \coordinate (u-5-28) at (5.938281,2.700000);
            \coordinate (u-5-29) at (6.127344,2.700000);
            \coordinate (u-5-31) at (6.505469,2.700000);
            \draw[gray, dashed, line width=0.6pt] (0.48,3.405) -- (6.85,3.405);
            \draw[updatepath] (u-8-23) -- (u-7-11) -- (u-6-5) -- (u-5-2) -- (u-4-1) -- (u-3-0) -- (u-2-0) -- (u-1-0) -- (u-0-0);
            \draw[updatepath] (u-8-55) -- (u-7-27) -- (u-6-13) -- (u-5-6) -- (u-4-3) -- (u-3-1) -- (u-2-0) -- (u-1-0) -- (u-0-0);
            \draw[updatepath] (u-8-87) -- (u-7-43) -- (u-6-21) -- (u-5-10) -- (u-4-5) -- (u-3-2) -- (u-2-1) -- (u-1-0) -- (u-0-0);
            \draw[updatepath] (u-8-119) -- (u-7-59) -- (u-6-29) -- (u-5-14) -- (u-4-7) -- (u-3-3) -- (u-2-1) -- (u-1-0) -- (u-0-0);
            \draw[updatepath] (u-8-151) -- (u-7-75) -- (u-6-37) -- (u-5-18) -- (u-4-9) -- (u-3-4) -- (u-2-2);
            \draw[updatepath] (u-8-183) -- (u-7-91) -- (u-6-45) -- (u-5-22) -- (u-4-11) -- (u-3-5);
            \draw[updatepath] (u-8-215) -- (u-7-107) -- (u-6-53) -- (u-5-26) -- (u-4-13) -- (u-3-6) -- (u-2-3) -- (u-1-1);
            \draw[updatepath] (u-8-247) -- (u-7-123) -- (u-6-61) -- (u-5-30) -- (u-4-15) -- (u-3-7) -- (u-2-3) -- (u-1-1);
            \draw[querypath] (u-8-170) -- (u-7-85) -- (u-6-42) -- (u-5-21) -- (u-4-10) -- (u-3-5) -- (u-2-2) -- (u-1-1) -- (u-0-0);
            \node[update] at (u-8-23) {};
            \node[leaflabel] at (u-8-23) {$V_1$};
            \node[update] at (u-8-55) {};
            \node[leaflabel] at (u-8-55) {$V_2$};
            \node[update] at (u-8-87) {};
            \node[leaflabel] at (u-8-87) {$V_3$};
            \node[update] at (u-8-119) {};
            \node[leaflabel] at (u-8-119) {$V_4$};
            \node[update] at (u-8-151) {};
            \node[leaflabel] at (u-8-151) {$V_5$};
            \node[update] at (u-8-183) {};
            \node[leaflabel] at (u-8-183) {$V_6$};
            \node[update] at (u-8-215) {};
            \node[leaflabel] at (u-8-215) {$V_7$};
            \node[update] at (u-8-247) {};
            \node[leaflabel] at (u-8-247) {$V_8$};
            \node[query] at (u-8-170) {};
            \node[font=\scriptsize\bfseries, text=black, anchor=north] at (u-8-170) {$X$};
            \node[meet] at (u-0-0) {};
            \node[meetlabel, anchor=south west] at (u-0-0) {$a_1=\cdots=a_4$};
            \node[meet] at (u-1-1) {};
            \node[meetlabel, anchor=west] at (u-1-1) {$a_7=a_8$};
            \node[meet] at (u-2-2) {};
            \node[meetlabel, anchor=east] at (u-2-2) {$a_5$};
            \node[meet] at (u-3-5) {};
            \node[meetlabel, anchor=west] at (u-3-5) {$a_6$};
            \foreach \depth in {0,...,8} {
                \node[levelmark] at (0.30,{5.05-0.47*\depth}) {$\depth$};
            }
            \node[font=\scriptsize, text=black, rotate=90] at (-0.12,3.17) {Level};
            \node[font=\scriptsize, align=center, text width=6.7cm] at (3.58,0.45)
                {All eight update paths diverge from $X$\\above the dashed line};
        \end{scope}
        \begin{scope}[xshift=7.85cm]
            \node[font=\bfseries] at (3.58,6.15) {Hard Updates};
            \node[font=\small] at (3.58,5.80) {$V_i[1{:}i-1]=X[1{:}i-1],\ V_i[i]\ne X[i]$};
            \coordinate (h-0-0) at (3.575000,5.050000);
            \coordinate (h-1-0) at (1.644689,4.580000);
            \coordinate (h-1-1) at (4.669689,4.580000);
            \coordinate (h-2-0) at (1.157032,4.110000);
            \coordinate (h-2-2) at (4.062657,4.110000);
            \coordinate (h-2-3) at (5.992968,4.110000);
            \coordinate (h-3-1) at (1.400861,3.640000);
            \coordinate (h-3-4) at (3.102582,3.640000);
            \coordinate (h-3-5) at (4.425861,3.640000);
            \coordinate (h-3-7) at (6.236797,3.640000);
            \coordinate (h-4-2) at (1.278946,3.170000);
            \coordinate (h-4-8) at (2.980668,3.170000);
            \coordinate (h-4-10) at (4.184572,3.170000);
            \coordinate (h-4-11) at (5.144647,3.170000);
            \coordinate (h-4-14) at (6.114882,3.170000);
            \coordinate (h-5-5) at (1.339903,2.700000);
            \coordinate (h-5-17) at (3.041625,2.700000);
            \coordinate (h-5-20) at (3.646117,2.700000);
            \coordinate (h-5-21) at (4.364903,2.700000);
            \coordinate (h-5-23) at (5.205605,2.700000);
            \coordinate (h-5-29) at (6.175840,2.700000);
            \coordinate (h-6-10) at (1.309425,2.230000);
            \coordinate (h-6-34) at (3.011146,2.230000);
            \coordinate (h-6-40) at (3.615638,2.230000);
            \coordinate (h-6-42) at (4.215050,2.230000);
            \coordinate (h-6-43) at (4.753505,2.230000);
            \coordinate (h-6-46) at (5.175126,2.230000);
            \coordinate (h-6-58) at (6.145361,2.230000);
            \coordinate (h-7-21) at (1.324664,1.760000);
            \coordinate (h-7-69) at (3.026385,1.760000);
            \coordinate (h-7-81) at (3.630877,1.760000);
            \coordinate (h-7-84) at (3.961062,1.760000);
            \coordinate (h-7-85) at (4.349664,1.760000);
            \coordinate (h-7-87) at (4.768745,1.760000);
            \coordinate (h-7-93) at (5.190365,1.760000);
            \coordinate (h-7-117) at (6.160600,1.760000);
            \coordinate (h-8-42) at (1.317045,1.290000);
            \coordinate (h-8-138) at (3.018766,1.290000);
            \coordinate (h-8-162) at (3.623258,1.290000);
            \coordinate (h-8-168) at (3.953442,1.290000);
            \coordinate (h-8-170) at (4.222670,1.290000);
            \coordinate (h-8-171) at (4.476658,1.290000);
            \coordinate (h-8-174) at (4.761125,1.290000);
            \coordinate (h-8-186) at (5.182746,1.290000);
            \coordinate (h-8-234) at (6.152981,1.290000);
            \coordinate (h-2-1) at (2.251721,4.110000);
            \coordinate (h-3-0) at (0.793829,3.640000);
            \coordinate (h-3-2) at (2.007893,3.640000);
            \coordinate (h-3-3) at (2.495550,3.640000);
            \coordinate (h-3-6) at (5.629765,3.640000);
            \coordinate (h-4-0) at (0.671914,3.170000);
            \coordinate (h-4-1) at (0.915743,3.170000);
            \coordinate (h-4-3) at (1.642149,3.170000);
            \coordinate (h-4-4) at (1.885978,3.170000);
            \coordinate (h-4-5) at (2.129807,3.170000);
            \coordinate (h-4-6) at (2.373636,3.170000);
            \coordinate (h-4-7) at (2.617464,3.170000);
            \coordinate (h-4-9) at (3.343871,3.170000);
            \coordinate (h-4-12) at (5.507851,3.170000);
            \coordinate (h-4-13) at (5.751679,3.170000);
            \coordinate (h-4-15) at (6.478086,3.170000);
            \coordinate (h-5-0) at (0.610957,2.700000);
            \coordinate (h-5-1) at (0.732872,2.700000);
            \coordinate (h-5-2) at (0.854786,2.700000);
            \coordinate (h-5-3) at (0.976700,2.700000);
            \coordinate (h-5-4) at (1.098615,2.700000);
            \coordinate (h-5-6) at (1.581192,2.700000);
            \coordinate (h-5-7) at (1.703107,2.700000);
            \coordinate (h-5-8) at (1.825021,2.700000);
            \coordinate (h-5-9) at (1.946935,2.700000);
            \coordinate (h-5-10) at (2.068850,2.700000);
            \coordinate (h-5-11) at (2.190764,2.700000);
            \coordinate (h-5-12) at (2.312678,2.700000);
            \coordinate (h-5-13) at (2.434593,2.700000);
            \coordinate (h-5-14) at (2.556507,2.700000);
            \coordinate (h-5-15) at (2.678421,2.700000);
            \coordinate (h-5-16) at (2.800336,2.700000);
            \coordinate (h-5-18) at (3.282914,2.700000);
            \coordinate (h-5-19) at (3.404828,2.700000);
            \coordinate (h-5-22) at (4.964316,2.700000);
            \coordinate (h-5-24) at (5.446893,2.700000);
            \coordinate (h-5-25) at (5.568808,2.700000);
            \coordinate (h-5-26) at (5.690722,2.700000);
            \coordinate (h-5-27) at (5.812636,2.700000);
            \coordinate (h-5-28) at (5.934551,2.700000);
            \coordinate (h-5-30) at (6.417128,2.700000);
            \coordinate (h-5-31) at (6.539043,2.700000);
            \draw[gray, dashed, line width=0.6pt] (0.48,3.405) -- (6.85,3.405);
            \draw[updatepath] (h-8-42) -- (h-7-21) -- (h-6-10) -- (h-5-5) -- (h-4-2) -- (h-3-1) -- (h-2-0) -- (h-1-0) -- (h-0-0);
            \draw[updatepath] (h-8-234) -- (h-7-117) -- (h-6-58) -- (h-5-29) -- (h-4-14) -- (h-3-7) -- (h-2-3) -- (h-1-1);
            \draw[updatepath] (h-8-138) -- (h-7-69) -- (h-6-34) -- (h-5-17) -- (h-4-8) -- (h-3-4) -- (h-2-2);
            \draw[updatepath] (h-8-186) -- (h-7-93) -- (h-6-46) -- (h-5-23) -- (h-4-11) -- (h-3-5);
            \draw[updatepath] (h-8-162) -- (h-7-81) -- (h-6-40) -- (h-5-20) -- (h-4-10);
            \draw[updatepath] (h-8-174) -- (h-7-87) -- (h-6-43) -- (h-5-21);
            \draw[updatepath] (h-8-168) -- (h-7-84) -- (h-6-42);
            \draw[updatepath] (h-8-171) -- (h-7-85);
            \draw[querypath] (h-8-170) -- (h-7-85) -- (h-6-42) -- (h-5-21) -- (h-4-10) -- (h-3-5) -- (h-2-2) -- (h-1-1) -- (h-0-0);
            \node[update] at (h-8-42) {};
            \node[leaflabel] at (h-8-42) {$V_1$};
            \node[update] at (h-8-234) {};
            \node[leaflabel] at (h-8-234) {$V_2$};
            \node[update] at (h-8-138) {};
            \node[leaflabel] at (h-8-138) {$V_3$};
            \node[update] at (h-8-186) {};
            \node[leaflabel] at (h-8-186) {$V_4$};
            \node[update] at (h-8-162) {};
            \node[leaflabel] at (h-8-162) {$V_5$};
            \node[update] at (h-8-174) {};
            \node[leaflabel] at (h-8-174) {$V_6$};
            \node[update] at (h-8-168) {};
            \node[leaflabel] at (h-8-168) {$V_7$};
            \node[update] at (h-8-171) {};
            \node[leaflabel] at (h-8-171) {$V_8$};
            \node[query] at (h-8-170) {};
            \node[font=\scriptsize\bfseries, text=black, anchor=north] at (h-8-170) {$X$};
            \node[meet] at (h-0-0) {};
            \node[meetlabel, anchor=south west] at (h-0-0) {$a_1$};
            \node[meet] at (h-1-1) {};
            \node[meetlabel, anchor=west] at (h-1-1) {$a_2$};
            \node[meet] at (h-2-2) {};
            \node[meetlabel, anchor=east] at (h-2-2) {$a_3$};
            \node[meet] at (h-3-5) {};
            \node[meetlabel, anchor=west] at (h-3-5) {$a_4$};
            \node[meet] at (h-4-10) {};
            \node[meetlabel, anchor=east] at (h-4-10) {$a_5$};
            \node[meet] at (h-5-21) {};
            \node[meetlabel, anchor=west] at (h-5-21) {$a_6$};
            \node[meet] at (h-6-42) {};
            \node[meetlabel, anchor=east] at (h-6-42) {$a_7$};
            \node[meet] at (h-7-85) {};
            \node[meetlabel, anchor=west] at (h-7-85) {$a_8$};
            \node[font=\scriptsize, align=center, text width=6.7cm] at (3.58,0.45)
                {Each $a_i=\operatorname{LCA}(X,V_i)$ lies\\at a different level};
        \end{scope}
    \end{tikzpicture}
    \caption{The dyadic structure over the key universe, with eight update keys $V_1,\ldots,V_8$ and the same query $X$ in both panels. The left panel shows exactly equally spaced updates, whose paths diverge from the query path at levels at most 3. In the hard distribution, the eight lowest common ancestors occupy distinct levels.}
    \label{fig:uniform-versus-hard-distribution}
\end{figure}
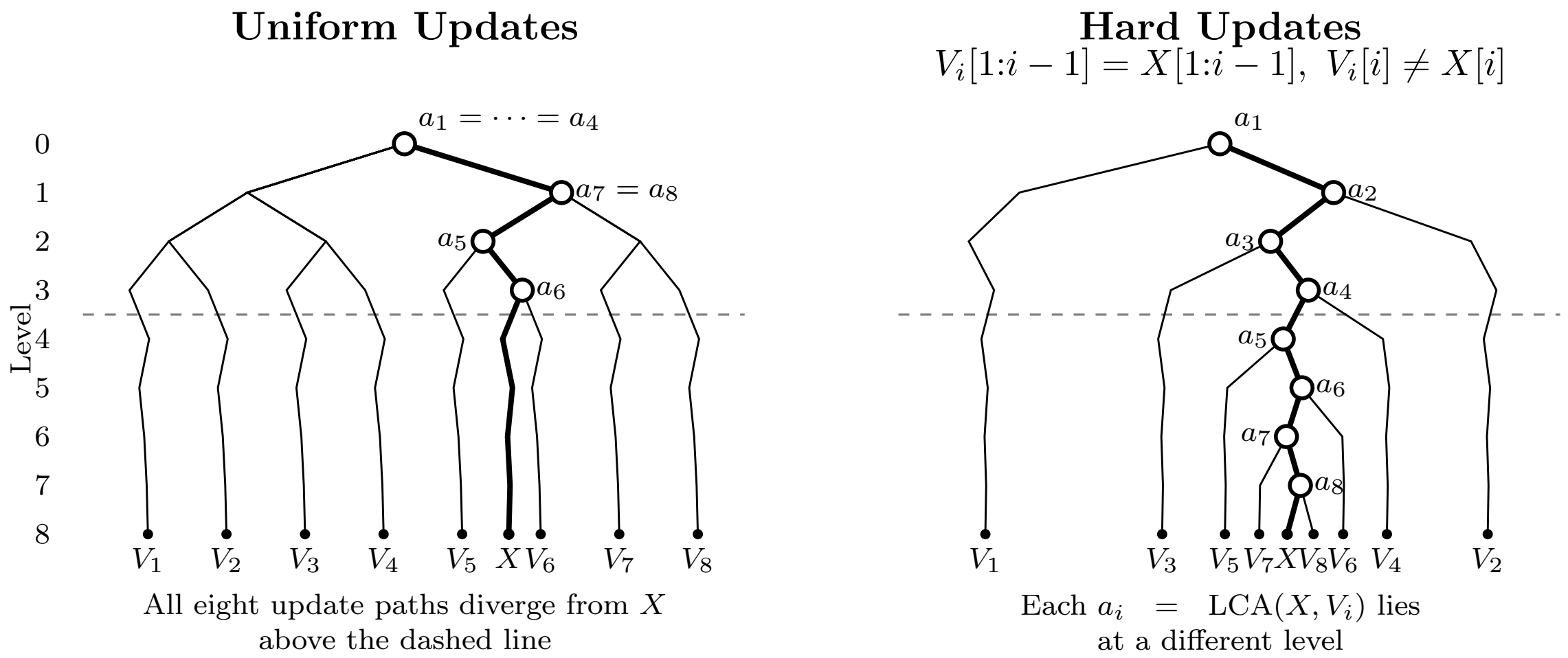

Formally, let $k$ be a fixed constant integer\footnote{The most natural choice is to let $k=2$, but we need it to be larger for technical reasons.}, and let $d:=\lfloor\log_k U\rfloor=\Theta(\log U)$. For an integer $y\in [k^d]$ and $1\le i\le d$, the \defn{$i$-th digit} of $y$ is defined as the $i$-th most significant entry in the $k$-ary representation of $y$, denoted as $y[i]$. The \defn{hard distribution} $\mathcal{D}_U$ consists of $d$ insertion updates $V_1,\dots,V_d$ and a query location $X$. We first sample the query $X$ uniformly from $[k^d]\subseteq [U]$, then for each $1\le i\le d$, update location $V_i$ is sampled uniformly from the set
\begin{align*}
    S_i=\{v\in [k^d]:v[1:i-1]=X[1:i-1],v[i]\ne X[i]\}.
\end{align*}
That is, $V_i$ agrees with $X$ on the largest $i-1$ digits, and differs from $X$ in the $i$-th digit.\footnote{In the formal proof (\cref{def:prefix-cylinder}), we actually use a slightly different set than $S_i$, where $V_i$ is only required to agree with $X$ on the first $i-1$ digits.}
In the $k$-ary trie, the keys $V_i$ branch off the path leading to $X$ at level $i-1$ (see \cref{fig:uniform-versus-hard-distribution}).
The task reduces to determining how many of them branch off from the left vs right approximately.

Over this distribution, many levels of the Dyadic CountSketch are useful, so the hack for the uniform distribution no longer works: The distance between $V_i$ and $X$ is roughly $k^{d-i+1}$, so during query, $V_i$ is found in some dyadic interval at level $\approx (\log k)\cdot i$. This means that each $V_i$ uses a distinct level, and if we want to keep a subset of levels that contains every $V_i$ with probability $0.99$, then this subset would have size $d=\Omega(\log U)$.

\subsubsection{Key Proof Step: Rank Estimation Requires Remembering $V_i[i]$ for Most $i$}

We now start proving the lower bound under the hard distribution. Recall that our goal is to use a linear sketch to estimate the rank of $X$, which can be written as $\Rk:=\sum_{i=1}^{d=\Theta(\log U)}\ind\{V_i< X\}$, up to an additive error of $\varepsilon\cdot d=O\bk*{ \sqrt{\log U}}$. By the definition of $\mathcal{D}_U$, the random variable $\ind\{V_i< X\}$ depends only on $V_i[i]$ and $X[i]$, so actually
\begin{align*}
    \Rk=\sum_{i=1}^{d}\ind\{V_i[i]\le X[i]\}.
\end{align*}

Let $\mathcal{A}$ denote a linear sketch. We will only use the fact that $\mathcal{A}$ is \emph{history independent} (that is, its memory state only depends on the final frequency vector). Other properties of linear sketches are not used here. We let $\mathcal{A}(a_1,\dots,a_m)$ denote the memory state of $\mathcal{A}$ after we start from the empty state and see insertions $a_1,\dots,a_m\in [U]$.

\paragraph{Intuition.} Our plan is to show that $\mathcal{A}$ contains $\Omega(\log^2U)$ bits of information about the input $V_1,\dots,V_d$. In particular, for each $V_i$ we only care about the digits starting from $V_i[i]$, as $V_i[1:i-1]=X[1:i-1]$ does not contain new information. Using a chain rule, we have (we also condition on $X$ for technical reasons)
\begin{align*}
    &\MIc{V_1,\dots,V_d}{\mathcal{A}(V_1,\dots,V_d)}{X}\\
    ={}&\sum_{i=1}^{d}\sum_{j=i}^{d}\underbrace{\MIc{V_i[j]}{\mathcal{A}(V_1,\dots,V_d)}{X,V_i[i:j-1],V_{i+1},\dots,V_d}}_{I_{i,j}}.
\end{align*}
We will use $I_{i,j}$ as a proxy for the amount of mutual information between the sketch $\mathcal{A}(V_1,\dots,V_d)$ and the $j$-th digit of $V_i$, $V_i[j]$.

In this section, we first show that $\mathcal{A}(V_1,\dots,V_d)$ must contain constant bits of information from $V_i[i]$ for an average $i\in [d]$ (that is, $I_{i,i}=\Theta(1)$ for a constant fraction of $1\le i\le d$). Intuitively, this is true because the additive error $\varepsilon\cdot d\le \sqrt{\log U}/\log k$ is sufficiently small compared to the standard deviation of $\Rk$ (which is $\Theta\bk*{\sqrt{\log U/\log k}}$) when $k$ is a large constant, so estimating $\Rk$ up to such precision is nearly as hard as computing $\Rk$ exactly, which requires remembering all of $V_i[i]$. A similar phenomenon also arises in communication complexity, e.g., for the Gap-Hamming-Distance problem~\citep{IndykW03,ChakrabartiR12}.

It may seem that what we are proving here is not very interesting: Even when this is true, it only gives an $\Omega(\log U)$ lower bound on the size of the linear sketch, whereas we want $\Omega(\log^2U)$. However, this is already quite nontrivial as it exploits many properties of linear sketches: Indeed, if $\mathcal{A}(V_1,\dots,V_d)$ can be any function, then by properties of our hard distribution, it can infer the value of $X$ when given $V_1,\dots,V_d$ (with possibly the exception of $X[d]$), and only has to store $O(\log\log U)$ bits of information, namely, the value of $R$.

\paragraph{Rephrasing the problem using conditional variance.} Following the technique in recent works in streaming algorithms and communication complexity \cite{BGW, SawettamalyaY26}, we use \defn{expected conditional variance} to measure the quality of the final memory state at answering the query. Specifically, if $\mathcal{A}(V_1,\dots,V_d)$ can be used to estimate $\Rk$ within an error of $\varepsilon\cdot d$ when given the query location $X$, then (see \cref{lem:accuracy-variance})
\begin{align*}
    \varbarc{\Rk}{X,\mathcal{A}(V_1,\dots,V_d)}\le O((\varepsilon\cdot d)^2).
\end{align*}

Here, $\varbarc{Z}{W}:=\E_{W}\Bk*{\Varc{Z}{W}}$. Intuitively, this is true because after learning $X,\mathcal{A}(V_1,\dots,V_d)$, we can estimate $R$ within an error of $\varepsilon\cdot d$ (with good probability), so the variance should be $O((\varepsilon\cdot d)^2)$.

As $\varepsilon= 1/\sqrt{\log U}=\Theta\bk*{1/\sqrt{d}}$, our goal in this section can be rephrased as: If
\begin{align}
    \varbarc{\Rk}{X,\mathcal{A}(V_1,\dots,V_d)}\le \eta\cdot d\label{equ:var_overview}
\end{align}
for some small constant $\eta>0$, then
\begin{align}
    I_{i,i}=\Omega(1) \text{ for a constant fraction of } i\label{equ:mi_overview}
\end{align}

\paragraph{The proof: Exploiting history independence.} We prove the lower bound by decomposing \eqref{equ:var_overview} into a summation of $d$ terms, then comparing each of the $d$ terms with the corresponding term in \eqref{equ:mi_overview}. We first simplify the definitions of $I_{i,i}$, using the fact that the linear sketches are history independent. That is, $\mathcal{A}(V_1,\dots,V_d),V_{i+1},\dots,V_{d}$ contains exactly the same information as $\mathcal{A}(V_1,\dots,V_{i}),V_{i+1},\dots,V_{d}$. Applying this to $I_{i,i}$ gives
\begin{align*}
    I_{i,i}&=\MIc{V_i[i]}{\mathcal{A}(V_1,\dots,V_d)}{X,V_{i+1},\dots,V_d} \\
    &=\MIc{V_i[i]}{\mathcal{A}(V_1,\dots,V_i)}{X,V_{i+1},\dots,V_d} \tag{history independence}\\
    &=\MIc{V_i[i]}{\mathcal{A}(V_1,\dots,V_i)}{X}.\tag{independence}
\end{align*}
For \eqref{equ:var_overview}, we use the law of total variance to decompose it into a summation. We stress that for general communication problems, this decomposition is normally quite complicated. However, we can obtain a rather clean decomposition using the history independence of the linear sketch (we omit the details):

\begin{align*}
    &\varbarc{\Rk}{X,\mathcal{A}(V_1,\dots,V_d)} \\
    ={}&\sum_{i=1}^{d}\varbarc{\ind\{V_i[i]\le X[i]\}+\textcolor{blue}{\Ec{\sum_{j=1}^{i-1}\ind\{V_j[j]\le X[j]\}}{X,\mathcal{A}(V_1,\dots,V_{i-1})}}}{X,\mathcal{A}(V_1,\dots,V_{i})}.
\end{align*}

\paragraph{Analyzing the variance using a weak sense of ``independence''.} If we do not have the complicated expectation term (marked in blue), then the proof is done: For each $1\le i\le d$, we have that
\begin{align}
    \varbarc{\ind\{V_i[i]\le X[i]\}}{X,\mathcal{A}(V_1,\dots,V_{i})}\ll 1\label{equ:var_single}
\end{align}implies $\MIc{V_i[i]}{\mathcal{A}(V_{1},\dots,V_{i})}{X}=\Omega(1)$. Intuitively, if $\mathcal{A}(V_{1},\dots,V_{i})$ carried no information about $V_i[i]$ conditioned on $X$, then even after seeing $X$ and $\mathcal{A}(V_{1},\dots,V_{i})$, $V_i[i]$ would remain uniform. The indicator $\ind\{V_i[i]\le X[i]\}$ would then have variance $\Theta(1)$ over a uniform $X[i]$, contradicting \eqref{equ:var_single}. Given this, \eqref{equ:mi_overview} then follows by a straightforward averaging argument: \eqref{equ:var_overview} implies that \eqref{equ:var_single} must hold for many $i$, which then implies \eqref{equ:mi_overview}.

Surprisingly, the pairwise comparison between the variance and the mutual information is possible \emph{even with the blue term present}. The key observation is that the blue term is independent of $X[i]$, while the first term in the variance $\ind\{V_i[i]\le X[i]\}$ heavily depends on $X[i]$. This creates a weak sense of ``independence'' between these two terms. Note that if these two terms were truly independent, then we could safely discard the blue term. For our weak ``independence'', it turns out that something of the same nature is possible (see \cref{cor:one-digit-info}). This concludes the key proof step.

\subsubsection{Remembering $V_i[i]$ Requires Also Remembering $V_j[i]$ for $1\le j<i$}
\label{sec:remember_prefix}

While the key proof step is a big step towards our goal, it does not complete the proof. Indeed, it cannot rule out the possibility that $\mathcal{A}(V_1,\dots,V_d)$ exactly stores $V_1[1],\dots,V_d[d]$ (using only $O(d)=O(\log U)$ bits), then uses them to answer the query. However, we can show that this is impossible for linear sketches. 

The intuition is that for each $1\le i\le d$, if the keys arrive in an arbitrary order (which does not affect the memory state at the end), a linear sketch $\mathcal{A}$ cannot tell which key is $V_i$. Therefore, the only way for $\mathcal{A}$ to remember $V_i[i]$ is to also remember the $i$-th digit of every other update key $V_1,\dots,V_{i-1}$ (for other keys $V_{i+1},\dots,V_d$, the $i$-th digit is equal to $X[i]$). Formally, we can show that $I_{j,i}\ge I_{i,i}$ for any $1\le j<i$ (recall that $I_{j,i}$ is a measure for the mutual information between the sketch $\mathcal{A}$ and $V_j[i]$).

This is proved again using the history independence of $\mathcal{A}$, and the fact that for $1\le j<i$, \emph{$V_j$ and $V_i$ are identically distributed conditioned on $V_1,\dots,V_{j-1}$}, so that $\mathcal{A}$ cannot tell the difference between $V_i$ and $V_j$ (when given $V_1,\dots,V_{j-1}$). See \cref{lem:information-transport} for the details.

Finally, since we have $I_{i,i}=\Omega(1)$ for a constant fraction of $1\le i\le d$ by \eqref{equ:mi_overview}, and since $I_{j,i}\ge I_{i,i}$ for any $1\le j<i$, we can conclude that
\begin{align}
    \label{equ:mi_special_case}
    \MI{V_1,\dots,V_d}{\mathcal{A}(V_1,\dots,V_d)\mid X}=\Omega(d^2),
\end{align}so $\mathcal{A}$ must use $\Omega(d^2)=\Omega(\log ^2U)$ bits of space. This concludes the lower bound proof for the key regime.

\subsection{Supporting Larger $n$: A White-Box Application of Augmented Indexing}
\label{sec:aug_index_overview}

Previously, we have shown that when $\varepsilon\le 1/\sqrt{\log U}$ and the $\ell_1$ norm is at most $\Theta(\log U)$, any linear sketch that solves rank estimation must use $\Omega(\log^2U)$ bits of space. In this section, we extend the lower bound to any $\ell_1$ norm $\log U\le n\le U^{\Omega(1)}$ (with asymptotically the same error), and prove a lower bound of $\Omega(\log^2U\cdot (\log n-\log\log U))$ bits. Later in \cref{sec:common-augmentation}, we will generalize this to arbitrarily large $n$.

\begin{figure}[H]
    \centering
    \begin{tikzpicture}[
        x=1cm,
        y=1cm,
        stream/.style={draw=black!70, line width=0.65pt,
            rounded corners=1.5pt},
        blocklabel/.style={font=\scriptsize\bfseries, align=center},
        rangelabel/.style={font=\scriptsize, align=center, text=black!62}
    ]
        \def\blockw{2.50}
        \def\blockh{1.22}

        \foreach \x/\range/\shade in {
            0/{$[0,U)$}/72,
            2.75/{$[U,2U)$}/58,
            5.50/{}/44,
            8.25/{$[i\cdot U,(i+1)U)$}/30,
            11.00/{}/17,
            13.75/{$[(m-1)\cdot U,mU)$}/5
        } {
            \draw[stream, fill=black!\shade]
                (\x,0.42) rectangle +(\blockw,\blockh);
            \node[rangelabel]
                at (\x+0.5*\blockw,0.12) {\range};
        }

        \node[blocklabel, text=white] at (1.25,1.03)
            {$\boldsymbol{2^{m-1}\cdot S^{(0)}}$};
        \node[blocklabel, text=white] at (4.00,1.03)
            {$\boldsymbol{2^{m-2}\cdot S^{(1)}}$};
        \node[blocklabel, text=white] at (6.75,1.03)
            {$\boldsymbol{\cdots}$};
        \node[blocklabel, text=black!85] at (9.50,1.03)
            {$\boldsymbol{2^{m-1-i}\cdot S^{(i)}}$};
        \node[blocklabel, text=black!85] at (12.25,1.03)
            {$\boldsymbol{\cdots}$};
        \node[blocklabel, text=black!85] at (15.00,1.03)
            {$\boldsymbol{S^{(m-1)}}$};
    \end{tikzpicture}
    \caption{The $m$-fold hard distribution.  Independent hard streams are
    shifted into disjoint sub-universes of size $U$.  Darker shades denote
    larger coefficients, which decrease geometrically from $2^{m-1}$ to
    $1$.}
    \label{fig:augmented_indexing}
\end{figure}
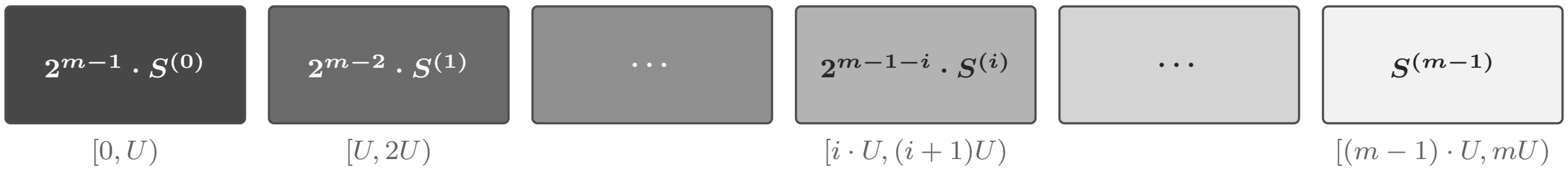

\paragraph{The $m$-fold hard distribution.} Recall that $\mathcal{D}_U$ is a hard distribution over universe $U$ that has $\ell_1$ norm $d\le \log U$, such that any linear sketch $\mathcal{A}$ solving rank estimation with error $\varepsilon\le 1/\sqrt{\log U}$ on $\mathcal{D}_U$ must use $\Omega(\log^2U)$ bits of space. Using $\mathcal{D}_U$, we show that for any $m\ge 1$, we can construct a new hard distribution $\mathcal{D}_{U,m}$ over universe $U\cdot m$ that has $\ell_1$ norm $d\cdot (2^m-1)$, such that any linear sketch solving rank estimation with error $\varepsilon/2$ must use $\Omega(m\cdot \log^2U)$ bits of space; see \cref{fig:augmented_indexing}. Given this, we can prove the lower bound for any universe size $U$ and $\ell_1$ norm $\log U\le n\le U^{\Omega(1)}$ by setting $m=\log n-\log\log U$, and considering the hard distribution $\mathcal{D}_{U/m,m}$.

Formally, a stream $S$ in the \defn{$m$-fold hard distribution} $\mathcal{D}_{U,m}$ is sampled as follows: We partition the universe $[U\cdot m]$ into $m$ sub-universes, each of size $U$, and sample a sub-stream on each sub-universe. For $0\le i\le m-1$, we sample a sub-stream $S^{(i)}$ from the hard distribution $\mathcal{D}_{U}$ with $X^{(i)}$ being the query location, then shift the value of each key by $U\cdot i$ (so that these keys fall in the $i$-th sub-universe $[U\cdot i,U\cdot (i+1))$). The final stream $S$ is the concatenation of $2^{m-1-i}$ copies of $S^{(i)}$ for each $0\le i\le m-1$.

\paragraph{The proof.} Let $\mathcal{A}$ denote the sketch, and let $\mathcal{A}(S)$ denote the final memory state of $\mathcal{A}$ after seeing a stream sampled from $\mathcal{D}_{U,m}$. Note that the $0$-th sub-universe takes up at least half of the total $\ell_1$ norm. Therefore, $\mathcal{A}(S)$ can be used to estimate the rank of $X^{(0)}$ amongst $S^{(0)}$ up to an additive error of $\varepsilon$ times the $\ell_1$ norm of $S^{(0)}$ (we assumed that $\mathcal{A}$ has error $\varepsilon/2$ over the entire stream $S$). Since $S^{(0)},X^{(0)}$ is sampled from the hard distribution $\mathcal{D}_{U}$, the lower bound \eqref{equ:mi_special_case} for the key regime gives us $\MI{S^{(0)}}{\mathcal{A}(S)}=\Omega(\log ^2U)$. Similarly, for every $0\le i\le m-1$, we can show that
\begin{align}
    \label{equ:info_aug_index}
    \MIc{S^{(i)}}{\mathcal{A}(S)}{S^{(<i)}}=\Omega(\log^2 U).
\end{align}
This holds because given $S^{(<i)}$, we can subtract the first $i$ sub-universes from $\mathcal{A}(S)$, then solve the $i$-th sub-universe using the resulting sketch (where the $i$-th sub-universe takes up at least half of the remaining mass). A chain rule over these bounds then gives \begin{align}
    \label{equ:mi_aug_index}
    \MI{S^{[m]}}{\mathcal{A}(S)}=\Omega(m\cdot \log ^2U),
\end{align}which means that $\mathcal{A}$ must use $\Omega(m\cdot \log ^2U)$ bits of space.

\paragraph{Intuition: Relations to Augmented Indexing.} Our proof borrows some intuition from \cite{Nelson26}. Their $\Omega(\varepsilon^{-1}\log(\varepsilon n)\log(\varepsilon U/\log(\varepsilon n)))$ lower bound on rank estimation is also shown by creating $\Theta(\log(\varepsilon n))$ independent copies of a hard stream, then concatenating the copies while assigning exponentially decreasing weights. The lower bound is then proven by studying a related one-way communication problem called Augmented Indexing~\citep{MiltersenNSW98,BarYossefJKK04}. In this problem, Alice sees $m$ items $x_0,\dots,x_{m-1}\in [U]$, and Bob sees an index $0\le i\le m-1$ and $x_0,\dots,x_{i-1}$ and wants to know $x_i$. Augmented Indexing is interesting in that, while Bob's answer is only $\log U$ bits, Alice has to send $\Omega(m\cdot \log U)$ bits of information because she does not know $i$.

Although our proof did not use communication complexity, it is essentially a white-box application of Augmented Indexing: When viewing each $S^{(i)}$ as $x_i$ and $\mathcal{A}(S)$ as Alice's message, \eqref{equ:info_aug_index} essentially says that Alice has to send $\Omega(\log^2U)$ bits of information about each $x_i$, and \eqref{equ:mi_aug_index} is essentially what the lower bound on Augmented Indexing gives us.

\subsection{Generalizing to Any $\varepsilon$}

We now have a lower bound of $\Omega(\log^2 U\cdot \log n)$ bits when $\varepsilon\le 1/\sqrt{\log U}$ (and $\log ^2 U\le n\le \poly U$). In this section, we generalize to any $\varepsilon$.

\paragraph{An algorithmic reduction for large $\varepsilon$.} When $1/\sqrt{\log U}\le  \varepsilon<1$, we prove the lower bound by presenting an algorithmic reduction: For any such $\varepsilon$ and any $n$, if there exists an algorithm $\mathcal{A}$ that solves rank estimation on parameters $\varepsilon$ and $n$ and uses $o(\varepsilon^{-2}\log U\log n)$ bits of space, then we can construct an algorithm $\mathcal{A}'$ that is correct on the same $n$ with an error of $ O(1/\sqrt{\log U})$ and uses $o(\log ^2U\log n)$ bits of space (contradicting the previous lower bound \eqref{equ:mi_aug_index}). We assume access to free randomness.

Our new algorithm $\mathcal{A}'$ is defined as follows: Randomly partition the universe into $K=\Theta\bk*{\varepsilon^2\log U}$ subsets. Create $K$ copies of $\mathcal{A}$, each responsible for a subset. When updating coordinate $x$, we update the copy of $\mathcal{A}$ responsible for $x$. A rank query on $x$ is answered by summing up the estimations to rank of $x$ in every copy of $\mathcal{A}$.

To see why $\mathcal{A}'$ is correct on the hard distribution, consider the case where $\mathcal{A}$ is an unbiased estimator (the general case is handled in \cref{sec:large-error-lb}). When the total $\ell_1$ norm is $n$, each copy of $\mathcal{A}$ has an expected $\ell_1$ norm of $n/K$, and thus contributes an error of magnitude $\varepsilon n/K$ (with constant probability). Since $\mathcal{A}$ is an unbiased estimator, these $K$ errors are independent, so their sum concentrates\footnote{We are actually cheating a little bit here. When all updates in the stream are on the same coordinate, only one copy of $\mathcal{A}$ is useful and the sum of errors does not concentrate. However, we only have to show that $\mathcal{A}'$ solves the hard distribution $\mathcal{D}_{U,m}$, which does not have this issue.}, giving us a total error of $\sqrt{K}\cdot \varepsilon n/K=\varepsilon n/\sqrt{K}\ll n/\sqrt{\log U}$.

Finally, if each copy of $\mathcal{A}$ uses $o(\varepsilon^{-2}\log U\log n)$ bits of space, then $\mathcal{A}'$ uses
\begin{align*}
    &K\cdot o(\varepsilon^{-2}\log U\log n)={}o(\log ^2U\log n)
\end{align*}
bits of space, which is a contradiction.

\paragraph{Concatenating copies of the hard distribution for small $\varepsilon$.} On the other side, when $\varepsilon\ll 1/\sqrt{\log U}$, we construct a lower bound distribution on which any correct sketch must use $\Omega(\varepsilon^{-1}\log ^{1.5}U\log n)$ bits of space.

The idea for the hard distribution is basically the same as \cref{sec:aug_index_overview}, except without the exponentially decreasing weights. We partition the universe $U$ evenly into $t=\Theta\bk*{1/\bk*{\varepsilon\sqrt{\log U}}}$ sub-universes, and place a hard stream $S^{(i)},X^{(i)}$ in each sub-universe (each hard stream has $\ell_1$ norm $n/t$).

Suppose that a sketch $\mathcal{A}$ is correct for parameters $\varepsilon$ and $n$. Since the first sub-universe now takes up $1/t$ of all the mass instead of half, the sketch $\mathcal{A}$ can estimate the rank of $X^{(1)}$ amongst $S^{(1)}$ up to an error of $\varepsilon\cdot t\ll 1/\sqrt{\log U}$. As such, \eqref{equ:mi_aug_index} says that it must store $\Omega(\log^2(U/t)\log (n/t))$ bits of information about the first sub-stream $S^{(1)}$, and the same holds for all other sub-streams. The total space lower bound is thus
\begin{align*}
    &\Omega(\log^2(U/t)\log (n/t))\cdot t.
\end{align*}
In the normal regimes where $\log (U/t)=\Omega(\log U)$ and $\log(n/t)=\Omega(\log n)$, this bound becomes
\begin{align*}
    \Omega(\varepsilon^{-1}\log ^{1.5}U\log n),
\end{align*}which is exactly what we need. The corner cases are handled in \cref{sec:generalizing-regimes}. They are slightly different, but do not contain fundamentally different ideas.

\paragraph{Paper organization.} We provide a technical overview of our lower bound in \cref{sec:overview}. We first formally prove the lower bound for linear sketches in the regime where $\varepsilon=\Theta(1/\sqrt{\log U})$ and $n=\Theta(\log U)$ in \cref{sec:key-regime}, then generalize this to any $0<\varepsilon<1/8$ and $n\ge \log U$ in \cref{sec:generalizing-regimes}. Our lower bounds are then lifted to general streaming algorithms in \cref{sec:streaming-transfer}. Finally, we formally present our upper bounds in \cref{sec:upper-bounds}, with additional details in \cref{app:derandomization-details}.
\section{Preliminaries}
\label{sec:preliminaries}

\subsection{Linear Sketches}
\label{sec:prelim-sketches}

Our lower bounds hold for linear sketches, and all of our algorithms are
linear sketches.  We first give the formal definition of a randomized linear sketch, following Jiang, Liu, and Yu~\citep[Definition~3.1.3]{JiangLY26}.

\begin{definition}[Linear sketch]
\label{def:additive-sketch}
A \emph{randomized linear sketch} draws a public random \emph{seed} $\rho$,
independent of the input and the query, which fixes an abelian group $G_\rho$,
an item map $\mathcal A_\rho:[U]\to G_\rho$ and a decoder $g_\rho$.  On a
frequency vector $f\in\Z^U$, it stores the \emph{state}
\[
  A_\rho(f)\coloneqq\sum_{u\in[U]}f_u\,\mathcal A_\rho(u)\in G_\rho,
\]
and answers a query $x$ with $g_\rho(A_\rho(f),x)$.  Its \emph{space} is the
least $s$ such that, for every seed $\rho$, the state takes at most $2^s$
values.  Its \emph{seed length} is the number of bits of $\rho$.
\end{definition}

An update $(i,\Delta)$ adds $\Delta\,\mathcal A_\rho(i)$ to the state, so a
linear sketch can be maintained over a turnstile stream.  The space of a linear
sketch does not count the seed, so our lower bounds hold even if random bits
are free.  In the streaming literature, however, the space of a randomized
algorithm includes its random bits~\citep[Definition~6]{LiNW14},
\citep[Section~2]{AHLW16}, since the algorithm has to store them to process
the updates.  So the seed length also matters, and we typically want it to be
smaller than the space.  All of our algorithms store their seeds, and the space
bounds we prove for them include the seed.

\subsection{Information Theory}
\label{sec:prelim-information}

Our lower bound proofs are based by information arguments. We now recall the notions and facts we use.

All logarithms are base two, so all quantities below are in
bits~\citep{CoverT06}.  The \emph{entropy} of a random variable $A$ and the
\emph{conditional entropy} of $A$ given $B$ are
\[
  H(A)\coloneqq\sum_a\Prb[A=a]\log\frac1{\Prb[A=a]},
  \qquad
  H(A\mid B)\coloneqq\sum_b\Prb[B=b]\,H(A\mid B=b),
\]
where $H(A\mid B=b)$ is the entropy of $A$ conditioned on $B=b$.  The
\emph{mutual information} of $A$ and $B$, and its conditional version given
$C$, are
\[
  \MI{A}{B}\coloneqq H(A)-H(A\mid B),
  \qquad
  \MIc{A}{B}{C}\coloneqq H(A\mid C)-H(A\mid B,C).
\]
As distance measures between two distributions $P,Q$ on a finite set, we use
the \emph{relative entropy} and the \emph{total variation distance},
\[
  \KL(P\,\|\,Q)\coloneqq\sum_xP(x)\log\frac{P(x)}{Q(x)},
  \qquad
  \TV(P,Q)\coloneqq\frac12\sum_x\bigl|P(x)-Q(x)\bigr|.
\]
We use the following standard facts.

\begin{fact}[Chain rule~{\citep[Theorem~2.5.2]{CoverT06}}]
\label{fact:chain-rule}
For random variables $A_1,\ldots,A_m,B,C$,
\[
  \MIc{A_1,\ldots,A_m}{B}{C}
  =\sum_{i=1}^m\MIc{A_i}{B}{C,A_1,\ldots,A_{i-1}}.
\]
\end{fact}

\begin{fact}[Data processing~{\citep[Theorem~2.8.1]{CoverT06}}]
\label{fact:data-processing}
For random variables $A,B,C$ and a function $B'$ of $(B,C)$,
$\MIc{A}{B'}{C}\le\MIc{A}{B}{C}$.
\end{fact}

\begin{fact}[Fano's inequality~{\citep[Theorem~2.10.1]{CoverT06}}]
\label{fact:fano}
Let $Z\in\{0,1\}$ be uniform and independent of $C$, and let $\widehat Z$ be a
function of $(B,C)$ with $\Prb[\widehat Z\ne Z]\le\delta\le1/2$.  Then
$\MIc{Z}{B}{C}\ge1-h_2(\delta)$, where
$h_2(\delta)\coloneqq-\delta\log\delta-(1-\delta)\log(1-\delta)$.
\end{fact}

\begin{fact}[Mutual information and relative entropy~{\citep[Section~2.3]{CoverT06}}]
\label{fact:mi-kl}
For random variables $A,B$, let $P$ be the distribution of $A$, and $P_b$ the
distribution of $A$ conditioned on $B=b$.  Then
\[
  \MI{A}{B}=\sum_b\Prb[B=b]\,\KL(P_b\,\|\,P).
\]
\end{fact}

\begin{fact}[Pinsker's inequality~{\citep[Lemma~11.6.1]{CoverT06}}]
\label{fact:pinsker}
For distributions $P,Q$ on a finite set,
$\TV(P,Q)^2\le\frac{\ln2}2\KL(P\,\|\,Q)$.
\end{fact}

\section{Lower Bound for the Key Regime}
\label{sec:key-regime}
This section and \cref{sec:generalizing-regimes} prove the optimal space lower bound for linear sketches that answer rank queries. We define rank queries formally below. 

\begin{definition}[Rank query]
\label{def:rank-query}
For a frequency vector $f$ on $[U]$ and $x\in\{0,\ldots,U\}$, the rank query at
$x$ asks for the mass of the prefix $[0,x)$, or  \[R_f(x)=\sum_{u<x}f_u. \]
\end{definition}

The main theorem for this section is that we need $\Omega(\log ^2 U)$ bits to approximate a rank query up to an additive error of $\varepsilon=\Theta(1/\sqrt{\log U})$ times the $\ell_1$ mass, given that there can be at most $\log U$ items, with sufficiently small constant failure probability. To support extension to other regimes, we state the theorem as a lower bound on the mutual information between the input and the memory state.

Note that throughout this section and \cref{sec:generalizing-regimes}, all
lower bounds are for (possibly) randomized linear sketches, formally defined
below, and are proved via information lower bounds conditioned on the public
random seed. As a result, the seed is not counted, so the lower bounds hold
even when random bits are free.

\subsection{Definitions and the Main Theorem}
\label{sec:hard-information}

As in the model, $0<\varepsilon\le1/8$.
Throughout this section, we fix $\eta_0=2^{-10}$ and a sufficiently large power of two $k$ with $k^{-1}\le\eta_0/8$.

In the following, we assume up to a constant loss that the universe size $U$ is at least $k^d$ for some variable $d$. We will consider each of the first $k^d$ integers in $[U]$ as a string of length $d$, with each element in $[k]$. The mapping is simply by lexicographical order, with smaller integers in $[U]$ mapped to smaller strings. We use $X[a:b]$ to represent the $a$-th to $b$-th elements in the string, indexed starting from one, where the most significant digits come first. 

We now introduce the hard distribution of the input for linear sketches.
\begin{definition}[Hard distribution]
\label{def:prefix-cylinder}
Choose $X$ uniformly from $[k]^d$. Given $X$, independently choose
$V_i$ uniformly among the strings agreeing with $X$ in their first
$i-1$ digits, for $i=1,\ldots,d$.
\end{definition}
Here $X$ is the query and $V_1,\ldots,V_d$ are the updates: the stream inserts one copy of each $V_i$, and the query is then made at $X$.

Instead of only considering the rank query at $X$, we prove our lower bound for a more general family of queries. This is to make it easier for us to generalize the lower bound beyond the key regime.

We now define the family of hard functions under this distribution. This family of functions is supposed to be hard for linear sketches of the frequency vector formed by adding one item each for $V_i(1\leq i \leq d)$. The (possibly randomized) linear sketches are explicitly described as follows. The formal definition of linear sketches contains noise, to better fit into our proof framework. Explicitly, the noise is always the contribution of other items that are not yet canceled out to the sketch, and can always be considered zero in the key regime. 

\begin{definition}[Linear sketch with noise]
\label{def:linear-sketch}
Let $\rho$ be the public randomness of the sketching algorithm independent of $(X,V_{1:d})$. Given $\rho$, we fix an
abelian group $G_\rho$, and a map $\mathcal A_\rho:[k]^d\to G_\rho$.
For technical reasons (specifically, we need this when we generalize the lower bound to support arbitrary $n\ge \log U$), the linear sketches may contain an independent noise. Explicitly, let $N\in G_\rho$ be a noise independent of $(X,V_{1:d})$ given $\rho$.
For any tuple of items, write the memory state of the linear sketch after seeing $v_1,\dots,v_m$ to be
\[
  A_\rho(v_1,\ldots,v_m;N)
  \coloneqq N+\sum_{t=1}^m\mathcal A_\rho(v_t).
\]
In particular, $A_\rho(V_{1:i};N)$ is the state of the first $i$ items
with noise $N$, and $A_\rho(V_{1:0};N)=N$.
\end{definition}

\paragraph{History independence.}
As in the overview, the proof never uses any property of the abelian group $G_\rho$ itself. It only uses that the sketch is \emph{history independent}, in the following two senses. First, the state does not depend on the order of the items: if $(v'_1,\ldots,v'_m)$ is a permutation of $(v_1,\ldots,v_m)$, then
\begin{equation}
  A_\rho(v'_1,\ldots,v'_m;N)=A_\rho(v_1,\ldots,v_m;N).
  \label{eq:hi-order}
\end{equation}
Second, a known item can be cancelled from or added to the state: there are functions $\mathrm{del}_\rho$ and $\mathrm{ins}_\rho$, depending only on $\rho$, such that
\begin{equation}
\begin{aligned}
  A_\rho(v_1,\ldots,v_{m-1};N)&=\mathrm{del}_\rho\bigl(A_\rho(v_1,\ldots,v_m;N),v_m\bigr),\\
  A_\rho(v_1,\ldots,v_m;N)&=\mathrm{ins}_\rho\bigl(A_\rho(v_1,\ldots,v_{m-1};N),v_m\bigr).
\end{aligned}
  \label{eq:hi-cancel}
\end{equation}
For linear sketches, $\mathrm{del}_\rho$ and $\mathrm{ins}_\rho$ subtract and add $\mathcal A_\rho(v_m)$, respectively.

We define the signed rank query below, which is similar to rank queries. We require the random sign for future applications in \cref{sec:generalizing-regimes}.

\begin{definition}[Sign function, signed rank and approximation]

For each $\rho$, we fix a function $\sigma_{\rho}: [k]^d \to \{-1,1\}$. We define the signed rank query at $X$ by
\[
  R\coloneqq\sum_{i=1}^d\sigma_\rho(V_i)\ind\{V_i<X\}
\]

Most of the time, $\ind\{V_i<X\}$ is exactly the $i$-th digit comparison defined by $B_i\coloneqq \ind\{V_i[i] \leq X[i]\}$. The easier to estimate form is the approximation of $R$ by these digit comparisons, defined as
\[\widetilde{R}\coloneqq\sum_{i=1}^d\sigma_\rho(V_i)B_i \]
\end{definition}

The main theorem, presented below, shows the hardness of approximation of $R$ for linear sketches.
\begin{theorem}[Decoding forces information]
\label{thm:hard-decoding-information}
For absolute constants $a_0,\delta_0,c>0$, if
\begin{equation}
  \Prb[|T-R|>a_0\sqrt d]\le\delta_0
  \label{eq:hard-decoding-hypothesis}
\end{equation}
then
\begin{equation}
  \MIc{V_{1:d}}{A_\rho(V_{1:d};N)}{\rho}\ge cd^2.
  \label{eq:hard-decoding-conclusion}
\end{equation}
Here, the probability averages over the hard instance, public randomness, and
noise.
\end{theorem}

Here we notice that the theorem holds for \emph{any chosen} functions $\sigma_{\rho}$. This also includes the normal rank query, where all the functions $\sigma_{\rho}$ take the sign $1$.

\paragraph{Differences from the overview.}
\Cref{thm:hard-decoding-information} differs from the overview in three respects. These differences only make the proof calculations more fluent or the statement more general, and all the intuitions stay essentially unchanged. For the key regime itself we only need $N=0$ and $\sigma_\rho\equiv1$ (\cref{cor:boundary-no-augmentation}); the general form is what allows us to reuse the theorem beyond the key regime.
\begin{itemize}
    \item \emph{Distribution.} In the overview, $V_i$ is required to differ from $X$ in the $i$-th digit, so that the rank is exactly $\sum_{i}\ind\{V_i[i]\le X[i]\}$. \Cref{def:prefix-cylinder} drops this requirement, so that $V_i[i]$ is uniform and independent of $X[i]$, and all the remaining digits of $V_i$ are uniform. The price is that $B_i$ may differ from $\ind\{V_i<X\}$ when $V_i[i]=X[i]$, which happens with probability $1/k$. We therefore work with $\widetilde R$ and lose $\sqrt{d/k}$ in the conditional standard deviation (\cref{lem:rank-to-digits}), which is acceptable because $k$ is a large constant.
    \item \emph{Signed rank.} The query is the signed rank $R$ for an arbitrary sign function $\sigma_\rho$, which may depend on the public randomness, instead of the rank itself.
    \item \emph{Noise.} The sketch may use public randomness $\rho$ and contain a noise $N$ that is independent of the input given $\rho$, whereas the overview considers a deterministic sketch without noise.
\end{itemize}

\subsection{Tools for Proving the Theorem}

\subsubsection{Digitwise Information}
Our main proof strategy of the main theorem is to partition the main mutual information into a summation of $\Omega(d^2)$ \emph{digitwise information} using chain rule (where each term intuitively measures the amount of mutual information between the memory state and $V_i[j]$ for some $1\le i\le j$), and prove that a constant fraction of those needs to be $\Omega(1)$. Explicitly, the chain rule says that
\begin{equation}
  \MIc{V_{1:d}}{A_\rho(V_{1:d};N)}{\rho}
  =\sum_{i=1}^d\sum_{j=1}^d
    \MIc{V_i[j]}{A_\rho(V_{1:d};N)}
        {\rho,V_i[1:j-1],V_{i+1:d}}.
\label{eq:information-chain-rule}
\end{equation}
Regarding the definition of the hard distribution, we notice that some terms in this expansion are $0$. Indeed, for $i<d$, revealing $V_d$ exposes the first $d-1$ digits of $X$ and therefore the first $i-1$ digits of each of the coordinates $V_i$. This indicates that the terms in the summation with $j<i<d$ are $0$, so we can ignore them. We also ignore the terms with $i=d$ and $j<d$ for technical reasons. The other terms play a vital role in our analysis, and we call them \emph{digitwise information} from now on.

\begin{definition}[digitwise information]
For each $i \leq j$, define the digitwise information $I_{i,j}$ as
\[
  \MIc{V_i[j]}{A_\rho(V_{1:d};N)}
      {\rho,V_i[1:j-1],V_{i+1:d}}.
\]
\end{definition}
Intuitively, we use $I_{i,j}$ to measure the information the sketch $A_\rho(V_{1:d};N)$ stores about $V_i[j]$. The following claim is the above chain rule plus the fact that the other terms are nonnegative, by the nonnegativity of mutual information.

\begin{claim}
\label{cor:information-triangle}
\[
  \MIc{V_{1:d}}{A_\rho(V_{1:d};N)}{\rho}
  \ge\sum_{i=1}^d\sum_{j=i}^d I_{i,j}.
\]
\end{claim}

For convenience, we simplify the definition of $I_{i,j}$ using the history independence of linear sketches and properties of the hard distribution.
\begin{lemma}[Suffix independence]
\label{lem:partial-state-information}
\[
  I_{i,j}=\MIc{V_i[j]}{A_\rho(V_{1:i};N)}{\rho,V_i[1:j-1]}.
\]
\end{lemma}
\begin{proof}
\[
  I_{i,j}
  =\MIc{V_i[j]}{A_\rho(V_{1:i};N)}
      {\rho,V_i[1:j-1],V_{i+1:d}}
  =\MIc{V_i[j]}{A_\rho(V_{1:i};N)}{\rho,V_i[1:j-1]}.
\]
Here the first identity uses history independence: given $V_{i+1:d}$, it allows us to cancel these items from $A_\rho(V_{1:d};N)$ and to add them back \eqref{eq:hi-cancel}, so the two states determine each other; the second one uses the fact that, conditional on $(\rho,V_i[1:j-1])$, $V_{i+1:d}$ is independent of (the joint distribution of) all the other terms.
Indeed, $V_i[1:j-1]$ fixes $X[1:i-1]$; the first $i$ items use their own independent suffixes, while the later items use the remaining query digits and their own suffixes. Conditioning further on the digits $V_i[i:j-1]$ only fixes some of that item's suffix digits. Noise independence gives the same joint independence with $N$ included.
\end{proof}

\subsubsection{From Accuracy to Conditional Variance}
In this section, we define conditional variance and show that the accuracy requirement is equivalent, after adjusting constants, to requiring that conditioned on the final memory state, the answer has small variance. We also show that the variance of $R$ and $\widetilde{R}$ conditioned on the memory state are close to each other.

\begin{definition}[conditional variance]
For a square-integrable $Z$
and observation $C$, write
\[
  \varbarc{Z}{C}:=\E[\Var(Z\mid C)]
  =\E\bigl[(Z-\Ec{Z}{C})^2\bigr].
\]
This is the mean squared error of the best decoder, averaged over $C$.
\end{definition}

The lemma below shows that small $\Theta(d)$ variance of $R$ conditional on the sketch is equivalent to a high probability estimator of $R$ within small $\Theta(\sqrt d)$ additive error. The proof only uses the fact that $R$ itself has concentration around its expected value, so that we can bound the variance of $R$ even if we don't have any guarantee on how the algorithm behaves when it fails.

\begin{lemma}[Accuracy and conditional variance]
\label{lem:accuracy-variance}
For every $\eta>0$, there exist $a,\delta>0$ such that
\[
  \Prb[|T-R|>a\sqrt d]\le\delta \Rightarrow \varbarc{R}{X,A_\rho(V_{1:d};N),\rho}\le\eta d.
\]
Conversely, for every $a,\delta>0$, this bound with sufficiently
small $\eta>0$ ensures the same accuracy for the conditional mean.
\end{lemma}

\begin{proof}
Fix $\eta>0$ and put $\mu=\Ec{R}{X,\rho}$.
Given $(X,\rho)$, the summands in $R$ are independent and lie in
$[-1,1]$, so Hoeffding's inequality gives
\[
  \Prb[|R-\mu|>t\sqrt d\mid X,\rho]\le 2e^{-t^2/2}.
\]
Integrating this tail, choose a constant $\lambda\ge1$, depending
only on $\eta$, such that
\[
  \E\bigl[(R-\mu)^2\ind\{|R-\mu|>\lambda\sqrt d\}\bigr]
  \le \eta d/12.
\]
Clip $T$ to $[\mu-\lambda\sqrt d,\mu+\lambda\sqrt d]$ and
denote the result by $\widehat T$.

On $|R-\mu|\le\lambda\sqrt d$, clipping cannot increase the
error. Successful answers therefore contribute at most $a^2d$
to the mean squared error. Failed answers have squared error
at most $4\lambda^2d$ and occur with probability at most $\delta$,
so their contribution is at most $4\lambda^2\delta d$.
On the complementary event,
$|\widehat T-R|\le2|R-\mu|$, so its contribution is at most
$\eta d/3$.
Choosing $a^2\le\eta/3$ and $\delta\le\eta/(12\lambda^2)$ gives
$\E[(\widehat T-R)^2]\le\eta d$.
Since $\widehat T$ is determined by
$(X,A_\rho(V_{1:d};N),\rho)$, the conditional
variance is no larger.

Conversely, the conditional mean has mean squared error equal to
the conditional variance. If this is at most $\eta d$, Markov's
inequality bounds its failure probability at error $a\sqrt d$
by $\eta/a^2$. Taking $\eta=a^2\delta$ proves the claim.
\end{proof}

The next lemma allows us to pass from $R$ to $\widetilde R$
while controlling conditional variance.
Where not displayed explicitly, $\rho$ is suppressed from conditioning.

\begin{lemma}[From rank to digit comparisons]
\label{lem:rank-to-digits}
For every message $M$,
\begin{equation}
  \sqrt{\varbarc{\widetilde R}{X,M}}
  \le \sqrt{\varbarc{R}{X,M}}+\sqrt{d/k}.
  \label{eq:rank-to-digits}
\end{equation}
\end{lemma}

\begin{proof}
The difference $\widetilde R-R$ is
$\sum_{i=1}^d\sigma_\rho(V_i)\ind\{V_i[i]=X[i],\,V_i\ge X\}$.
Given $(X,\rho)$, its summands are independent and each has
variance at most $1/k$. Their sum therefore has expected
conditional variance at most $d/k$, and revealing $M$ can only
decrease this quantity. The triangle inequality for the residuals
$Z-\Ec{Z}{X,M}$ proves the claim.
\end{proof}

\subsection{Proof of \cref{thm:hard-decoding-information}}
By \cref{lem:accuracy-variance} and \cref{lem:rank-to-digits}, it suffices to prove the following
variance form of the information theorem.

\begin{lemma}[Variance form]
\label{lem:common-map-information}
There exist absolute constants $\eta,c>0$ such that if
\[
  \varbarc{R}{X,A_\rho(V_{1:d};N),\rho}
  \le\eta d,
\]
then
\begin{equation}
  \MIc{V_{1:d}}{A_\rho(V_{1:d};N)}{\rho}
  \ge cd^2.
  \label{eq:hard-variance-interface}
\end{equation}
\end{lemma}

\subsubsection{Partitioning the Variance into Digitwise Variance}
In this section, we partition the objective
\[\varbarc{\widetilde R}{X,A_\rho(V_{1:d};N),\rho}\]
into $d$ terms. Similar to digitwise information, we call these terms digitwise variance.

\begin{definition}[digitwise variance]
    For each level $i$, define the digitwise variance $\Var_i$ as
\[
\Var_i = \varbarc{
      \sigma_\rho(V_i) B_i+\Ec{\sum_{t=1}^{i-1}\sigma_\rho(V_t) B_t}{X,A_\rho(V_{1:i-1};N)}
    }{X,A_\rho(V_{1:i};N)}.
\]
\end{definition}

The next lemma formalizes this partition. While the digitwise variances seem very complicated, they have the nice property that the $i$-th term actually only depends on the first $i$ terms of the sequence $V$. This allows us to view it as one kind of chain rule, so that we can relate each term to the digitwise information later.

\begin{lemma}[Variance partition]
\label{lem:variance-partition}
We have the exact identity
\begin{equation}
  \varbarc{\widetilde R}{X,A_\rho(V_{1:d};N)}=\sum_{i=1}^d
    \Var_i.
  \label{eq:variance-partition}
\end{equation}
\end{lemma}

To prove \cref{lem:variance-partition}, we use the law of total variance. Normally, the law of total variance gives us something complicated and impossible to work with. Here, however, history independence allows us to cancel a known item from the sketch. This means that when we know (or condition on) some $V_i$ already, we can cancel $V_i$ from the sketch and remove the fixed term involving $V_i$ from the variance. After cancelling $V_i$, the preceding sum and state are independent of $V_i$ given $X$ and $\rho$, so $V_i$ can be further removed from the condition.

\begin{proof}
The law of total variance gives that, for random variables $Z,C,D$,
\begin{equation}
  \varbarc{Z}{C}
  =\varbarc{Z}{C,D}+\varbarc{\Ec{Z}{C,D}}{C}.
  \label{eq:total-variance-reveal}
\end{equation}

Fix $i\in\{1,\ldots,d\}$.  Apply \eqref{eq:total-variance-reveal} to
\[
  Z=\sum_{t=1}^{i}\sigma_\rho(V_t) B_t,\qquad
  C=\left(X,A_\rho(V_{1:i};N)\right),\qquad
  D=V_i.
\]
Notice that once $V_i$ is given, the signed comparison $\sigma_\rho(V_i) B_i$ is known, and history independence allows us to cancel and add
$V_i$ \eqref{eq:hi-cancel}, which makes the random variables
\[
  \left(X,A_\rho(V_{1:i};N),V_i\right)
  \quad\text{and}\quad
  \left(X,A_\rho(V_{1:i-1};N),V_i\right)
\]
equivalent.

Another observation is that given $X$ and the public variables, all the preceding items and noise are
independent of $V_i$, or written formally as
\begin{equation}
  \CIc{\left(\sum_{t=1}^{i-1}\sigma_\rho(V_t) B_t,
                 A_\rho(V_{1:i-1};N)\right)}
       {V_i}{X}.
  \label{eq:previous-pair-independent}
\end{equation}
Thus revealing $V_i$ does not change the preceding sum's conditional law
given $X$ and the preceding state.

Using these two observations, the first term of \eqref{eq:total-variance-reveal} becomes
\begin{equation}
  \begin{aligned}
  \varbarc{\sum_{t=1}^{i}\sigma_\rho(V_t) B_t}
      {X,A_\rho(V_{1:i};N),V_i}
  &=\varbarc{\sum_{t=1}^{i-1}\sigma_\rho(V_t) B_t}
      {X,A_\rho(V_{1:i-1};N),V_i}\\
  &=\varbarc{\sum_{t=1}^{i-1}\sigma_\rho(V_t) B_t}
      {X,A_\rho(V_{1:i-1};N)}.
  \end{aligned}
  \label{eq:variance-after-revelation}
\end{equation}
Here the first equality uses history independence, and the second uses \eqref{eq:previous-pair-independent}.

The inner prediction (the $\mathbb E[Z \mid C,D]$ term) in the second term of \eqref{eq:total-variance-reveal} is
\begin{equation}
  \begin{aligned}
  \Ec{\sum_{t=1}^{i}\sigma_\rho(V_t) B_t}
      {X,A_\rho(V_{1:i};N),V_i}
  &=\sigma_\rho(V_i) B_i+\Ec{\sum_{t=1}^{i-1}\sigma_\rho(V_t) B_t}
      {X,A_\rho(V_{1:i-1};N),V_i}\\
  &=\sigma_\rho(V_i) B_i+\Ec{\sum_{t=1}^{i-1}\sigma_\rho(V_t) B_t}
      {X,A_\rho(V_{1:i-1};N)}.
  \end{aligned}
  \label{eq:prediction-after-revelation}
\end{equation}
Here the first equality uses linearity of expectation, and the second one uses \eqref{eq:previous-pair-independent}.

Substituting \eqref{eq:variance-after-revelation} and \eqref{eq:prediction-after-revelation}
into \eqref{eq:total-variance-reveal} thus gives
\begin{multline*}
  \varbarc{\sum_{t=1}^{i}\sigma_\rho(V_t) B_t}{X,A_\rho(V_{1:i};N)}
  =\varbarc{\sum_{t=1}^{i-1}\sigma_\rho(V_t) B_t}
      {X,A_\rho(V_{1:i-1};N)}\\
  +\varbarc{
    \sigma_\rho(V_i) B_i+\Ec{\sum_{t=1}^{i-1}\sigma_\rho(V_t) B_t}{X,A_\rho(V_{1:i-1};N)}
  }{X,A_\rho(V_{1:i};N)}.
\end{multline*}

That is, for every $i\in\{1,\ldots,d\}$,
\begin{equation}
  \varbarc{\sum_{t=1}^{i}\sigma_\rho(V_t) B_t}{X,A_\rho(V_{1:i};N)}=\varbarc{\sum_{t=1}^{i-1}\sigma_\rho(V_t) B_t}
      {X,A_\rho(V_{1:i-1};N)}+\Var_i.
  \label{eq:ltv-step}
\end{equation}
To prove \eqref{eq:variance-partition}, note that by definition of $\widetilde{R}$, the LHS is exactly equal to
\[
    \varbarc{\sum_{t=1}^{d}\sigma_\rho(V_t) B_t}{X,A_\rho(V_{1:d};N)}.
\]
Repeatedly applying \eqref{eq:ltv-step} thus gives
\[
    \varbarc{\sum_{t=1}^{d}\sigma_\rho(V_t) B_t}{X,A_\rho(V_{1:d};N)}=\varbarc{\sum_{t=1}^{d-1}\sigma_\rho(V_t) B_t}{X,A_\rho(V_{1:d-1};N)}+\Var_d
    =\dots=\sum_{i=1}^{d}\Var_i.\qedhere
\]

\end{proof}

\subsubsection{Key Proof Step: Small Digitwise Variance Implies Large Digitwise Information on the Diagonal}
After partitioning the variance into $d$ terms, our remaining goal is to show that: If $\sum_{i=1}^{d}\Var_i\le \eta d$, then $\sum_{i=1}^{d}\sum_{j=i}^{d}I_{i,j}\ge cd^2$. In this section, we prove that $\Var_i \leq \eta_0$ implies diagonal digitwise information $I_{i,i} \geq c_0$ for absolute constants $\eta_0,c_0>0$ (we later relate $\eta_0,c_0$ to $\eta,c$). We start with lower bounds on the diagonal digitwise information $I_{i,i}$ because these essentially describe the mutual information between the memory state and the digits $V_i[i]$, and $\widetilde R$ only depends on those digits of $V$. This lower bound is later extended to every term $I_{i,j}$ for $i\le j$.

\begin{theorem}
\label{cor:one-digit-info}
\label{eq:one-digit-info}
    Suppose that $\Var_i \leq \eta_0$. Then $I_{i,i} \geq c_0$.
\end{theorem}

\begin{proof}
We seek to rewrite the expressions of $\Var_i$ and $I_{i,i}$ so that both are conditioned on the same observation. Recall, from the definition and \cref{lem:partial-state-information}, that
\[
  \Var_i=\varbarc{
    \sigma_\rho(V_i) B_i+\Ec{\sum_{t=1}^{i-1}\sigma_\rho(V_t) B_t}
      {X,A_\rho(V_{1:i-1};N)}
  }{X,A_\rho(V_{1:i};N)},
\]
\[
  I_{i,i}=\MIc{V_i[i]}{A_\rho(V_{1:i};N)}
      {\rho,V_i[1{:}i-1]}.
\]
We first give a name to the prediction from the first $i-1$ items. Let
\[
  Y=\Ec{\sum_{t=1}^{i-1}\sigma_\rho(V_t) B_t}
       {X,A_\rho(V_{1:i-1};N)},
\]
so that $\Var_i=\varbarc{\sigma_\rho(V_i)B_i+Y}{X,A_\rho(V_{1:i};N)}$.
We compress this prediction into one letter simply because the following proof only requires $Y$ to be some function of $\rho$, the first $i-1$ query digits $X[1{:}i-1]$, and the state $A_\rho(V_{1:i-1};N)$ of the first $i-1$ items; its exact form does not matter.

The next two paragraphs simplify the two objectives so that they condition on almost the same object, namely $\bigl(\rho,X[1{:}i-1],A_\rho(V_{1:i};N)\bigr)$: we remove the query digits after $X[i]$ from the conditioning of $\Var_i$, and use the chain rule to move the conditioning of $I_{i,i}$ into the observation. The only remaining difference is that $\Var_i$ is also conditioned on the query digit $X[i]$.

\paragraph{Independence of the query suffix.}
Given $(\rho,X[1{:}i-1])$, the suffix $X[i:d]$ is independent of $(V_{1:i},N)$; the terms defining $Y$ use only the earlier query digits, so $Y$ depends on $X$ only through $X[1:i-1]$. Thus $X[i:d]$ is independent of $(V_i,A_\rho(V_{1:i};N),Y)$ given $(\rho,X[1{:}i-1])$. In particular, $X[i+1:d]$ can be removed from the conditioning of $\Var_i$:
\[
  \Var_i=\varbarc{\sigma_\rho(V_i)B_i+Y}{\rho,X[1{:}i-1],A_\rho(V_{1:i};N),X[i]}.
\]

\paragraph{Chain rule.}
Since $V_i[1{:}i-1]=X[1{:}i-1]$, and $V_i[i]$ and $(\rho,X[1{:}i-1])$ are independent (they are on completely different uniformly generated digits), $\MI{V_i[i]}{\rho,X[1{:}i-1]}=0$, and the chain rule gives
\[
  I_{i,i}=\MIc{V_i[i]}{A_\rho(V_{1:i};N)}{\rho,X[1{:}i-1]}
  =I\bigl(V_i[i];\rho,X[1{:}i-1],A_\rho(V_{1:i};N)\bigr).
\]

\paragraph{The common observation.}
Let $\Gamma=\bigl(\rho,X[1{:}i-1],A_\rho(V_{1:i};N)\bigr)$. Then what we are considering can be simplified as
\[
  \Var_i=\varbarc{\sigma_\rho(V_i)B_i+Y}{\Gamma,X[i]},
  \qquad I_{i,i}=\MI{V_i[i]}{\Gamma},
\]
where $B_i=\ind\{V_i[i]\le X[i]\}$. This allows us to view both $\Var_i$ and $I_{i,i}$ by conditioning on $\Gamma$.

Note that here we use one-letter notation for $Y$ and $\Gamma$ on purpose to indicate that we will not look deeper into their structures. However, we still want to extract out the minimal relationships used to prove this.
\paragraph{Some further facts we have.}
By definition, $V_i[i]$ and $X[i]$ are uniformly randomly chosen from $[k]$.

To deal with $Y$, the key step is that $X[i]$ is very new relative to it. Indeed, by the independence above, and since $X[i]$ is independent of $(\rho,X[1{:}i-1])$, the digit $X[i]$ is independent of $(\Gamma,V_i,Y)$.

We will then show how to intuitively solve the problem now.
\paragraph{Intuition.} As explained above, we can analyze what happens conditioned on $\Gamma$ (so we will omit it from the conditioning in this paragraph). We are going to show that a small variance indicates that the entropy of $V_i[i]$ is, on average over $\Gamma$, a constant smaller than its maximum $\log k$.

To begin with, let's see what happens when $Y=0$. It is simply $\varbarc{\sigma_\rho(V_i)B_i}{X[i]}\geq \varbarc{B_i}{X[i]}$ (since $B_i$ is always nonnegative, adding a sign only increases variance). However, we notice that $X[i]$ is still uniform now, so the small variance indicates that $V_i[i]$ is mostly concentrated on a smaller interval, or has much less entropy than the uniform distribution. This step can be finished by Pinsker's inequality, presented in the formal proof.

To handle $Y$, we simply try to eliminate it. This can be done by multiplying by a sign that depends only on $X[i]$ to make the term involving $Y$ have expectation zero (using independence), and eliminating it by Cauchy--Schwarz.
\paragraph{Proof of the key inequality.}
Fix $\Gamma$ and suppress it from the conditioning until we average over it below; by the facts above, $X[i]$ is still uniform on $[k]$ and independent of $(V_i,Y)$.
Define the sign $h(q)=1$ when $q\ge\frac k2$ and $-1$ otherwise. We have $\mathbb E[h(X[i])]=0$.
Define, for $v\in[k]$,
\[
  f(v)=\mathbb E\bigl[h(X[i])\ind\{v\le X[i]\}\bigr]
      =\frac{\min\{v,k-v\}}k\in[0,1/2].
\]
Write $\mathbb E_{X[i]}$ for the average over $X[i]$ alone, with $(V_i,Y)$ fixed.
Notice that
\[
\begin{aligned}
    \varbarc{\sigma_\rho(V_i)B_i+Y}{X[i]}
    &=\varbarc{(\sigma_\rho(V_i)B_i+Y)h(X[i])}{X[i]}\\
    &\ge \Var\!\left(\mathbb E_{X[i]}[(\sigma_\rho(V_i)B_i+Y)h(X[i])]\right)\\
    &=\Var(\sigma_\rho(V_i)f(V_i[i]))
    \ge\Var(f(V_i[i])).
\end{aligned}
\]
For the first inequality, put $W=(\sigma_\rho(V_i)B_i+Y)h(X[i])$ and apply Cauchy--Schwarz in $X[i]$:
\[
  \Var(\mathbb E_{X[i]} W)
  =\mathbb E\!\left[\left(\mathbb E_{X[i]}[W-\mathbb E[W\mid X[i]]]\right)^2\right]
  \le\mathbb E\bigl[(W-\mathbb E[W\mid X[i]])^2\bigr].
\]
The next equality uses $\mathbb E_{X[i]} h(X[i])=0$. The last inequality follows from $f\ge0$ and $|\sigma_\rho(V_i)|=1$, since
\[
  \Var(\sigma_\rho(V_i)f(V_i[i]))
  =\mathbb E[f(V_i[i])^2]-(\mathbb E[\sigma_\rho(V_i)f(V_i[i])])^2
  \ge\mathbb E[f(V_i[i])^2]-(\mathbb E[f(V_i[i])])^2.
\]
Averaging over $\Gamma$ therefore gives
\[
  \varbarc{f(V_i[i])}{\Gamma}\le\Var_i\le\eta_0.
\]

Finally, Pinsker's inequality finishes the proof. Since $V_i[i]$ is uniform,
\[
  \mathbb E[f(V_i[i])]=\frac14,\qquad
  \Var(f(V_i[i]))=\frac1{48}+\frac1{6k^2}\ge\frac1{48}.
\]
For each $\gamma$, let $P_\gamma$ be the distribution of $V_i[i]$ conditioned on $\Gamma=\gamma$. The bound $0\le f\le1/2$ and Pinsker's inequality in bits give
\[
  \left|\mathbb E[f(V_i[i])\mid \Gamma=\gamma]-\frac14\right|^2
  \le\frac14\TV\bigl(P_\gamma,\mathrm{Unif}([k])\bigr)^2
  \le\frac{\ln2}{8}\KL\bigl(P_\gamma\,\|\,\mathrm{Unif}([k])\bigr).
\]
Averaging this inequality gives
\[
  \Var(\mathbb E[f(V_i[i])\mid \Gamma])
  \le\frac{\ln2}{8}\MI{V_i[i]}{\Gamma}=\frac{\ln2}{8}I_{i,i}.
\]
The law of total variance now implies
\[
  \frac1{48}
  \le\Var(f(V_i[i]))
  =\varbarc{f(V_i[i])}{\Gamma}
    +\Var(\mathbb E[f(V_i[i])\mid \Gamma])
  \le\eta_0+\frac{\ln2}{8}I_{i,i}.
\]
Thus $I_{i,i}\ge\frac8{\ln2}(\frac1{48}-\eta_0)=:c_0>0$ for $\eta_0=2^{-10}$.
\end{proof}

\subsubsection{Data Processing for Off-Diagonal Information}
We use data processing to prove that certain off-diagonal information is no less than diagonal information, namely $I_{i,j} \geq I_{j,j}$ for $i \leq j$. Combined with the previous lower bound (\cref{cor:one-digit-info}), we are able to finish the proof. Intuitively, the argument treats $\MI{A}{B}$ as the view of the distribution of $A$ knowing $B$, and then simulates one view by another. This is a bit vague with how the conditions are treated, but data processing makes everything formal, as explained below.

\begin{lemma}[Diagonal to off-diagonal]
\label{lem:information-transport}
$I_{i,j} \geq I_{j,j}$ whenever $i \leq j$.
\end{lemma}

\begin{proof}
We first recall $I_{i,j}$ and $I_{j,j}$ in the form of \cref{lem:partial-state-information}:
\[
I_{i,j}
  = \MIc{V_i[j]}{A_\rho(V_{1:i};N)}{\rho,V_i[1:j-1]},\qquad
I_{j,j}
  = \MIc{V_j[j]}{A_\rho(V_{1:j};N)}{\rho,V_j[1:j-1]}.
\]

The key idea here is that we can construct a new set of random variables $V'_1,\dots,V'_d$ (as a function of $V_1,\dots,V_d$), such that $(V'_1,\dots,V'_d)$ exactly follows the distribution of $(V_1,\dots,V_d)$, and that the digitwise information $I_{i,i}$ defined on $(V'_1,\dots,V'_d)$ is exactly the same as the information $I_{i,j}$ defined on $(V_1,\dots,V_d)$. Since the value of $I_{j,j}$ must be the same whether defined on $(V'_1,\dots,V'_d)$ or $(V_1,\dots,V_d)$, data processing proves that $I_{j,j}\le I_{i,j}$.

The construction works as follows. Given $\rho,V_i[1:j-1]$, we want to put the current $V_i$ at the $j$-th position of the hard distribution. To do that, we view $V_i[1:j-1]$ as the first $j-1$ digits of a fresh auxiliary query $X'$, sample $X'$ accordingly, and sample $J_t$ ($i \leq t \leq j-1$) according to the hard distribution. Specifically, $J_t$ has the first $t-1$ digits equal to those of $X'$, and has all the other digits sampled uniformly randomly. All these samples use fresh randomness independent of the original variables conditional on $(\rho,V_i[1:j-1])$.

The key observation is that
$(V_1,V_2,\cdots,V_{i-1},J_i,J_{i+1},\cdots,J_{j-1},V_i)$
given $(\rho,V_i[1:j-1])$ has the same law as
$(V_1,V_2,\cdots,V_j)$ given $(\rho,V_j[1:j-1])$.
Indeed, fixing the prefix already determines the prescribed digits of every old $V_t$ with $t<i$, while their remaining digits and the remaining digits of $V_i$ are independent and uniform. The new $J_t$ supply exactly the missing samples. Since $N$ is independent of the inputs conditional on $\rho$, the following identity follows from equality in conditional law:
\[
\MIc{V_i[j]}{A_\rho(V_{1:i-1},J_{i:j-1},V_i;N)}{\rho,V_i[1:j-1]}
=
\MIc{V_j[j]}{A_\rho(V_{1:j};N)}{\rho,V_j[1:j-1]}
=I_{j,j}.
\]

Notice that the sampling rule of $J_{i:j-1}$ ensures that, given
$\rho,V_i[1:j-1]$, $J_{i:j-1}$ is independent of the pair
$(V_i[j],A_\rho(V_{1:i};N))$, so we get
\[
\begin{aligned}
\MIc{V_i[j]}{A_\rho(V_{1:i-1},J_{i:j-1},V_i;N)}{\rho,V_i[1:j-1]}
&=
\MIc{V_i[j]}{A_\rho(V_{1:i},J_{i:j-1};N)}{\rho,V_i[1:j-1]}\\
&\leq
\MIc{V_i[j]}{A_\rho(V_{1:i};N),J_{i:j-1}}{\rho,V_i[1:j-1]}\\
&=
\MIc{V_i[j]}{A_\rho(V_{1:i};N)}{\rho,V_i[1:j-1]}
=I_{i,j}.
\end{aligned}
\]
Here the first equality uses history independence: the order of the items does not matter \eqref{eq:hi-order}. The inequality is data
processing, since history independence allows us to add the items $J_{i:j-1}$ to $A_\rho(V_{1:i};N)$ \eqref{eq:hi-cancel}, so $A_\rho(V_{1:i},J_{i:j-1};N)$
is a deterministic function of $(A_\rho(V_{1:i};N),J_{i:j-1})$
once $\rho$ is given. The equality removing $J_{i:j-1}$ uses the
conditional independence stated above.
\end{proof}

\subsubsection{Completing the Main Theorem and the Key Lower Bound}

We first prove the intermediate goal \cref{lem:common-map-information}.
\begin{proof}[Proof of \cref{lem:common-map-information}]

Take $\eta=\eta_0/8$. By \cref{lem:rank-to-digits} and
$k^{-1}\le\eta_0/8$, our assumption that the conditional variance of $\widetilde R$ is small becomes
\begin{equation}
  \varbarc{\widetilde R}{X,A_\rho(V_{1:d};N),\rho}
  \le(\sqrt\eta+k^{-1/2})^2d\le\eta_0d/2.
  \label{eq:digit-rank-variance}
\end{equation}
By \cref{lem:variance-partition}, we can partition the variance into a summation: $\sum_i\Var_i\le\eta_0d/2$.
Let $\mathcal J=\{j:\Var_j\le\eta_0\}$ be a set of levels, and note that $\sum_i\Var_i\le\eta_0d/2$ implies that $|\mathcal J|\ge d/2$. 

The chain rule, first over items in reverse order
and then over digits (\cref{cor:information-triangle}), gives
\[
  \MIc{V_{1:d}}{A_\rho(V_{1:d};N)}{\rho}
  \ge\sum_{i=1}^d\sum_{j=i}^d I_{i,j}.
\]
By \cref{cor:one-digit-info}, $I_{j,j}\ge c_0$ for every $j\in\mathcal J$.
\Cref{lem:information-transport} bounds every term with $j\in\mathcal J$ and $i\le j$ below
by $c_0$; the other terms are nonnegative. Thus, the sum is at least
$c_0\sum_{j\in\mathcal J}j\ge c_0d^2/8$.
\end{proof}

Finally, we prove the main theorem of this section.

\begin{proof}[Proof of \cref{thm:hard-decoding-information}]
Use the constants $\eta,c$ of \cref{lem:common-map-information}.
Choose $a_0,\delta_0>0$ using \cref{lem:accuracy-variance} for this $\eta$.
The average decoding hypothesis \eqref{eq:hard-decoding-hypothesis} and
\cref{lem:accuracy-variance} give conditional variance at most
$\eta d$. Applying \cref{lem:common-map-information} proves
\eqref{eq:hard-decoding-conclusion}.
\end{proof}

In \cref{thm:hard-decoding-information}, we require that the linear sketch successfully estimates $R$ with probability $1-\delta_0$, where $\delta_0$ is a small constant. In the following, we show that the same lower bound holds if we set the failure probability to be $1/3$.

\begin{lemma}[Key regime lower bound]
\label{lem:amplification}\label{cor:boundary-no-augmentation}
For $\varepsilon\le1/\sqrt{\log U}$, every linear sketch
correct with failure probability at most $1/3$ on inputs of mass at most $\log U$ uses $\Omega(\log^2U)$
bits.
\end{lemma}

\begin{proof}
Choose $d=\lfloor\kappa\log U\rfloor$, where
$0<\kappa\le\min\{1/\log k,a_0^2\}$ is absolute. Embed $[k]^d$
in $[U]$ and stream the $d$ items. Then
$\varepsilon d\le a_0\sqrt d$.

We first amplify the success probability. Run $m=\lceil18\ln(1/\delta_0)\rceil$ independent copies of the sketch, and answer with the median of their answers. On every fixed input, the median errs by more than the allowed error only if at least $m/2$ copies do, which by Hoeffding's inequality happens with probability at most $e^{-m/18}\le\delta_0$. The copies together form a linear sketch of $ms$ bits over the product of their groups, where $m$ is an absolute constant. We denote it again by $A_\rho$, where $\rho$ now collects the public randomness of all copies.

Now the pointwise guarantee implies the
average decoding hypothesis of \cref{thm:hard-decoding-information}
with $N=0$ and $\sigma_\rho\equiv1$, giving
\[
  ms\ge H\bigl(A_\rho(V_{1:d};0)\mid\rho\bigr)
   \ge\MIc{V_{1:d}}{A_\rho(V_{1:d};0)}{\rho}
   =\Omega(d^2).
\]
Since $m$ is an absolute constant, $s=\Omega(\log^2U)$.
For bounded $U$, one bit is necessary: the unit-mass inputs at
$0$ and $U-1$, queried at $1$, have disjoint success intervals. With zero bits, the answer has the same law on both inputs, so one of them fails with probability at least $1/2>1/3$.
\end{proof}
\section{Lower Bounds for All Regimes}
\label{sec:generalizing-regimes}

This section generalizes the key regime lower bound to all regimes.  Recall that a
hard instance always means the hard instance of the key regime
(\cref{def:prefix-cylinder}), $\varepsilon$ is an error parameter with
$0<\varepsilon\le\frac18$, $U$ is the universe size, and $n$ is an upper bound
on the $\ell_1$ norm of the frequency vector.

\paragraph{To larger $\varepsilon$.}
As mentioned in the overview, the main hope for larger $\varepsilon$ is that the
error behaves like a normal distribution, so that using $K$ such large-error
sketch copies properly makes the error $\sqrt K$ times smaller, and choosing a
proper $K$ reduces this to the key regime at a cost of $K$.  The strategy we
use here is to randomly split the items of one hard instance into $K$ buckets,
and use a separate copy of the large-error sketch for each bucket.  However,
directly adding up the estimates of the copies may not work, since their
errors may be correlated.  The solution is to put a random sign in front of
each estimate, which makes the scale right.
\Cref{thm:hard-decoding-information}, which allows an arbitrary sign function,
is exactly designed for this, and \cref{sec:large-error-lb} explains the
details of the proof.

\paragraph{To smaller $\varepsilon$ and larger $n$.}
\Cref{sec:common-augmentation} then extends both bounds to every mass bound
$n$.  Smaller $\varepsilon$ is done by placing $r$ independent instances side
by side in one layer, so that $\varepsilon$ can go down by a factor of $r$ while
the space bound increases by a factor of $r$.  For larger $n$, we use the idea
of augmented indexing.  We stack $q$ independent layers of the same
distribution, with weights $2^{q-1},2^{q-2},\ldots,1$.  Using the chain rule, it
suffices to show that, given the heavier layers, the current layer still
requires the information bound of a single layer.  After cancelling the heavier
layers by history independence, the lighter layers are exactly the noise
allowed in \cref{thm:hard-decoding-information,lem:large-error-layer}.

\subsection{Large Error}
\label{sec:large-error-lb}

The following lemma proves an information lower bound for the larger $\varepsilon$ case. The main idea is simple: we just hope that the errors of sketches cancel out when we have several of them. However, we have some technical issues to deal with, and these are all settled in the proof below. We will define a block as a set of consecutive coordinates in the universe ($k^d$ of them for a hard instance), and partition the universe into different blocks. We need the blocks when we apply the next lemma in \cref{sec:common-augmentation}. The next lemma mainly focuses on one of the blocks. A hard instance placed on a block is identified with $[k]^d$ in lexicographic order, and has weight $w$ if each of its items is inserted with weight $w$.

\begin{lemma}[Large error]
\label{lem:large-error-layer}\label{lem:bucket-information}\label{cor:one-layer-large-error}
There are absolute constants $\varepsilon_0,c>0$ such that the following holds
for $\varepsilon\le\varepsilon_0$.  Suppose a linear sketch of $s$ bits errs by
more than $\varepsilon\norm f_1$ with probability at most $1/3$ on every fixed
input $f$ of mass at most $n$.  Then there is a linear sketch of
$O((1+\varepsilon^2d)s)$ bits, with public randomness $\rho'$, whose state $M$
on $V+Z$ satisfies $\MIc{V_{1:d}}{M}{\rho'}\ge cd^2$ whenever $V$ is a hard
instance with weight $w\le n/(2d)$, and $Z$ is an input independent of $V$ and
$\rho'$, of mass at most $dw$, at most $\varepsilon dw$ of it in the block of
$V$.
\end{lemma}

\begin{proof}
Let $K=\ceil{C\varepsilon^2d}$, where $C\ge1$ is an absolute constant and
$\varepsilon_0=C^{-1/2}$, so that $K\le d$.  Split the universe into $K$
\emph{buckets} by a uniformly random function $\pi:[U]\to\{1,\ldots,K\}$.  Simple
amplification yields a linear sketch of $O(s)$ bits with failure probability
$\delta$; run it separately on the items of each bucket.  The $K$ states form a linear sketch with public
randomness $\rho'=(\rho,\pi)$ and $O(Ks)=O((1+\varepsilon^2d)s)$ bits. Note that since we are proving information bounds, $\pi$ never has to be stored.

We now construct a target $R$ from the ranks within each of the buckets, and show that it has small conditional variance. Also, $R$ will have the form in \cref{lem:common-map-information}, so that small conditional variance gives the required information bound. The detailed construction is as follows. 

Let $w$, $V$ and $Z$ be as in the statement, let $X$ be the query of $V$,
and let $M=(M_1,\ldots,M_K)$ be the state on $V+Z$.  Let $R_b$ be the number of
items of $V$ in bucket $b$ that are at most $X-1$, and let
$\sigma_1,\ldots,\sigma_K$ be independent uniform signs.  Then
\[
  R=\sum_{b=1}^K\sigma_bR_b=\sum_{i=1}^d\sigma_{\pi(V_i)}\ind\{V_i<X\}
\]
is the signed rank of \cref{thm:hard-decoding-information} with the sign
function $v\mapsto\sigma_{\pi(v)}$, and $M$ is a linear sketch of $V_{1:d}$ with
public randomness $(\rho',\sigma)$ and the bucket states of $Z$ as noise.  By
\cref{lem:common-map-information}, it suffices to show
$\varbarc{R}{X,M,\rho',\sigma}\le\eta d$, where $\eta$ is its constant. An upper bound of this conditional variance is the expected squared error from the most intuitive estimator. Specifically, for any
estimates $T_b$ of $R_b$ from $(X,M_b,\rho')$, the estimate
$T=\sum_b\sigma_bT_b$ gives, since the signs are independent of everything else,
\[
  \varbarc{R}{X,M,\rho',\sigma}\le\E\bigl[(T-R)^2\bigr]=\sum_{b=1}^K\E\bigl[(T_b-R_b)^2\bigr].
\]
So the lemma follows from the following claim.

\begin{claim}[Bucket variance]
\label{clm:bucket-variance}
There are estimates $T_b$ of $R_b$ from $(X,M_b,\rho')$ with
$\sum_b\E[(T_b-R_b)^2]\le\eta d$.
\end{claim}
\end{proof}

\begin{proof}[Proof of \cref{clm:bucket-variance}]
We estimate each bucket rank by two queries, clip the estimate, and then use the concentration of the targets $R_b$ to prove the result. The proof is similar to that of \cref{lem:accuracy-variance}, with a bit more detail.

\paragraph{Queries.}
Up to the sketch error, $Z$ only shifts the answers in the block of $V$, by its
mass before the block and by its part $Z^\circ$ inside the block, of mass at
most $\varepsilon dw$.  So querying $M_b$ at $X$ and just before the block of
$V$, subtracting, and dividing by $w$ estimates $R_b$, except with probability
$2\delta$ given $(X,\pi,V,Z)$, within the mass $e_b$ in bucket $b$ of
$c=(2\varepsilon(V+Z)+Z^\circ)/w$; here $\sum_be_b=\norm c_1\le5\varepsilon d$.

\paragraph{Clipping.}
Given $(X,\pi)$, $R_b$ is a sum of independent indicators; let $\mu_b$ be its
mean and $\nu_b=\sum_i\sum_v\Prb[V_i=v\mid X]\,\ind\{\pi(v)=b\}$ the expected
number of items of $V$ in bucket $b$.  As in the proof of
\cref{lem:accuracy-variance}, let $T_b$ be the estimate clipped to
$\mu_b\pm W_b$, where $W_b=\lambda(\sqrt{\nu_b}+1)$.  A correct estimate stays
within $e_b$ of $R_b$, and within $2W_b$ after clipping, so its squared error is
at most $\min\{e_b,2W_b\}^2\le2W_be_b$.  The other cases cost only a small
fraction of the original variance: an incorrect estimate, which occurs with
probability at most $2\delta$, has squared error at most
$4W_b^2\le8\lambda^2(\nu_b+1)$ when $R_b$ lies in the window, and when $R_b$
lies outside, the error is at most $2|R_b-\mu_b|$, whose expected square there
is at most $f(\lambda)(\nu_b+1)$ by Bernstein's inequality, where
$f(\lambda)\to0$ as $\lambda\to\infty$.  We fix $\lambda\ge1$ with
$f(\lambda)\le\eta/8$.  As $\sum_b(\nu_b+1)=d+K\le2d$,
\[
  \sum_b\E\bigl[(T_b-R_b)^2\bigr]
  \le2\lambda\E\Bigl[\sum_b(\sqrt{\nu_b}+1)e_b\Bigr]
   +\Bigl(32\lambda^2\delta+\frac\eta4\Bigr)d.
\]

\paragraph{Correct estimates.}
It remains to bound $2\lambda\E[\sum_b(\sqrt{\nu_b}+1)e_b]$.  If every $\nu_b$
were about $d/K$, this would be about
$2\lambda\cdot5\varepsilon d\sqrt{d/K}\le10\lambda d/\sqrt C$; we check that
$e_b$ cannot concentrate on buckets with large $\nu_b$.  For every point $p$, as
$\pi$ is independent of $(X,V,Z)$, $V_i$ is uniform on $k^{d-i+1}$ strings
given $X$, and $K\le d$,
\[
  \Ec{\nu_{\pi(p)}}{X,V,Z}
  \le\sum_{i=1}^d\Bigl(\frac1K+\Prb[V_i=p\mid X]\Bigr)
  \le\frac dK+\sum_{i=1}^dk^{-(d-i+1)}\le\frac{2d}K.
\]
As $c$ is determined by $V$ and $Z$, and $\sum_be_b\le5\varepsilon d$,
Cauchy--Schwarz then gives
\[
  \E\Bigl[\sum_b\sqrt{\nu_b}\,e_b\Bigr]
  \le\Bigl(\E\Bigl[\sum_pc_p\,\nu_{\pi(p)}\Bigr]\Bigr)^{1/2}(5\varepsilon d)^{1/2}
  \le\Bigl(\frac{10\varepsilon d^2}K\cdot5\varepsilon d\Bigr)^{1/2}
  \le\frac{8d}{\sqrt C},
\]
using $K\ge C\varepsilon^2d$.  Hence
\[
  2\lambda\E\Bigl[\sum_b(\sqrt{\nu_b}+1)e_b\Bigr]
  \le\Bigl(\frac{16\lambda}{\sqrt C}+10\lambda\varepsilon\Bigr)d.
\]

\paragraph{Constants.}
Take $\delta=\eta/(128\lambda^2)$ and $C=(64\lambda/\eta)^2$, so that
$\varepsilon\le\varepsilon_0=\eta/(64\lambda)$.  Then $32\lambda^2\delta$,
$16\lambda/\sqrt C$ and $10\lambda\varepsilon$ are each at most $\eta/4$, and
the total is at most $\eta d$.
\end{proof}

\begin{remark}[Single items]
\label{rem:one-bit}\label{lem:one-bit}\label{lem:one-item-layers}\label{lem:constant-error-completion}
For $\varepsilon>\varepsilon_0$ the buckets are too coarse, and single items
suffice instead.  Let $Y$ be a uniform point of $\{0,1\}^d$, placed on a block
of $2^d$ consecutive coordinates in lexicographic order with weight
$w\le n/2$, and let $Z$ be an input independent of $Y$ and $\rho$, of mass less
than $w$, at most $w/4$ of it in the block of $Y$, and with a fixed mass before
that block.  Then every linear sketch correct with failure probability at
most $1/3$ on inputs of mass at most $n$ has $\MIc{Y}{M}{\rho}\ge(1-h_2(1/3))d$
for its state $M$ on $Y+Z$.  Indeed, for each digit $\ell$, query
$x=(Y[1{:}\ell-1],1,0,\ldots,0)$, subtract the fixed mass of $Z$ before the
block and $w/8$, and divide by $w$.  This estimates $\ind\{Y[\ell]=0\}$ within
$2\varepsilon+1/8<1/2$, so rounding recovers $Y[\ell]$ from $M$, $\rho$ and
$Y[1{:}\ell-1]$ with failure probability at most $1/3$, and Fano's inequality
gives $\MIc{M}{Y[\ell]}{Y[1{:}\ell-1],\rho}\ge1-h_2(1/3)$.  The chain rule over
$\ell$ finishes.
\end{remark}

\subsection{Small Error and Larger $n$}
\label{sec:common-augmentation}\label{sec:small-error-lb}\label{sec:master}

We first introduce the construction that packs multiple instances of the key regime into one input. The first intuition is that, when the universe $U$ is at least $r$ times larger than that needed for a single hard instance, and $\varepsilon$ is $r$ times smaller, then we can just pack $r$ hard instances together, each occupying a contiguous interval.

The second concern is how to make $n$ larger. This is done by using geometric weights for different layers, so that the lighter ones only slightly affect the heavier ones, and we can subtract the heavier ones once they are conditioned as in the chain rule. This takes ideas from \cite{Nelson26}. In practice, $U$ might not be large enough for all the layers, so we just circle around the construction, and make the layers coming back to the same block small enough to not disturb the answers.

\paragraph{Packed input.}
The construction uses nothing about the cells except that they are independent
and have fixed sizes and weights.  Fix integers $q\ge1$, $1\le h\le q$ and
$r,t,b\ge1$.  Take $h$ disjoint \emph{groups} $P_0<\cdots<P_{h-1}$ of
consecutive coordinates of $[U]$, each consisting of $r$ consecutive blocks of
$b$ coordinates, its \emph{positions} $1,\ldots,r$; this needs $hrb\le U$.  The
input consists of independent random \emph{cells} $(j,a)$, for $0\le j<q$ and
$1\le a\le r$, ordered lexicographically: cell $(j,a)$ is a tuple
$\mathcal V_{j,a}$ of $t$ items in position $a$ of $P_{j\bmod h}$, each with
weight $w_j=2^{q-1-j}$.  The cells $(j',a)$ with $j'>j$ and $j'\equiv j\pmod h$
\emph{return} to the position of cell $(j,a)$; their weights sum to at most
$\sum_{s\ge1}2^{-sh}w_j\le\theta w_j$, where $\theta=2^{1-h}$ if $h<q$, and
$\theta=0$ if $h=q$, when no cell returns.

\begin{theorem}[Packed lower bound]
\label{thm:packed-lower}\label{thm:master-lower}\label{lem:hashing-composition}\label{lem:layers}\label{lem:common-augmentation}\label{lem:layer-decoding}
Let $A$ be a linear sketch of $s$ bits with public randomness $\rho$.  Suppose
that for every cell $c=(j,a)$ and every input $Z$ independent of cell $c$ and
$\rho$, of mass less than $(2r-1)tw_j$, at most $\theta tw_j$ of it in the
position of $c$, and with a fixed mass before that position, the state $M$ of
$A$ on $\mathcal V_c+Z$ satisfies $\MIc{\mathcal V_c}{M}{\rho}\ge B$.  Then
$s\ge qrB$.
\end{theorem}

\begin{proof}
\label{app:hashing-layers}
Let $M$ be the state of $A$ on the whole input.  By the chain rule over the $qr$
cells,
\begin{equation}
  s\ge H(M\mid\rho)\ge\sum_c\MIc{\mathcal V_c}{M}{\mathcal V_{<c},\rho}.
  \label{eq:master-one-term}
\end{equation}
Fix a cell $c=(j,a)$, and let $Z$ be the input of the later cells and $M^c$ the
state on $\mathcal V_c+Z$.  Given $\mathcal V_{<c}$, history independence allows
us to cancel the earlier cells from $M$ and to add them back
\eqref{eq:hi-cancel}, so the term equals
$\MIc{\mathcal V_c}{M^c}{\mathcal V_{<c},\rho}=\MIc{\mathcal V_c}{M^c}{\rho}$, as the
earlier cells are independent of everything else.  The later cells are
independent of cell $c$ and $\rho$.  They have mass less than
$(r-a)tw_j+\sum_{j'>j}rtw_{j'}<(2r-1)tw_j$, at most $\theta tw_j$ of it in the
position of $c$, from the returning cells, and a fixed mass before that
position, as every cell has $t$ items.  So the term is at least $B$.
\end{proof}

For small error, the cells are hard instances, $r$ of them per layer, and
\cref{thm:hard-decoding-information} gives the bound for each of them.

\begin{lemma}[Small error]
\label{lem:small-error-cell}
Let $r\ge1$ with $\varepsilon r\sqrt d\le a_0/3$, and suppose a linear sketch
errs by more than $\varepsilon\norm f_1$ with probability at most $\delta_0$ on
every fixed input $f$ of mass at most $n$, where $a_0,\delta_0$ are the
constants of \cref{thm:hard-decoding-information}.  Let $V$ be a hard instance
with weight $w\le n/(2rd)$, and let $Z$ be an input independent of $V$ and
$\rho$, of mass less than $(2r-1)dw$, at most $\varepsilon rdw$ of it in the
block of $V$, and with a fixed mass before that block.  Then the state $M$ on
$V+Z$ satisfies $\MIc{V_{1:d}}{M}{\rho}=\Omega(d^2)$.
\end{lemma}

\begin{proof}
Let $X$ be the query of $V$ and $R$ its rank at $X$.  The input $V+Z$ has mass
less than $2rdw\le n$, so one query at $X$ errs by at most $2\varepsilon rdw$,
except with probability $\delta_0$.  The true answer is the fixed mass of $Z$
before the block of $V$, plus $wR+G$, where $0\le G\le\varepsilon rdw$ is the
mass of $Z$ in the block up to $X$.  So subtracting the fixed mass and
$\varepsilon rdw/2$, and dividing by $w$, estimates $R$ within
$2\varepsilon rd+\varepsilon rd/2<a_0\sqrt d$.  \Cref{thm:hard-decoding-information}
with $\sigma_\rho\equiv1$, the map $v\mapsto w\mathcal A_\rho(v)$, and the state
of $Z$ as noise gives the claim.
\end{proof}

\begin{corollary}[Lower bound for linear sketches]
\label{cor:sketch-lower}\label{cor:large-error-lower}\label{cor:small-error-lower}\label{cor:sub-layer}
Let $U\ge4$ and $n\ge\log^2U$.  Then every linear sketch correct with failure
probability at most $1/3$ on inputs of mass at most $n$ uses
$\Omega(\mathbf{RQ}(\varepsilon,n,U))$ bits.
\end{corollary}

\begin{proof}
We bound $s$ from below by one of the three terms of
$\mathbf{RQ}(\varepsilon,n,U)$, up to a constant factor.  Write
$L=\log(U\varepsilon)$, and let $L'=\log(2+U\varepsilon)$ be the $L$ of
\cref{thm:intro-main}; for $\varepsilon\ge2/U$, we have $L\le L'\le2L$.

\paragraph{Large error: $\varepsilon\ge(\log U)^{-1/2}$.}
The assumptions force $\log U\ge64$.  Let $U_0$ be a large absolute constant,
and take
\[
  d=\ceil{\frac{\log U}{4\log k}},\qquad q=\floor{\log\frac nd}.
\]
As $d\le\log U\le n/2$, we have $q\ge\max\{1,\frac12\log(n/\log U)\}$ and
$d2^q\le n$, so every weight satisfies $w_j\le2^{q-1}\le n/(2d)$.

If $\varepsilon\le\varepsilon_0$ and $U\ge U_0$, use the packed input with
$r=1$, hard instances as cells ($b=k^d$, $t=d$), and
$h=\min\{q,\ceil{\log(2/\varepsilon)}\}$, so that $\theta\le\varepsilon$.  It
fits in the universe, as $k^d\le kU^{1/4}$ and
$h\le\log(4/\varepsilon)\le\log(4\sqrt{\log U})$ give $hk^d\le U$.  As
$(2r-1)tw_j=dw_j$ and $\theta tw_j\le\varepsilon dw_j$, the sketch of
\cref{lem:large-error-layer}, of $O((1+\varepsilon^2d)s)$ bits, satisfies the
hypothesis of \cref{thm:packed-lower} with $B=\Omega(d^2)$, so
$(1+\varepsilon^2d)s=\Omega(qd^2)$.  As
$\varepsilon^2d\ge1/(4\log k)$, this gives $s=\Omega(\varepsilon^{-2}qd)$.

Otherwise $\varepsilon^{-2}\le\max\{\varepsilon_0^{-2},\log U_0\}=O(1)$.  Use
instead single uniform points of $\{0,1\}^{d'}$ as cells, with
$d'=\floor{\frac12\log U}$ ($b=2^{d'}$, $t=1$), and $h=\min\{q,3\}$, so that
$\theta\le1/4$; it fits as $3\cdot2^{d'}\le U$.  By \cref{rem:one-bit},
\cref{thm:packed-lower} applies with $B=\Omega(d')$, so $s=\Omega(qd')$.
In both cases, $s=\Omega(\varepsilon^{-2}\log U\log(n/\log U))$, which is
$\Omega(\varepsilon^{-2}\log U\log n)$ as $n\ge\log^2U$.

In the remaining cases $\varepsilon\le(\log U)^{-1/2}$, the
lower bound uses a single layer of mass about $n$ when
$n\varepsilon\le2\sqrt L$, and many layers otherwise.  In both, simple
amplification lets us assume that the sketch has failure probability at most
$\delta_0$, at the cost of a constant factor in $s$.

\paragraph{One layer: $2/U\le\varepsilon\le(\log U)^{-1/2}$ and $n\varepsilon\le2\sqrt L$.}
Let $c_1>0$ be small, let $c_2=\sqrt{c_1}\,a_0/12$, and let $U_0$ be large.  If
$U<U_0$, then $n\le2\sqrt L/\varepsilon\le U$ is $O(1)$, and one single item
($q=r=h=d=1$) gives $\Omega(1)$ by \cref{rem:one-bit,thm:packed-lower}.
Otherwise, use the packed input with $q=h=1$, so that $\theta=0$, hard
instances as cells, and
\[
  d=\ceil{c_1\log(2+U/n)},\qquad r=\floor{\frac{c_2n}d}.
\]
As $n\ge\log U\ge\log(2+U/n)$ and $c_1\log U\ge1$, we have $d\le2c_1n$, so
$c_2n/d\ge2$ for $c_2\ge4c_1$; hence $c_2n/(2d)\le r\le c_2n/d$, and the weight
$w_0=1$ is at most $n/(2rd)$.  As $n\le2\sqrt L/\varepsilon$, we have
$U/n\ge2^{L-1}/\sqrt L$, so $\log(2+U/n)\ge L/4$ and $d\ge c_1L/4$; hence
$\varepsilon r\sqrt d\le c_2\varepsilon n/\sqrt d\le4c_2/\sqrt{c_1}=a_0/3$.
Finally, $n\le U$ and $c_1\log k\le1/8$ give $k^d\le k(3U/n)^{1/8}$, so
$rk^d\le c_2kn(3U/n)^{1/8}\le U$ as $c_2$ is small.  So
\cref{lem:small-error-cell,thm:packed-lower} give
$s=\Omega(rd^2)=\Omega(nd)=\Omega(n\log(2+U/n))$, which is
$\Omega(\log\binom{U+n}{n})$ as $n\le2\sqrt L/\varepsilon\le U$.

\paragraph{Many layers: $2/U\le\varepsilon\le(\log U)^{-1/2}$ and $n\varepsilon\ge2\sqrt L$.}
Let $c\le\min\{a_0/3,1/2\}$ be small enough that $kc\ceil{\log(8/c)}\le1$, let
$\alpha>0$ be small, and let $\log U_0\ge1/\alpha$.  If $U<U_0$, then
$\varepsilon^{-1}\le U/2$ and $L$ are $O(1)$, so the claim is $O(\log n)$, and
single points ($d=r=1$, $q=\floor{\log n}$, $h=\min\{q,3\}$) give
$\Omega(\log n)$ by \cref{rem:one-bit,thm:packed-lower}, as
$6\le2/\varepsilon\le U$.  Otherwise, use the packed input with hard instances
as cells and
\[
  d=1+\floor{\alpha L},\quad
  r=\floor{\frac c{2\varepsilon\sqrt d}},\quad
  q=\floor{\log\frac n{rd}},\quad
  h=\min\bigl\{q,\ceil{\log(2/(\varepsilon r))}\bigr\},
\]
so that $\theta\le\varepsilon r$.  As $\varepsilon^2\le1/\log U\le\alpha$ and
$L\le\log U$, we have $\varepsilon^2d\le2\alpha\le c^2/16$, so
$c/(4\varepsilon\sqrt d)\le r\le c/(2\varepsilon\sqrt d)$; hence
$\varepsilon r\sqrt d\le a_0/3$.  As $d\le2L$ and $c\le1/2$, we have
$rd\le c\sqrt d/(2\varepsilon)\le\sqrt L/\varepsilon\le n/2$, so $q\ge1$ and
$rd2^q\le n$, which gives $w_j\le n/(2rd)$.  For the universe,
$\varepsilon r\ge c/(4\sqrt d)$ gives $h\le\ceil{\log(8\sqrt d/c)}$, and
$\alpha\log k\le1/4$ gives $k^d\le k(U\varepsilon)^{1/4}$.  If $d=1$, then
$hrk^d\le kc\ceil{\log(8/c)}/(2\varepsilon)\le U$.  If $d\ge2$, then
$L\ge1/\alpha$, and $r\le1/\varepsilon$ gives
$hrk^d/U\le k\ceil{\log(8\sqrt{2L}/c)}\,2^{-3L/4}\le1$ for small $\alpha$.  So
\cref{lem:small-error-cell,thm:packed-lower} give $s=\Omega(rqd^2)$, where
$rd^2\ge cd^{3/2}/(4\varepsilon)$ with $d\ge\alpha L$, and
$q\ge\frac12\log(n\varepsilon/\sqrt L)$ as $n/(rd)\ge n\varepsilon/\sqrt L\ge2$.
So $s=\Omega(\varepsilon^{-1}L^{3/2}\log(n\varepsilon/\sqrt L))$, which is
$\Omega$ of the term $\varepsilon^{-1}L'^{3/2}\log(2+n\varepsilon/\sqrt{L'})$,
as $L'\le2L$ and $\log(2+n\varepsilon/\sqrt{L'})\le2\log(n\varepsilon/\sqrt L)$.

\paragraph{Extremely small error: $\varepsilon<2/U$.}
If $U\ge16$, the sketch also has error $2/U\le1/8$, so the two
cases above at $2/U$, where $L=1$, give $\Omega(n\log(2+U/n))$ for $n\le U$ and
$\Omega(U\log(2+n/U))$ for $n\ge U$, both $\Omega(\log\binom{U+n}{n})$.  If
$4\le U<16$, then $\log\binom{U+n}{n}=\Theta(\log n)$.  Use the packed input
with single points of $\{0,1\}$ as cells ($d=r=t=1$, $b=2$),
$q=\floor{\log n}$ and $h=\min\{q,2\}$, so that $2h\le U$ and $\theta\le1/2$.
The argument of \cref{rem:one-bit}, subtracting $w/4$ instead of $w/8$, allows
mass $w/2$ of $Z$ in the block, as the error is then less than
$2\varepsilon+1/4\le1/2$.  So \cref{thm:packed-lower} gives
$\Omega(\log n)$.
\end{proof}
\section{From turnstile algorithms to linear sketches}
\label{sec:streaming-transfer}

The lower bounds of \cref{sec:key-regime,sec:generalizing-regimes} hold for
linear sketches.  The lower bound of \cref{thm:intro-main} is for streaming
algorithms, so we transfer them to arbitrary algorithms on polynomial-length
turnstile streams, using the following reduction of Jiang, Liu, and Yu~\cite{JiangLY26}.  The
main difference is that we have $U+1$ different
possible queries, while the theorem only supports one. To circumvent this, we
give an additional coordinate and put the query on this extra coordinate. This
adds at most $\ceil{\log(U+1)}$ bits to the algorithm
by explicitly storing this coordinate, and this is absorbed in all but trivial
cases that are already handled in \cref{rem:one-bit}.

\begin{theorem}[{\cite[Corollary~6.1.7(2) and Remark~6.2.5]{JiangLY26}}]
\label{thm:jly-exact}
Let $\mathcal X\subseteq\Z^N$ be a finite set of vectors of Euclidean norm at
most $D$, let $\mathcal R\subseteq\mathcal X\times\mathcal O$ be a relation, let
$0<\delta<1$, and let $R\ge\max\{N^8,D^3,\delta^{-3}\}$ be a sufficiently large
integer.  Suppose an $S$-bit randomized algorithm, on every unit-update
turnstile stream of length at most $R^4$ whose final vector $y$ lies in
$\mathcal X$, outputs $o$ with $(y,o)\in\mathcal R$ with probability at least
$1-\delta$.  Then there is a randomized linear sketch of space $O(S)$,
with an absolute constant, that outputs such an $o$ with probability at least
$1-3\delta$ for every $y\in\mathcal X$.  The same holds for algorithms promised
strict-turnstile streams, when $\mathcal X$ consists of nonnegative
vectors.
\end{theorem}

\begin{proof}[Proof of the lower bound in \cref{thm:intro-main}]
Consider an $S$-bit randomized algorithm as in
\cref{thm:intro-main}.  Put the query in the extra coordinate $U$,
so that the final vector is $(f,x)\in\Z^{U+1}$.  The query is only used at the
end, so it does not matter when this coordinate is updated: run the median of
$O(1)$ copies of the algorithm on the first $U$ coordinates, keep the last
coordinate modulo $U+1$, and query at it at the end.  On every stream whose
final vector lies in
$\mathcal X=\{(f,x):f\in\Z_{\ge0}^U,\ \norm f_1\le n,\ 0\le x\le U\}$, this
answers the rank query $x$ on $f$ with failure probability at most $1/9$, using
$O(S)+\ceil{\log(U+1)}=O(S)$ bits, as $S=\Omega(\log U)$ by the argument of
\cref{rem:one-bit} with $w=1$ and $Z=0$, which uses no linearity.
\Cref{thm:jly-exact}, with $D=n+U$, $\delta=1/9$ and
$R=R_0+(U+1)^8+(n+U)^3$ for a large constant $R_0$, which only
uses streams of length at most $R^4$, turns it into a randomized
linear sketch $\mathcal A_\rho$ of space $O(S)$ that is correct with
probability at least $2/3$ on every vector of $\mathcal X$; for strict streams,
$\mathcal X$ is nonnegative.  Restricted to the first $U$ coordinates, with
$x\mathcal A_\rho(e_U)$ added before decoding, it is a linear
sketch of space $O(S)$ that answers every rank query with failure
probability at most $1/3$ on every input of mass at most $n$.  So
\cref{cor:sketch-lower} gives
$\Omega(\mathbf{RQ}(\varepsilon,n,U))$ bits.
\end{proof}
\section{Upper Bounds}
\label{sec:upper-bounds}

The previous sections proved lower bounds for rank estimation. In this section, we present the algorithms that match them, and prove \cref{thm:intro-main} together with its word-model version, \cref{thm:main-upper}. We start by informally recalling what the two theorems say.

First, we recall the problem setting. A turnstile stream updates a frequency vector $f$ over the universe $[U]$, and the $\ell_1$ norm $\norm f_1$ at the end of the stream is promised to be at most $n$ ($f$ is guaranteed to be nonnegative entrywise at the end of the stream). After the stream, a query arrives, which is of one of the following two types. A \emph{rank query} $x\in\{0,\ldots,U\}$ asks for an estimate of the rank $R_f(x)=\sum_{i<x}f_i$, the mass of the prefix $[0,x)$, within additive error $\varepsilon\norm f_1$. A \emph{select query} is the inverse of a rank query: given a target rank $r\in[0,\norm f_1]$, it asks for a coordinate $\widehat x\in[U]$ whose rank is $r$ up to an additive error of $\varepsilon\norm f_1$, that is, $R_f(\widehat x+1)\ge r-\varepsilon\norm f_1$ and $R_f(\widehat x)\le r+\varepsilon\norm f_1$. Our algorithms answer a fixed rank query and a fixed select query, both correct with probability at least $2/3$.

\Cref{thm:intro-main} assumes $n\ge\log^2U$, measures the space in bits, and determines the optimal space up to a constant factor.  \Cref{thm:main-upper}, stated in \cref{sec:upper-assembly}, considers the standard setting where $n$ is polynomial in $U$ and the space is measured in words of $\Theta(\log U)$ bits. It states that $O(U)$, $O(\varepsilon^{-1}\log^{3/2}(U\varepsilon))$, and $O(\varepsilon^{-2}\log U)$ words suffice for rank queries in the extremely-small-error, small-error, and large-error regimes, respectively. Both theorems are stated for rank queries, but all of our algorithms also answer select queries within the same space bounds.

Both bounds have three branches, and each branch is achieved by a different algorithm:
\begin{itemize}
  \item \textbf{Exact recovery} (\cref{sec:exact-upper}). For $\varepsilon\le2/U$, the allowed error is so small that we cannot do much better than storing the frequency vector $f$ itself. This takes $O(\log_2 \binom{U+n}{n})$ bits and, with probability at least $99/100$, answers both rank and select queries exactly. On a word RAM, its fast implementation uses $O(\min\{n\log U,\ U\log n\})$ bits, with $O(\log^4U)$ update time and $O(\log U)$ query time.
  \item \textbf{The Dyadic CountSketch} (\cref{sec:dyadic-upper}). For $2/U\le\varepsilon\le(\log U)^{-1/2}$, we use a variant of the DCS~\citep{LuoWY16,Nelson26}. It consists of $O(\varepsilon^{-1}L^{3/2})$ counters, where $L=\ceil{\log(U\varepsilon)}$, and uses $O(\varepsilon^{-1}L^{3/2}\log(2+n\varepsilon/\sqrt L))$ bits. It answers both rank and select queries. On a word RAM, its fast implementation uses $O(\varepsilon^{-1}L^{3/2}\log n)$ bits, with $O(\log^2U)$ update time and $O(\log^2U)$ query time.
  \item \textbf{The priority sketch} (\cref{sec:shared-seed}). For $(\log U)^{-1/2}\le\varepsilon\le1/8$, we use a sketch based on priority sampling~\citep{DuffieldLT07}. It consists of $O(\varepsilon^{-2}\log U)$ counters and uses $O(\varepsilon^{-2}\log U\log n)$ bits. It answers both rank and select queries. Moreover, it has update time $O(\log U)$ and query time $O(\varepsilon^{-4}\log^2U)$.
\end{itemize}
When $n$ is polynomial in $U$, every coordinate of $f$ and every counter of the two sketches fits in $O(1)$ words, so the three algorithms use $O(U)$, $O(\varepsilon^{-1}\log^{3/2}(U\varepsilon))$, and $O(\varepsilon^{-2}\log U)$ words, respectively. These are exactly the three bounds of \cref{thm:main-upper}.

For general $n$, the three bounds in bits listed above are the three terms in the upper bound of \cref{thm:intro-main}, with one difference: which algorithm to use now depends on $n$ as well as on $\varepsilon$. Indeed, when $n$ is small, storing $f$ exactly takes only $O(n\log(2+U/n))$ bits. So for $2/U\le\varepsilon\le(\log U)^{-1/2}$, the bound of \cref{thm:intro-main} is achieved by exact recovery when $n\varepsilon\le2\sqrt L$, and by the Dyadic CountSketch otherwise. The matching lower bounds were proved in \cref{sec:generalizing-regimes,sec:streaming-transfer}.

All three algorithms are linear sketches (\cref{def:additive-sketch}). We now briefly recall this model. A randomized linear sketch draws a public random seed $\rho$, independent of the input and the query, which fixes an abelian group $G_\rho$, an item map $\mathcal A_\rho:[U]\to G_\rho$ and a decoder $g_\rho$. On a frequency vector $f\in\Z^U$, it stores the state $A_\rho(f)=\sum_{u\in[U]}f_u\,\mathcal A_\rho(u)\in G_\rho$, and answers a query $x$ with $g_\rho(A_\rho(f),x)$. Its space is the least $s$ such that, for every seed $\rho$, the state takes at most $2^s$ values, and its seed length is the number of bits of $\rho$. The space of a linear sketch does not count the seed, but all three of our algorithms store their seeds, and the space bounds above include the seed length.

The rest of this section is organized as follows. We present exact recovery, the Dyadic CountSketch, and the priority sketch in \cref{sec:exact-upper,sec:dyadic-upper,sec:shared-seed}, respectively. Finally, \cref{sec:upper-assembly} combines the three algorithms to prove \cref{thm:intro-main,thm:main-upper}.

\subsection{The Exact Recovery Algorithm for Extremely Small Error}
\label{sec:exact-upper}

This subsection handles the extremely-small-error regime
$\varepsilon\le2/U$. We present an algorithm that proves the upper bound of
\cref{thm:intro-main} in this regime. The same algorithm also gives the upper
bound in the small-error regime when $n\varepsilon\le2\sqrt L$. In this regime, the allowed error $\varepsilon$ is so small that we directly recover the final frequency vector $f$ exactly with \emph{zero} approximation error with high probability.

\begin{proposition}[Exact recovery]
\label{prop:exact-storage}
Let $n\ge1$. On turnstile streams of final $\ell_1$ norm at most $n$,
there exists a randomized linear sketching algorithm that, with
probability at least $99/100$, answers every rank query and every
select query exactly. It uses
\[
  O\!\left(\log\binom{U+n}{n}\right)
  =O\bigl(\min\{U\log(2+n/U),\ n\log(2+U/n)\}\bigr)
  \qquad\text{bits}
\]
with random seed
included, throughout updates and answering queries. Moreover, on a
word {\rm{RAM}} machine with $O(\log nU)$-bit words,
$O(\min\{n\log U,\ U\log n\})$ bits of space can be achieved with
$O(\log^4U)$ update time and $O(\log U)$ query time.
\end{proposition}

For a constant $\eta>0$, the space $O(\min\{n\log U,\ U\log n\})$ of
the fast implementation is the same as $O(\log\binom{U+n}{n})$ when
$n\le U^{1-\eta}$ or $n\ge U^{1+\eta}$.

\begin{proof}
We store a random fingerprint of the final vector $f$, long enough to tell apart all possible final vectors, and recover $f$
from it at the end of the stream. The final vector $f$ is a
nonnegative integer vector in $\Z^U$ with $\norm f_1\le n$, and there
are $K\coloneqq\binom{U+n}{n}$ such vectors. We fix a prime $p$ with
$100UK<p\le200UK$, draw a uniformly random evaluation point
$\alpha\in\mathbb F_p$, and maintain the fingerprint
\[
  H\coloneqq\sum_{j\in[U]}f_j\,\alpha^{j}\bmod p.
\]
Storing the prime $p$, the
evaluation point $\alpha$, and the fingerprint $H$ takes
$O(\log p)=O(\log K)$ bits.

At the end of the stream, we enumerate the $K$ candidate vectors and
output one whose fingerprint equals $H$. A candidate $f'\ne f$ has the
same fingerprint only if $\alpha$ is a root of the polynomial
$\sum_j(f_j-f'_j)z^{j}$, which is nonzero modulo $p$ since $p>2n$
and has degree less than $U$. This happens with probability less than
$U/p$, so by a union bound over the $K$ candidates, $f$ is recovered
with probability at least $99/100$. The prime $p$ can be found by trial
division, and the candidates can be enumerated one at a time in
lexicographic order; we keep only the index of the matching candidate,
from which each coordinate of $f$ can be recomputed. All of this takes
$O(\log K)$ working bits. Then, a rank query and a select query can be answered exactly using the recovered vector $f$.

The fast implementation stores the counters separately rather than
jointly, using modular counters when $n\ge U$ and sparse recovery techniques when $n\le U$. See \cref{app:exact-fast} for the
details and for the comparison of the two space bounds.
\end{proof}

\subsection{The Sketching Algorithm for Small $\varepsilon\le(\log U)^{-1/2}$}
\label{sec:dyadic-upper}

In the small-error regime $2/U\le\varepsilon\le(\log U)^{-1/2}$, the previous exact recovery algorithm covers only the case $n\varepsilon\le2\sqrt L$, because
its $O(n\log(2+U/n))$ bits grow linearly with the $\ell_1$ norm bound
$n$. For larger $n$, we use the Dyadic CountSketch, whose space grows only
logarithmically with $n$. It gives the upper bound of
\cref{thm:intro-main} in this regime when $n\varepsilon\ge2\sqrt L$.

Our sketch is a variant of the Dyadic CountSketch~\citep{LuoWY16}, recently analyzed by~\citep{Nelson26}. We start from the bound in~\cite{Nelson26}.

\begin{theorem}[{\cite[Theorem~4.2]{Nelson26}}]
\label{thm:bkm-dyadic}
On turnstile streams, there exists a randomized linear sketching
algorithm that answers a fixed rank query, correct with probability at
least $2/3$. It uses
\[
  O\bigl(\varepsilon^{-1}\log^{3/2}U\bigr)
  \qquad\text{words}.
\]
\end{theorem}

We improve \cref{thm:bkm-dyadic} as follows. First, \citep{Nelson26} measures space in words. Since a word contains at least $\Omega(\log U)$ bits, this does not determine the space in bits. Second, we treat the finer regime of very small $\varepsilon$. In this regime, we improve the factor $\log^{3/2}U$ to $L^{3/2}$, where $L\coloneqq\ceil{\log(U\varepsilon)}$. This matches the lower bound in \cref{cor:small-error-lower} whenever $n\varepsilon\ge2\sqrt L$. The optimal-space implementation may take exponential time to answer queries. However, with $O(\varepsilon^{-1}L^{3/2}\log n)$ bits instead, updates and queries take only $O(\log^2U)$ time. Third, \citep[Theorem~4.2]{Nelson26} answers a fixed select query using $O(\varepsilon^{-1}\log^{3/2}U\log(1/\varepsilon)+\log\log U)$ words. We remove this extra factor of $\log(1/\varepsilon)$ over rank queries, so our sketch answers select queries within the same space as rank queries.

\begin{theorem}[Upper bound for small $\varepsilon$]
\label{thm:small-error-upper}
Let $2/U\le\varepsilon\le(\log U)^{-1/2}$ and
$L=\ceil{\log(U\varepsilon)}$. On turnstile streams of final $\ell_1$
norm at most $n$, there exists a randomized linear sketching algorithm
that answers a fixed rank query and a fixed select query, both
correct with probability at least $2/3$. It uses
\begin{equation}
  O\!\left(
    \varepsilon^{-1}L^{3/2}
    \log\!\left(2+\frac{n\varepsilon}{\sqrt L}\right)
  \right)
  \qquad\text{bits}
\end{equation}
with random seed
included, throughout updates and answering queries. Moreover, on a
word {\rm{RAM}} machine with $O(\log nU)$-bit words,
$O(\varepsilon^{-1}L^{3/2}\log n)$ bits of space can be achieved with
$O(\log^2U)$ update time and $O(\log^2U)$ query time.
\end{theorem}

For a constant $\eta>0$, the space $O(\varepsilon^{-1}L^{3/2}\log n)$ of
the fast implementation is the same as the optimal space $  O\!\left(
    \varepsilon^{-1}L^{3/2}
    \log\!\left(2+\frac{n\varepsilon}{\sqrt L}\right)
  \right)$ bits when
$n\ge(\sqrt L/\varepsilon)^{1+\eta}$.

\paragraph{Intuition.} We first walk through the Dyadic CountSketch~\citep{LuoWY16,Nelson26}. The overview below then explains the two changes we make to obtain \cref{thm:small-error-upper}. Without loss of generality, we assume that $U$ is a power of two (padding to the next power of two if necessary), and we write $f_S=\sum_{i\in S}f_i$ for the mass of a subset $S\subseteq[U]$. The DCS builds a perfect binary tree over $[U]$, whose nodes are called \emph{dyadic intervals}. For each \emph{level} $j=0,1,\ldots,\log U$, the level-$j$ dyadic intervals are the $2^j$ intervals
\[
  \bigl[a\,U/2^j,\ (a+1)\,U/2^j\bigr),\qquad a=0,1,\ldots,2^j-1,
\]
each of length $U/2^j$. Thus level $0$ consists of $[U]$ itself (the root), level $\log U$ consists of the single coordinates (the leaves), the intervals at each level partition $[U]$, and each level-$j$ interval with $j<\log U$ is the union of its two children, which are level-$(j+1)$ intervals. Every prefix $[0,x)$ is a disjoint union of at most one dyadic interval per level: if $x=2^{k_1}+2^{k_2}+\cdots$ with $k_1>k_2>\cdots$, then $[0,x)$ is the union of consecutive dyadic intervals of lengths $2^{k_1},2^{k_2},\ldots$, which lie at distinct levels. Therefore, the rank $R_f(x)=f_{[0,x)}$ is the sum of at most $\log U+1$ interval masses, one per level, and it suffices to estimate the mass of every dyadic interval.

Next, we sketch these masses level by level. For each level $j$, let $y^{(j)}$ be the vector whose coordinates are the masses $f_I$ of the $2^j$ level-$j$ intervals $I$. At the end of the stream, each $y^{(j)}$ has $\ell_1$ norm $\norm f_1$, and an update $(i,\Delta)$ to $f$ adds $\Delta$ to exactly one of $y^{(j)}$'s coordinates, namely the mass of the level-$j$ interval containing $i$. We maintain each vector $y^{(j)}$ with a CountSketch~\citep{CharikarCF04}, a classical streaming data structure that is also the main building block of the priority sketch in \cref{sec:shared-seed}. We now state it.
\begin{definition}[CountSketch~\citep{CharikarCF04}]
\label{def:countsketch}
A \emph{CountSketch} with width $b$ and $r$ rows is a streaming data
structure for a vector $v\in\mathbb R^N$ that is updated over time.
Each row $s\in[r]$ independently draws a hash function
$h_s:[N]\to[b]$ and signs $\sigma_s:[N]\to\{-1,1\}$, and maintains the
$b$ counters
\[
  C_{s,a}=\sum_{i:\,h_s(i)=a}\sigma_s(i)\,v_i,
  \qquad a\in[b].
\]
An update $(i,\Delta)$, which adds $\Delta$ to $v_i$, adds
$\sigma_s(i)\Delta$ to the counter $C_{s,h_s(i)}$ of every row $s$.
At any time, the CountSketch estimates $v_i$ by the median of its $r$
row estimates:
\[
  \mathsf{CS}_{b,r}(v)_i\coloneqq
  \operatorname{median}_{s\in[r]}\ \sigma_s(i)\,C_{s,h_s(i)}.
\]
\end{definition}
Back to the Dyadic CountSketch. For each level $j$, we maintain $y^{(j)}$ by a CountSketch with $r=5$ rows and width $b$, independently across levels, as in the original DCS of~\citep{LuoWY16}. To store the hashes and signs with few random bits, \citep{Nelson26} generates them from a short seed, with signs that fool intersections of halfspaces~\citep[Theorem~2.2]{Nelson26}. We instead generate the signs by Nisan's generator~\citep{Nisan92}, which needs a slightly longer seed but lets us evaluate every sign in $O(\log b)$ time~\citep{KachamPTW23}. The errors satisfy the bounds in the following lemma. We defer the construction and the proof to \cref{app:dyadic-details}.

\begin{lemma}[Errors of a CountSketch]
\label{lem:prg-countsketch}
Let $n\ge1$, $b\ge2$, and $y\in\Z^N$ with $\norm y_1\le n$. The hashes $h_s$ and signs $\sigma_s$ of a CountSketch of $y$ with width
$b$ and $5$ rows can be generated from a seed of $O(\log N+\log b\log(nb))$
bits such that every $i\in[N]$ satisfies
\[
  \E\bigl[(\mathsf{CS}_{b,5}(y)_i-y_i)^2\bigr]
  =O\!\left(\frac{\norm y_1^2}{b^2}\right),
  \qquad
  \bigl|\E[\mathsf{CS}_{b,5}(y)_i-y_i]\bigr|
  =O\!\left(\frac{\norm y_1}{b^3}\right).
\]
Moreover, on a word RAM with $O(\log(nNb))$-bit words, each $h_s(i)$
can be evaluated in $O(1)$ time and each $\sigma_s(i)$ in $O(\log b)$
time.
\end{lemma}
Using the lemma, each such CountSketch estimates every coordinate of $y^{(j)}$ with an error of variance $O(\norm f_1^2/b^2)$ and mean only $O(\norm f_1/b^3)$. To answer a rank query $x$, we add up the estimates of the at most $\log U+1$ dyadic intervals in $[0,x)$. Since the errors of different levels are independent, their variances add up: the total error has variance $O(\log U\cdot\norm f_1^2/b^2)$, and its mean $O(\log U\cdot\norm f_1/b^3)$ is negligible. Therefore, with width $b=\Theta(\sqrt{\log U}/\varepsilon)$, Chebyshev's inequality shows that the estimate is within $\varepsilon\norm f_1$ of $R_f(x)$ with constant probability. For space complexity, the sketch has $5b(\log U+1)=O(\varepsilon^{-1}\log^{3/2}U)$ counters, which matches the word bound of \cref{thm:bkm-dyadic} (we use one word for each counter).

\paragraph{Overview of the sketching algorithm.} We make two changes to the sketch above, which give \cref{thm:small-error-upper}:
\begin{itemize}
  \item \textbf{sketching only $L$ levels}: Recall $L=\ceil{\log(U\varepsilon)}$, and let $j_0\coloneqq\log U-L$. There are only $2^{j_0}=U/2^L\le1/\varepsilon$ intervals at level $j_0$, each of length $2^L$, which we call \emph{blocks}, so we store the mass of every block exactly. A prefix $[0,x)$ then consists of whole blocks, whose masses are exact, and  a remaining disjoint union of at most one interval at each of the $L$ finest levels $j=j_0+1,\ldots,\log U$. Hence, only these remaining $L$ levels need to be sketched, and we use one CountSketch per level as above. The calculation above with $L$ levels in place of $\log U+1$ gives width $b=\Theta(\sqrt L/\varepsilon)$, and $5Lb+U/2^L=O(\varepsilon^{-1}L^{3/2})$ counters in total. To answer a rank query $x$, we sum the exact masses of the whole blocks in $[0,x)$, then the estimates of the at most $L$ dyadic intervals in its remaining partial block.
  \item \textbf{counting the space in bits}: The sketch above uses one word per counter. We instead count the space in bits explicitly, using the same idea as exact recovery (\cref{prop:exact-storage}). We observe that all used $D\coloneqq5Lb+U/2^L=O(\varepsilon^{-1}L^{3/2})$ counters are integer combinations of coordinates of $f$ and have absolute values summing to at most $R\coloneqq(5L+1)n$ (see \cref{app:dyadic-proof} for details). So, as in \cref{prop:exact-storage}, we store a single random fingerprint of the counter vector, long enough to tell apart all vectors in $\Z^D$ of $\ell_1$ norm at most $R$, and recover all counters from it at the end of the stream. There are at most $\binom{2D+R}{R}$ such vectors, so the fingerprint takes the desired
  \[
    O\!\left(\log\binom{2D+R}{R}\right)
    =O\bigl(D\log(2+R/D)\bigr)
    =O\!\left(\varepsilon^{-1}L^{3/2}\log\!\left(2+\frac{n\varepsilon}{\sqrt L}\right)\right)
    \qquad\text{bits.}
  \]
\end{itemize}

See \cref{app:dyadic-proof} for the full proof of \cref{thm:small-error-upper}, including the fast implementation.

\subsection{The Sketching Algorithm for Large $\varepsilon\geq (\log U)^{-1/2}$}
\label{sec:shared-seed}

Once $\varepsilon>(\log U)^{-1/2}$, the Dyadic CountSketch is no longer
optimal, since $\varepsilon^{-1}\log^{3/2}U$ is more than
$\varepsilon^{-2}\log U$. The large-error regime
$(\log U)^{-1/2}\le\varepsilon\le1/8$ thus needs a different idea, for
which we turn to priority sampling~\citep{DuffieldLT07}.

The resulting priority sketch answers rank and select queries with success probability at least $2/3$ using at most $O\bigl(\varepsilon^{-2}\log U\,\log n\bigr)$ bits (random bits are charged in the space count). This gives the upper bound of \cref{thm:intro-main} in the large-error regime and the third branch of \cref{thm:main-upper}, and matches the large-error lower bound when $n\ge\log^2U$, see \cref{cor:large-error-lower}. We first state the full result.
\begin{theorem}[Full upper bound for large $\varepsilon$]
\label{thm:priority-bits}
Let $(\log U)^{-1/2}\le\varepsilon\le1/8$ and $n\ge2$.  On turnstile streams of final $\ell_1$ norm at most $n$, there exists a randomized linear sketching algorithm that answers a fixed rank query
and a fixed select query, both correct with probability at least $2/3$. It uses
\[
  O\bigl(\varepsilon^{-2}\log U\,\log n\bigr)
  \qquad\text{bits}
\]
throughout updates and answering queries, including a
random seed of $O(\log nU)$ bits. On a word {\rm{RAM}} machine with
$O(\log nU)$-bit words, an update takes $O(\log U)$ time, and
answering a query takes $O(\varepsilon^{-4}\log^2U)$ time.
\end{theorem}

In the following, we present a proof sketch of \cref{thm:priority-bits} where we assume access to free randomness and allow $O(U)$ query time. See \Cref{app:derandomization-details} for the full proof.

\paragraph{The strengths and limitations of $\ell_1$ samplers.} In the standard regime where $n=U^{\Theta(1)}$ and we allow accessing randomness for free, \cref{thm:priority-bits} is easy using a powerful tool called \emph{$\ell_1$ samplers}.

Given a stream of updates, an \emph{$\ell_1$
sampler} outputs a random coordinate $i\in[U]$ with
probability approximately $f_i/\norm f_1$, where $f$ is the frequency vector at the end of the stream and $\norm f_1=\sum_i f_i$ is its $\ell_1$ norm. Given this, we can solve \cref{thm:priority-bits} by maintaining $K$ independent $\ell_1$ samplers, which are used at the end to draw $K=\Theta(\varepsilon^{-2})$ independent samples $i_1,\ldots,i_K\in [U]$. Note that each sample falls in $[0,x)$ with probability $\sum_{i<x} f_i/\norm f_1$ independently. Hence, we can estimate the rank $R_f(x)=\sum_{i<x}f_i $ using the proportion of samples falling in $[0,x)$:
\[
  \frac{R_f(x)}{\norm f_1} \quad \approx  \quad \frac 1K\cdot\#\{k\in[K]:\ i_k<x\}.
\]
One can check that the right-hand side above is an unbiased estimator with variance at most $1/K$. Thus, by Chebyshev's inequality, it is within an additive error of
$O(1/\sqrt{K}) = O(\varepsilon)$ of the answer $R_f(x)/\norm f_1$ with probability at least $2/3$, as desired.

In the general regime, however, $\ell_1$ samplers in the current literature do not match the desired space or the number of random bits.
\begin{itemize}
  \item Approximate $\ell_1$ samplers~\citep{MonemizadehW10,AndoniKO11,JowhariST11}
output $i$ with probability $(1\pm\nu)f_i/\norm f_1$ and use $\tilde{O}(\nu^{-1}\log^2U)$ bits per sample. We need $\nu=O(\varepsilon)$ and $K=\Theta(\varepsilon^{-2})$ independent samples, which is
$\widetilde O(\varepsilon^{-3}\log^2U)$ bits of space in total. This is larger than the desired space bound.
\item Perfect $\ell_1$ samplers~\citep{JayaramW21, SwartworthWZ25} output $i$ with probability $f_i/\norm f_1\pm U^{-c}$ for any constant
$c$, and use $O(\log^2U)$ bits per sample.
With $K$ independent copies, this is $O(\varepsilon^{-2}\log^2U)$ bits of space in total,
which is what we want when $n=U^{\Theta(1)}$.  However, this bound holds only
when random bits are free; when random bits are charged, it has an $O((\log\log U)^2)$ overhead.  Moreover, it assumes frequencies polynomial in
$U$, whereas we allow any $\ell_1$ norm bound $n$.
\end{itemize}
Therefore, if we want to obtain the space bound of $O(\varepsilon^{-2}\log U \log n)$ bits in total (including the regime where $n = U^{o(1)}$), with random seeds included, we need a different approach.

Instead of directly applying $\ell_1$ samplers, we prove \cref{thm:priority-bits} by opening the black box and looking inside the process. The techniques in \cite{MonemizadehW10,AndoniKO11,JowhariST11,JayaramW21,SwartworthWZ25} come from a rescaling trick, which is closely related to \emph{threshold sampling}~\citep{DuffieldLT05} and \emph{priority sampling}~\citep{DuffieldLT07}. We now discuss the intuition as follows.

\paragraph{Intuition.} We first consider a weaker problem which motivates priority sampling. In this problem, each coordinate $i \in [U]$ arrives \emph{only once} in the stream (whereas in turnstile streaming, items can come and get deleted multiple times), together with its frequency $f_i$. For simplicity, we also assume that the algorithm knows the total frequency $\norm f_1=\sum_i f_i$ in advance. Finally, a subset query $S \subseteq [U]$ arrives at the end of the stream, and the goal is to estimate the subset sum $f_S=\sum_{i\in S}f_i$ up to an additive error $\pm \varepsilon\norm f_1$.

Instead of sampling $K$ independent coordinates $i$ with probability $f_i/\norm f_1$ as the previous $\ell_1$ sampler does, we maintain a sparse subset $H$ of $[U]$ of size $O(K)$, which is used to estimate the subset sum at the end of the stream. To maintain
this subset efficiently, we want the decision of whether $i\in [U]$ is placed in the subset to depend only on $f_i$, so that it can be
made as soon as $i$ arrives. Such a subset is selected with the help of
\emph{priorities}, defined as follows.

We fix a threshold $\tau=\norm f_1/K$ in advance (which is possible since we assume $\norm f_1$ is known) and draw an independent $u_i$ uniform in $(0,1]$ when seeing each
item $i\in [U]$. Then, we assign item $i$ the \emph{priority} $p_i := f_i/u_i$, and place $i$ in the subset if and only
if its priority is above the threshold: $p_i>\tau$. We will store the index $i$ and its frequency $f_i$ for each placed item.

Then, at the end of the stream, one wants to construct an unbiased estimator for the subset sum $f_S$ using only the items in the subset. This is done by assigning each kept item $i$ the weight $\max\{f_i,\tau\}$ and taking the sum:
\begin{equation}
  \widehat f_S\coloneqq\sum_{i\in S}\max\{f_i,\tau\}\,
    \ind\{p_i>\tau\}=\sum_{i\in S\cap H}\max\{f_i,\tau\}.
  \label{eq:priority-ideal}
\end{equation}
Since $\Prb[p_i>\tau]=\Prb[u_i<f_i/\tau]=\min\{1,f_i/\tau\}$ (because $u_i$ is sampled uniformly in $(0,1]$), one can verify that
\begin{equation}
\begin{aligned}
  \E\Bk*{\widehat f_S}&=\sum_{i\in S}\max\{f_i,\tau\}
    \min\{1,\frac{f_i}{\tau}\}=\sum_{i\in S}f_i=f_S,\\
  \Var\bk*{\widehat f_S}&=\sum_{\substack{i\in S\\f_i<\tau}}\tau^2\,\frac{f_i}{\tau}
    \Bigl(1-\frac{f_i}{\tau}\Bigr)\le\tau\sum_{i\in S}f_i\le\tau\norm f_1.
\end{aligned}
\label{eq:priority-mean-var}
\end{equation}
Since $\tau=\norm f_1/K$ and $K={\Theta (\varepsilon^{-2})}$, Chebyshev's inequality shows that $\widehat f_S$ is within $\varepsilon\norm f_1$ of $f_S$ with
constant probability. 

For space complexity, recall that we store, for each
kept item, its index $i$ ($O(\log U)$ bits) and $f_i$
($O(\log n)$ bits). The number of kept items is at most $\sum_i\min\{1,f_i/\tau\}\le
\norm f_1/\tau=O(\varepsilon^{-2})$ in expectation. This is $O(\varepsilon^{-2}\log nU)$ bits in total.

\paragraph{Overview of the sketching algorithm.} We now turn to turnstile streams. Unlike in the previous weaker problem, in a turnstile stream, deletions move
coordinates in and out so that the frequency vector $f$ and the threshold $\tau=\norm f_1/K$ change over time. It is thus difficult to decide whether to place a coordinate in the subset when it appears in the stream, since the priority $f_i / u_i$ is neither fixed nor known in advance.

The surrogate here is to fix the priority random variables $u_i$ before the
stream and treat both the frequency vector $f$ and the priority vector $p$ as updated simultaneously over time (when $f_i$ is updated, we also update $p_i$ using $p_i = f_i / u_i$). Then, we maintain the threshold $\tau = \norm f_1 / K$ and use turnstile streaming data structures to maintain approximations of the subset $H$ and weights $\max\{f_i,\tau\}$ for $i\in H$, so that at the end we can use this to estimate $\widehat f_S$ for any query set $S$.

\begin{itemize}
    \item \textbf{maintaining $H$}: Recall that by definition, $i\in H$ if and only if $p_i>\tau$. Since $\Prb[f_i/u_i\ge \tau]\le f_i/\tau$, at most $O(\norm f_1 /\tau) = O(K)$ of the priorities $p_i $ exceed $\tau$ in
expectation. Therefore, the set $H$ is the set of $\ell_2$-heavy hitters of the priority vector $p$. Hence, we can approximately maintain $H$ using a CountSketch, even though $\tau$ may change over time.

We introduced the CountSketch in \cref{def:countsketch} for the Dyadic CountSketch. Here we use a different guarantee of it: with width $\Theta(k)$ and $r$ rows, that is, $O(kr)$ counters, it approximates every coordinate of a vector $v$ within additive error
$O(\gamma/\sqrt k)$ with success probability $1-2^{-r}$, where $\gamma$ is the $\ell_2$ norm of $v$ after
removing its $k$ largest coordinates.  For the vectors that we sketch, each counter can be stored in $O(\log n)$ bits, where $n$ is the bound on the final $\ell_1$ norm (see \cref{app:priority-counters}).  Here, we use a CountSketch to maintain the priority vector $p$, with parameters
$k=\Theta(K)$ and $r = \Theta(\log U)$. It can be checked that $\gamma=O(\tau\sqrt K)$. Therefore, the CountSketch estimates every $p_i$ within additive error $O(\gamma/\sqrt k)=O(\tau)$ with high probability, with space $O(\varepsilon^{-2}\log U\log n)$ bits. This tells us whether $i\in H$ for every coordinate $i$ whose priority $p_i$ is not too close to the threshold $\tau$. For the few coordinates whose priority is close to $\tau$, we decide whether they are in $H$ using one more, finer CountSketch for $p$. This CountSketch has only one row, but its width is larger by a factor of $1/\varepsilon^2$, so it estimates each priority much more precisely (see \cref{app:priority-free} for details).
    \item \textbf{maintaining $\max\{f_i,\tau\}$}: 
To maintain $\max\{f_i,\tau\}$ for $i\in H$, another CountSketch with parameters $k = \Theta(K^2)$ and $r=1$ for the frequency vector $f$ is used to approximate each $\max\{f_i,\tau\}$ within additive error $O(\tau \sqrt K / \sqrt{K^2}) = O(\tau/\sqrt K)$, with space $O(\varepsilon^{-4}\log n) \leq O(\varepsilon^{-2} \log U\log n)$ bits, where we used the fact that $\varepsilon\geq (\log U)^{-1/2}$. Since $H$ contains $O(K)$ coordinates, the errors of the weights of all coordinates in $H$ sum to $O(\tau\sqrt K)=O(\varepsilon\norm f_1)$.
\end{itemize}

In total, our data structure uses $O(\varepsilon^{-2}\log U\log n)$ bits, as desired.

\subsection{Proofs of the Two Assembled Theorems}
\label{sec:upper-assembly}

We now combine the three algorithms of this section, namely the Dyadic CountSketch (\cref{thm:small-error-upper}), exact recovery (\cref{prop:exact-storage}), and the priority sketch (\cref{thm:priority-bits}), with the lower bounds of \cref{sec:streaming-transfer} to prove \cref{thm:intro-main,thm:main-upper}. All three algorithms also answer a fixed select query, so the same assembly gives these upper bounds for select queries as well.

We prove \cref{thm:intro-main} separately in the three error regimes. Let $L=\log(U\varepsilon)$; for $\varepsilon\ge2/U$, it is within a constant factor of the $L$ in \cref{thm:intro-main}. Up to constant factors, the minimum in \cref{thm:intro-main} then reads as follows.
\begin{enumerate}
\item[(i)] For $0<\varepsilon\le2/U$: $\log\binom{U+n}{n}=\Theta\bigl(\min\{n\log(2+U/n),\ U\log(2+n/U)\}\bigr)$.
\item[(ii)] For $2/U\le\varepsilon\le(\log U)^{-1/2}$: $n\log(2+U/n)$ when $n\varepsilon\le2\sqrt L$ (the first branch), and $\varepsilon^{-1}L^{3/2}\log(n\varepsilon/\sqrt L)$ when $n\varepsilon\ge2\sqrt L$ (the second branch).
\item[(iii)] For $(\log U)^{-1/2}\le\varepsilon\le1/8$, where $n\ge\log^2U$: $\varepsilon^{-2}\log U\log(n/\log U)$.
\end{enumerate}

\begin{proof}[Proof of \cref{thm:intro-main}]
The lower bound is proved in \cref{sec:streaming-transfer}, so we prove the
upper bound, in the three regimes listed above.

\emph{Part (ii).} For the first branch, \cref{prop:exact-storage}
uses $O(n\log(2+U/n))$ bits. For the second,
\cref{thm:small-error-upper} uses
$O(\varepsilon^{-1}L^{3/2}\log(2+\tfrac{n\varepsilon}{\sqrt L}))$ bits.

\emph{Part (i).} The upper bound is \cref{prop:exact-storage}.

\emph{Part (iii).} The upper bound is \cref{thm:priority-bits}.
When $n\ge\log^2U$, the upper bound has the stated form since
$\log n\le2\log(n/\log U)$.

At $\varepsilon=2/U$, where $L=1$, the two branches of (ii) agree with (i).
At $n\varepsilon=2\sqrt L$, both branches of (ii) read
$\Theta(\varepsilon^{-1}L^{3/2})$, and at
$\varepsilon=(\log U)^{-1/2}$, where $L=\Theta(\log U)$ and
$\varepsilon^{-1}L^{3/2}=\Theta(\varepsilon^{-2}\log U)
=\Theta(\log^2U)$, the second branch of (ii) and part (iii) agree for
every $n\ge\log^2U$.
\end{proof}

\begin{remark}[Space and time of the fast implementations]
On a word RAM with $O(\log nU)$-bit words, each part also has a fast
implementation, with the following space and time.
\begin{itemize}
  \item Part (i) and the first branch of part (ii): $O(\min\{n\log U,\ U\log n\})$
    bits, $O(\log^4U)$ update time, and $O(\log U)$ query time
    (\cref{prop:exact-storage}).
  \item The second branch of part (ii): $O(\varepsilon^{-1}L^{3/2}\log n)$ bits,
    $O(\log^2U)$ update time, and $O(\log^2U)$ query time
    (\cref{thm:small-error-upper}).
  \item Part (iii): $O(\varepsilon^{-2}\log U\log n)$ bits, $O(\log U)$ update
    time, and $O(\varepsilon^{-4}\log^2U)$ query time (\cref{thm:priority-bits}).
\end{itemize}
For a constant $\eta>0$, this space is the same as the optimal space in part (i)
and the first branch of part (ii) when $n\le U^{1-\eta}$ or $n\ge U^{1+\eta}$, in
the second branch of part (ii) when $n\ge(\sqrt L/\varepsilon)^{1+\eta}$, and
throughout part (iii).
\end{remark}

In the standard setting where $n$ is polynomial in $U$, we measure the space in
words. In the \emph{polynomial word model}, the stream length, update
magnitudes, and numerical precision are bounded by polynomials in $U$, and a
word has $\Theta(\log U)$ bits.

\begin{theorem}[Word-model upper bound]
\label{thm:main-upper}
In the polynomial word model, fixed-query rank estimation on $[U]$ admits
a randomized sketching algorithm that succeeds with probability at least $2/3$
on every fixed query and whose working memory, including all random seeds, is
$O(U)$ words if $0<\varepsilon\le2/U$,
$O(\varepsilon^{-1}\log^{3/2}(U\varepsilon))$ words if
$2/U<\varepsilon\le(\log U)^{-1/2}$, and $O(\varepsilon^{-2}\log U)$ words if
$(\log U)^{-1/2}<\varepsilon\le1/8$.  In the
large-error branch, updates take $O(\log U)$ time and decoding takes
$O(\varepsilon^{-4}\log^2U)$ time; the construction uses no random oracle or
uncharged random table.
\end{theorem}

\begin{proof}[Proof of \cref{thm:main-upper}]
Work in the word model with words of $\Theta(\log U)$ bits, so that the
$\ell_1$ norm bound satisfies $n=U^{O(1)}$; the parameters $U$ and $\varepsilon$
are known in advance, so the sketch selects its branch accordingly.

For $0<\varepsilon\le2/U$, store $f$ exactly, keeping each coordinate
as its residue modulo a power of two larger than $2n$, as in
\cref{app:exact-fast}: $U$ counters of $O(\log n)=O(\log U)$
bits each, i.e.\ $O(U)$ words.

For $2/U<\varepsilon\le(\log U)^{-1/2}$, keep each counter of the
Dyadic CountSketch of \cref{thm:small-error-upper} as its residue modulo a
power of two larger than twice its final range:
$O(\varepsilon^{-1}\log^{3/2}(U\varepsilon))$ counters of
$O(\log U)$ bits each, i.e.\ that many words.

For $(\log U)^{-1/2}<\varepsilon\le1/8$, use the fast priority sketch
of \cref{thm:priority-bits}. Its $O(\varepsilon^{-2}\log^2U)$ bits
occupy $O(\varepsilon^{-2}\log U)$ words.

In each branch the count is the bound of \cref{thm:main-upper}, and the
success probability is at least $2/3$ on every fixed query.  All
three constructions are linear sketches conditioned on their stored
seeds, and the seed lengths have already been included in the displayed
space bounds.
\end{proof}

\section{Acknowledgment}

The authors are grateful to Jelani Nelson for valuable discussions and for pointing out that the $O(\varepsilon^{-2}\log ^2 U)$ space bound can be achieved using $\ell_1$ samplers.

\section*{AI Methodology}
We report AI assistance using the categories recommended by the Simons Institute AI + TCS working group~\cite{SimonsAITCS26}.

\begin{center}
\renewcommand{\arraystretch}{1.15}
\begin{tabular}{@{}l@{\qquad}c@{}}
\multicolumn{1}{c}{\textbf{Activity}} & \textbf{AI used} \\
Asking the research question & No \\
Coming up with the approach & No \\
Proof development & No \\
Writing and exposition & Yes \\
Checking for bugs & Yes \\
Other supporting tasks & Yes \\
\end{tabular}
\end{center}

\textbf{Details.}
We used OpenAI Codex and Anthropic Claude to help draft and revise technical exposition, check technical arguments for potential errors, and assist with language polishing, copy-editing, and LaTeX formatting. The authors supplied the research questions, proof strategy, and mathematical arguments, and reviewed and revised the AI suggestions. The authors take responsibility for the correctness and presentation of the paper.

\newcommand{\etalchar}[1]{$^{#1}$}

\appendix
\section{Omitted Details for the Upper Bounds}
\label{app:derandomization-details}

This appendix completes the proofs of \cref{sec:upper-bounds}.
\Cref{app:exact-fast} gives the fast implementation of exact recovery
(\cref{prop:exact-storage}). \Cref{app:dyadic-proof} proves
\cref{thm:small-error-upper} for the Dyadic CountSketch, including its fast
implementation, and \cref{app:priority-sketch} proves
\cref{thm:priority-bits} for the priority sketch, first with free
randomness, then with a seed, and finally with fast updates and
queries. Throughout, $C$ is a sufficiently large absolute constant.

\subsection{Fast Implementation of Exact Recovery}
\label{app:exact-fast}

This subsection proves the last claim of \cref{prop:exact-storage}: on
a word RAM with $O(\log nU)$-bit words, $O(\min\{n\log U,\ U\log n\})$
bits of space suffice with $O(\log^4U)$ update time and $O(\log U)$
query time. Unlike the fingerprint in the proof of
\cref{prop:exact-storage}, which stores all counters jointly, the
constructions below store the counters separately: as modular counters
when $n\ge U$, and in buckets decoded by sparse recovery when $n\le U$.

\paragraph{Modular counters for $n\ge U$.}
To store an integer counter $z$ whose final value lies in $[-B,B]$, we
keep only its residue $z\bmod M$, where $M=2^{\ceil{\log(2B+1)}}>2B$.
An increment $\Delta$ updates the residue modulo $M$, and at the end of
the stream, $z$ is the representative of the residue in
$[-M/2,M/2)$. Each counter takes $O(\log(B+1))$ bits, and an update or
a read takes $O(1)$ word operations.

For exact recovery, we keep one such counter, with $B=n$, for each node
of a binary tree over $[U]$; it records the mass of $f$ over the leaves
of the node. An update, a rank query, and a select query each visit
$O(\log U)$ nodes, and the $O(U)$ counters take $O(U\log n)$ bits.

\paragraph{Buckets and sparse recovery for $n\le U$.}
Here we follow the bucketing of~\citep[Section~3.1]{NelsonW09}. Let $t=\min\{n,\,C\log U\}$, and hash
the coordinates into $B=\ceil{n/\log U}$ buckets by a $t$-wise
independent hash $h:[U]\to[B]$, with a seed of $O(t\log U)=O(n\log U)$
bits. Then $tB\le2Cn$. The final vector $f$ has at most $n$ nonzero
coordinates, so if $t=n$, every bucket ends with at most $t$ nonzero
coordinates. Otherwise, a fixed set of $t$ nonzero coordinates lands in
a given bucket with probability $B^{-t}$ by $t$-wise independence, so
every bucket ends with at most $t$ nonzero coordinates except with
probability at most
$B\binom ntB^{-t}\le U(en/(tB))^t\le U(e/C)^{C\log U}\le1/200$. Hence in
either case, with probability at least $1-1/200$, every bucket ends with
at most $t$ nonzero coordinates.
Fix a prime $2U+1<p\le4U+2$, so that $p>U\ge n$ and $p>2t+1$. For
each bucket $b$, we maintain the $2t$ \emph{syndromes} in $\mathbb{F}_p$:
\[
  S_{b,j}\coloneqq\sum_{i:\,h(i)=b}f_i\,(i+1)^j\bmod p,
  \qquad j=1,\ldots,2t.
\]
Since $p>U$, the points $i+1$ for $i\in[U]$
are distinct and nonzero in $\mathbb F_p$, and since $p>n$, every
nonzero final value $f_i$ stays nonzero modulo $p$. So if a bucket
ends with at most $t$ nonzero coordinates, the sparse recovery
algorithm of~\citep[Theorem~1]{NelsonW09} recovers these coordinates
and their masses from the final syndromes of the bucket. It
first computes from the syndromes the polynomial $\Lambda$ of degree at
most $t$ whose roots are the points $i+1$ of the nonzero coordinates,
then finds the roots of $\Lambda$, and finally solves for the masses.
Only the root finding is randomized, and we use the folklore
algorithm~\citep[Chapter~14]{GathenG13} with $8\ceil{\log t}$ rounds.
A round reads $\ceil{\log p}$ random bits, lets $a\in\mathbb F_p$ be
their value modulo $p$, and splits every factor $\Lambda'$ of $\Lambda$
found so far into $\gcd\bigl(\Lambda',(X+a)^{(p-1)/2}-1\bigr)$ and its
cofactor. The run \emph{fails} if some factor is not linear after the
last round. So a run reads
$s'\coloneqq8\ceil{\log t}\ceil{\log p}=O(\log t\log U)$ random bits,
and with fast polynomial arithmetic, it takes $O(\log^3U)$ field
operations. For a bucket with at most $t$ nonzero coordinates, two
roots $\alpha\ne\beta$ of $\Lambda$ are separated in a round if
$(\alpha+a)(\beta+a)$ is a quadratic nonresidue. This holds for
$(p-1)/2$ values of $a$, each of probability more than $1/(2p)$, so
with probability more than $1/5$. By a union bound over the fewer than
$t^2/2$ pairs of roots, a run fails with probability at most
$(t^2/2)(4/5)^{8\ceil{\log t}}\le1/2$.

\paragraph{Updates and queries.}
Besides the syndromes, we keep for each bucket a candidate list of at
most $t$ pairs $(i,f_i)$, and a balanced binary search tree that holds
the candidate pairs of all buckets, keyed by coordinate, in which every node
also stores the total mass of its subtree.
The random bits of the recovery runs come from the seed.
Independently of $h$, the seed contains a stationary random
walk $w_1,\ldots,w_{C\log U}$ on the Margulis--Gabber--Galil
expander~\citep{HooryLW06} with vertex set $\{0,1\}^{s'}$ (recall that in our case here, $s'=O(\log t\log U)$), which has
constant degree and a constant spectral gap, and in which a step takes
$O(1+s'/\log U)$ word operations. Each $w_k$ is uniform on
$\{0,1\}^{s'}$, and the walk takes $s'+O(\log U)$ bits.
An update $(i,\Delta)$ proceeds in three steps. First, it adds
$\Delta(i+1)^j$ to $S_{h(i),j}$ for all $j\le2t$, in $O(t)$ field
operations. Second, it runs the sparse recovery algorithm on the
current syndromes of the bucket $h(i)$ up to $C\log U$ times,
the $k$-th run with the random bits $w_k$, in $O(\log^4U)$ field operations in total. Third, it
replaces the candidate list of the bucket $h(i)$ by the output of the
first run that succeeds, or by the empty list if no run succeeds or the
output has more than $t$ pairs, and it deletes the old pairs of this
bucket from the search tree and inserts the new ones, in $O(t\log U)$
time. So an update takes $O(\log^4U)$ time. A rank query $x$ follows
the search path of $x$ along the binary search tree and, at every node with coordinate less than
$x$, adds the mass of the node and the subtree mass of its left
child. A select query descends from the root by comparing the
target rank with the subtree masses. Both take $O(\log U)$ time.

\paragraph{Correctness and space.}
The recovery may be wrong in the middle of the stream, when a bucket
may have more than $t$ nonzero coordinates. But the candidate list of a
bucket is recomputed at every update of the bucket, so at the end of
the stream, it comes from the final syndromes. With probability at
least $1-1/200$, every bucket ends with at most $t$ nonzero
coordinates. Then the last recovery of a bucket fails only if all of
its $C\log U$ runs fail. Their input, the final
syndromes of the bucket, is independent of the walk, and each $w_k$ is
uniform, so each run fails with probability at most $1/2$. The expander
Chernoff bound~\citep{Gillman98}
then shows that all of them fail with probability at most
$e^{-\Omega(C\log U)}\le1/(200U)$ for large $C$. By a union bound
over the $B\le U$ buckets, with probability at least $99/100$, the
search tree holds exactly the nonzero coordinates of $f$ with their
masses, and every query is answered exactly.

Since $n\le U$, a word has $O(\log U)$ bits. The $2tB$ syndromes, the
at most $tB$ candidate pairs, and the search tree on them take
$O(tB)=O(n)$ words, that is, $O(n\log U)$ bits. So do the prime $p$
and the seed of $h$. The walk takes
$s'+O(\log U)=O(t\log U)=O(n\log U)$ bits (recall that
$s'=O(\log t\log U)$), and its $C\log U$ steps
take $O(s'+\log U)=O(\log^2U)$ time per update.
Hence, for $n\le U$, the algorithm uses $O(n\log U)$ bits of space with
$O(\log^4U)$ update time and $O(\log U)$ query time. Together with the
modular counters for $n\ge U$, this proves the last claim of
\cref{prop:exact-storage}.

\subsection{The Dyadic CountSketch: Proof of
\texorpdfstring{\cref{thm:small-error-upper}}{the small-error upper bound}}
\label{app:dyadic-proof}

This subsection proves \cref{thm:small-error-upper}. We first prove
\cref{lem:prg-countsketch}, which bounds the error of a single
CountSketch. We then prove \cref{thm:small-error-upper} for rank
queries, together with its space and seed bounds. We then show how
the sketch answers a select query by a binary search. Finally, we
give the fast implementation.

\label{app:dyadic-details}
\begin{proof}[Proof of \cref{lem:prg-countsketch}]
We first give the construction. Let $m$ be the least power of two
with $m\ge b^3$, so that $m<2b^3$.
Each row $s$ independently draws pairwise independent hashes
$h_s:[N]\to[b]$ and $g_s:[N]\to[m]$, and a pseudorandom string $z_s\in\{-1,1\}^m$ from Nisan's pseudorandom generator~\citep{Nisan92} (see
also~\citep[Appendix~A]{KachamPTW23}) for read-once branching programs
of length $m$ with $O(\log n)$ bits of memory and error
$\delta\coloneqq b^{-3}$. The sign of a coordinate $i$ in row $s$ is then $\sigma_s(i)\coloneqq z_{s,g_s(i)}$. The hashes take $O(\log N)$ bits,
and the generator takes a seed of $O(\log m\cdot\log(nm/\delta))
=O(\log b\log(nb))$ bits. A hash is evaluated in constant time, and a
sign in $O(\log m)=O(\log b)$ time, since Nisan's generator computes
any requested block of its output with $O(\log m)$ evaluations of
pairwise independent hash functions on $O(\log(nb))$-bit
blocks~\citep[Appendix~A]{KachamPTW23}.

Fix a coordinate $i\in[N]$. The error
$\err(i)\coloneqq\mathsf{CS}_{b,5}(y)_i-y_i$ of the estimate of $y_i$
is the median of the row errors
\[
  Y_s\coloneqq\sigma_s(i)\,C_{s,h_s(i)}-y_i
  =\sum_{k\ne i:\,h_s(k)=h_s(i)}\sigma_s(i)\sigma_s(k)\,y_k,
  \qquad s=1,\ldots,5.
\]

\emph{The second moment.} Following~\citep{LarsenPT21}, we show
$\E[\err(i)^2]=O(\norm y_1^2/b^2)$.
Each collision $h_s(k)=h_s(i)$ with $k\ne i$ has probability $1/b$ by
pairwise independence, so $\E|Y_s|\le\norm y_1/b$. The median exceeds $t$ in absolute value
only if at least three of the five rows do, so Markov's inequality and
the independence of the rows give
\[
  \Prb[|\err(i)|>t]
  \le\min\{1,\,
      \binom{5}{3}\left(\frac{\norm y_1}{bt}\right)^3\}.
\]
Integrating this tail bound against $2t\,dt$ gives
$\E[\err(i)^2]=O(\norm y_1^2/b^2)$.

\emph{The mean.} Next, we show $|\E[\err(i)]|=O(\norm y_1/b^3)$. We
first consider fully random strings $z_s$, and then compare them with
the pseudorandom strings of Nisan's generator. Condition on the hashes
$h_s$ and $g_s$, and let $Y_s'$ be $Y_s$
without the terms $k$ such that $g_s(k)=g_s(i)$. With fully random strings,
flipping the bit $z_{s,g_s(i)}$ of every row $s$ negates all $Y_s'$ and
preserves the distribution of the strings, so the median of
$Y_1',\ldots,Y_5'$ has mean zero. Replacing each $Y_s'$ by $Y_s$
changes the median by at most
$\max_s|Y_s-Y_s'|\le\sum_s\sum_{k\ne i:\,g_s(k)=g_s(i)}|y_k|$, whose
expectation over the hashes $g_s$ is at most
$5\norm y_1/m\le5\norm y_1/b^3$. So with fully random strings,
$|\E[\err(i)]|\le5\norm y_1/b^3$.

Now fix the hashes $h_s$ and $g_s$. For each $a\in[m]$, merge the coefficients $y_k$ of the coordinates $k\ne i$ with $h_s(k)=h_s(i)$ and $g_s(k)=a$ into
$c_{s,a}\coloneqq\sum_{k\ne i:\,h_s(k)=h_s(i),\,g_s(k)=a}y_k$, so that
$Y_s=z_{s,g_s(i)}\sum_{a\in[m]}c_{s,a}z_{s,a}$. Since the hashes and
the coefficients $c_{s,a}$ are fixed, they can be built into the
transitions of a read-once branching program that reads only $z_s$, in
order, and keeps $\sigma_s(i)=z_{s,g_s(i)}$ and the running sum
$\sum_{a'\le a}c_{s,a'}z_{s,a'}\in[-\norm y_1,\norm y_1]$. This program
computes $Y_s$ with $O(\log n)$ bits of memory. Hence, Nisan's generator changes the distribution of $Y_s$ by at most $b^{-3}$ in total variation distance.
As the rows are independent, the distribution of $(Y_1,\ldots,Y_5)$
changes by at most $5b^{-3}$, and since $|\err(i)|\le\norm y_1$, the
mean of $\err(i)$ changes by at most $10\norm y_1/b^3$. Together,
$|\E[\err(i)]|\le15\norm y_1/b^3$.
\end{proof}

Next, we prove \cref{thm:small-error-upper}. Recall the Dyadic CountSketch
of \cref{sec:dyadic-upper}. It stores the exact masses of the $U/2^L$
\emph{blocks} and, for each of the $L$ finest levels
$j=\log U-L+1,\ldots,\log U$, an independent CountSketch of the level
vector $y^{(j)}$ with $5$ rows and width $b=\Theta(\sqrt L/\varepsilon)$,
whose hashes and signs are those of \cref{lem:prg-countsketch}. We write
$\widehat R_0(x)$ for its rank estimate at $x$, which adds up the exact
masses of the whole blocks in $[0,x)$ and the CountSketch estimates of
the at most $L$ dyadic intervals that cover the rest of $[0,x)$.

\begin{proof}[Proof of \cref{thm:small-error-upper}]
Without loss of generality assume $n\ge1$. Write $F=\norm f_1$, and
choose the width $b=\ceil{C\sqrt L/\varepsilon}$.

\emph{Rank queries.} Fix a rank query $x\in\{0,\ldots,U\}$. The whole
blocks in $[0,x)$ contribute their exact masses, so the error
$\widehat R_0(x)-R_f(x)$ is the sum of the errors $e_j$ of the
CountSketch estimates of the at most $L$ dyadic intervals that cover
the rest of $[0,x)$, one per fine level $j$. Different levels have
independent CountSketches, and each level vector $y^{(j)}$ has $\ell_1$
norm $F$, so by \cref{lem:prg-countsketch} the errors $e_j$ are
independent, with $\E[e_j^2]=O(F^2/b^2)$ and means
$\mu_j\coloneqq\E[e_j]$ of absolute value $O(F/b^3)$. Their variances
add up, so
\[
  \E\bigl[(\widehat R_0(x)-R_f(x))^2\bigr]
  \le\sum_j\E[e_j^2]+\Bigl(\sum_j|\mu_j|\Bigr)^2
  =O\!\left(\frac{LF^2}{b^2}+\frac{L^2F^2}{b^6}\right)
  =O\!\left(\frac{LF^2}{b^2}\right),
\]
where the last step uses $b\ge\sqrt L$. Since $b\ge C\sqrt L/\varepsilon$
with $C$ sufficiently large, this is at most
$\varepsilon^2F^2/10$, so the rank query fails
with probability at most $1/10$.

\emph{Counters and space.} The sketch has
$D=5Lb+U/2^L=\Theta(\varepsilon^{-1}L^{3/2})$ integer counters: the
masses of the $U/2^L$ blocks, and the counters $C_{s,a}$ of the $5L$
CountSketch rows, five for each of the $L$ fine levels. All of them are
integer combinations of coordinates of $f$. We bound the $\ell_1$ norm
of the final counter vector $z\in\Z^D$. The block masses are
nonnegative and sum to $F\le n$. In a row $s$ of the CountSketch of
level $j$, we have $C_{s,a}=\sum_{i:\,h_s(i)=a}\sigma_s(i)\,y^{(j)}_i$,
so by the triangle inequality,
\[
  \sum_{a\in[b]}|C_{s,a}|\le\norm{y^{(j)}}_1=F\le n.
\]
Summing over the blocks and the $5L$ rows gives
$\norm z_1\le R\coloneqq(5L+1)n$. We store $z$ using the fingerprint method in
the proof of \cref{prop:exact-storage}, applied to $z$ in place of $f$
and with an independent seed. The candidates are now the at most
$K'\coloneqq\binom{2D+R}{R}$ vectors in $\Z^D$ of $\ell_1$ norm at
most $R$, and we take a prime $100DK'<p\le200DK'$.
The same bound shows that, conditional on any CountSketch seeds,
the fingerprint recovers $z$ except with probability $1/100$, using
$O(\log K')=O(D\log(2+R/D))
=O(\varepsilon^{-1}L^{3/2}\log(2+n\varepsilon/\sqrt L))$ bits. On this
event, both queries are computed from the exact counters $z$.

\emph{Seed.} By \cref{lem:prg-countsketch} with $N\le U$, applied to
level vectors of $\ell_1$ norm $F\le n$, each of the $L$ levels takes a
seed of $O(\log U+\log b\log(nb))$ bits. This is absorbed by the space
bound. First, $L\log U\le\log^2U=O(D)$ because
$D=\Omega(\varepsilon^{-1}L^{3/2})=\Omega(\log^2U)$ (consider whether
$L\ge(\log U)/2$). Second,
$\log b\log(nb)=O(b\log(2+n/b))$ (consider whether $n\le b^2$), and
$D=\Theta(Lb)$ and $R=\Theta(Ln)$, so
$L\log b\log(nb)=O(D\log(2+R/D))$.

\emph{Select queries.}
Finally, we show that the sketch answers a fixed select query with
target rank $r\in[0,F]$. The sketch knows the norm $F$ exactly, as
the sum of the block masses, so it can compute the threshold
$\theta\coloneqq\varepsilon F/2$. The query runs a binary search as follows. It
maintains an interval $[s,t]$ of candidate
prefix endpoints, initially $[1,U]$. While $s<t$, it takes the midpoint
$m=\floor{(s+t)/2}$ and
\begin{itemize}
  \item stops and returns $\widehat x=m-1$ if
    $|\widehat R_0(m)-r|\le\theta$;
  \item continues with $[s,m]$ if $\widehat R_0(m)>r+\theta$;
  \item continues with $[m+1,t]$ if $\widehat R_0(m)<r-\theta$.
\end{itemize}
When $s=t$, it returns $\widehat x=s-1$. We first show that the answer
is correct whenever the rank estimates are accurate on a fixed search
path, and then bound the errors on this path.

Let the target endpoint $s^*$ be the least $s\in\{1,\ldots,U\}$ with
$R_f(s)\ge r$, which exists since $R_f(U)=F\ge r$, and let the fixed
path $P$ be the set of midpoints that the binary search visits if it
always moves toward $s^*$. We claim that the answer is correct whenever the rank estimates
satisfy $|\widehat R_0(m)-R_f(m)|\le\theta$ for all $m\in P$. Indeed,
every move of the search then agrees with $P$: going left means
$\widehat R_0(m)>r+\theta$, hence $R_f(m)>r$ and $s^*\le m$, and going
right means $\widehat R_0(m)<r-\theta$, hence $R_f(m)<r$ and $s^*>m$.
So the search either stops at some midpoint $m\in P$, where
$|R_f(m)-r|\le2\theta=\varepsilon F$, or ends at the endpoint $s^*$. In
the first case, $\widehat x=m-1$ is correct, since
$R_f(m)\ge r-\varepsilon F$ and $R_f(m-1)\le R_f(m)\le r+\varepsilon F$.
In the second case, $\widehat x=s^*-1$ is correct, since
$R_f(s^*)\ge r$ and, by the minimality of $s^*$ and $R_f(0)=0$,
$R_f(s^*-1)\le r$.

It remains to show that some rank estimate on the fixed path $P$ has
error more than $\theta$ with probability at most
$1/10$. The first $\log U-L$ midpoints of $P$ are boundaries of the
$U/2^L$ blocks, so their rank estimates are exact. Let $m_1,\ldots,m_\ell$, with $\ell\le L$, be the
remaining midpoints of $P$. At step $j$, let $e_j$ be
the error of the CountSketch estimate of the current left child, the
dyadic interval of step $j$ that ends at $m_j$, and let
$d_j\coloneqq\ind\{s^*>m_j\}$ indicate that $P$ turns right at $m_j$.
Apart from blocks, whose masses are exact, the prefix $[0,m_j)$
is the disjoint union of the left children at the earlier right turns
and the current left child, so the error at $m_j$ is
\[
  \delta_j\coloneqq\widehat R_0(m_j)-R_f(m_j)=\sum_{i<j}d_ie_i+e_j.
\]
Since $P$ is fixed and 
CountSketches are independent across different levels, the errors $e_1,\ldots,e_\ell$ are
independent. By \cref{lem:prg-countsketch}, applied to level vectors of
$\ell_1$ norm $F$, they satisfy $\E[e_j^2]=O(F^2/b^2)$, and their means
$\mu_j\coloneqq\E[e_j]$ satisfy $|\mu_j|=O(F/b^3)$. 

We then separate $\delta_j$ into its mean and centered parts by writing $e_j=\mu_j+\xi_j$, where the centered errors $\xi_j$ have mean zero. Then
\[
  \delta_j=\Bigl(\sum_{i<j}d_i\mu_i+\mu_j\Bigr)+\sum_{i<j}d_i\xi_i+\xi_j.
\]
Since
$b\ge C\sqrt L/\varepsilon$ with $C$ sufficiently large, the mean part has absolute value at most $\sum_{i\le\ell}|\mu_i|=O(LF/b^3)\le\theta/3$. So if $|\delta_j|>\theta$ for some $j$, then $|\sum_{i<j}d_i\xi_i|>\theta/3$ or $|\xi_j|>\theta/3$ for some $j$. We show that each of these two events has probability at most $1/20$. For the first event, we use Kolmogorov's
maximal inequality~\citep[Section~2.5]{Durrett19}: for independent
random variables $X_1,\ldots,X_\ell$ with mean zero and every
$\lambda>0$,
\[
  \Prb\Bigl[\max_{j\le\ell}\Bigl|\sum_{i\le j}X_i\Bigr|\ge\lambda\Bigr]
  \le\frac1{\lambda^2}\sum_{i\le\ell}\E[X_i^2].
\]
Applying it with $X_i=d_i\xi_i$ and $\lambda=\theta/3$ bounds the
probability of the first event by $9\sum_{i\le\ell}\E[\xi_i^2]/\theta^2 \leq 9\sum_{i\le\ell}\E[e_i^2]/\theta^2 = O(L/(b^2\varepsilon^2)) \le 1/20$, where we use $\E[\xi_i^2]\le\E[e_i^2]$, $\ell\le L$, and $b\ge C\sqrt L/\varepsilon$ with $C$ sufficiently large. For the second event, Chebyshev's inequality and a union bound over $j$ give probability at most $\sum_{i\le\ell}\Prb[|\xi_i|>\theta/3]\leq 9\sum_{i\le\ell}\E[e_i^2]/\theta^2 \le 1/20$ for the same reason.

Hence, with probability at least $9/10$, every rank estimate on $P$ has
error at most $\theta$, and the select query is answered correctly.
Together with the failure probabilities of the rank query and of the
fingerprint, a fixed rank query and a fixed select query are both
correct except with probability $1/10+1/10+1/100<1/3$.
\end{proof}

\paragraph{Fast implementation.}
Finally, we prove the last claim of \cref{thm:small-error-upper}: with
$O(\varepsilon^{-1}L^{3/2}\log n)$ bits, updates and queries take
$O(\log^2U)$ time. To do so, we replace the fingerprint by modular
counters as in \cref{app:exact-fast}.

\emph{Counters and space.} We store every counter separately as its
residue modulo a power of two larger than $2n$, and add a binary tree over the blocks whose nodes store the masses of
their blocks in the same way. Since the tree has fewer than $2U/2^L$
nodes, this gives $O(D)$ counters. At the end of the stream, every
counter has absolute value at most $n$, so each residue determines its counter and has
$O(\log n)$ bits, and the total space is $O(D\log n)=O(\varepsilon^{-1}L^{3/2}\log n)$ bits. The seeds of \cref{lem:prg-countsketch} take $O(D\log(2+R/D))=O(D\log n)$ bits, as shown above.

\emph{Updates and queries.} An update touches $O(\log U)$ counters: one node on each level of the
tree and five counters on each fine level. A rank query $x$ reads the
$O(\log U)$ tree nodes that cover the \emph{blocks} in $[0,x)$ and the
$5L$ counters of the at most $L$ dyadic intervals that cover the rest.
A select query runs the binary search above and keeps the sum $A$ of
the estimated masses of the left children skipped at earlier right
turns, hence also reads $O(\log U)$ counters. By \cref{lem:prg-countsketch} with $N\le U$
and $\log b=O(\log U)$, on $O(\log nU)$-bit words, each hash takes
constant time and each sign takes $O(\log U)$ time. As an update or a
query touches $O(\log U)$ counters, each needing one hash and one sign,
it takes $O(\log^2U)$ time. This completes the proof of
\cref{thm:small-error-upper}.

\subsection{The Priority Sketch: Proof of
\texorpdfstring{\cref{thm:priority-bits}}{the large-error upper bound}}
\label{app:priority-sketch}

This subsection proves \cref{thm:priority-bits}. We first describe the
priority sketch and prove its correctness with free randomness
(\cref{app:priority-free}). We then replace the free randomness by a
seed of $O(\log nU)$ bits: we list the properties of the randomness
that the proof uses, and construct a seed with these properties
(\cref{app:priority-seeded}). Finally,
we give the fast implementation (\cref{app:counter-implementations}).

\subsubsection{With Free Randomness}
\label{app:priority-free}

We proceed in four steps: we first recall the unbiased estimator of the
rank from \cref{sec:shared-seed}, then design a sketch that approximates
this estimator, then describe how the sketch answers queries, and
finally prove its correctness.

\paragraph{The unbiased estimator.}
Recall the estimator $\widehat f_S$ from \eqref{eq:priority-ideal}, where for
a rank query $x$, we take $S=[0,x)$.

Formally, we set the parameter $K\coloneqq\ceil{400\varepsilon^{-2}}=O(\log U)$, which is fixed, and
the threshold $\tau\coloneqq\norm f_1/K$, which is maintained by the sketch and changes over time. If the final norm is zero, return $0$ for rank queries and $1$ for select queries. Hence from now on, we assume $\norm f_1\ne0$.
Before the stream, we draw
independent random variables $u_i$ uniformly from $(0,1]$, define the
\emph{priorities} $p_i\coloneqq f_i/u_i$ (which depend on the current frequencies $f_i$ and the fixed $u_i$), and set the estimator as
\begin{equation}
  \widehat f_{[0,x)}=\sum_{i<x}\max\{f_i,\tau\}\,\ind\{p_i>\tau\}.
  \label{eq:priority-rank-ideal}
\end{equation}
By \eqref{eq:priority-mean-var}, it is an unbiased estimate of the rank
$R_f(x)=f_{[0,x)}$, and its variance is at most
$\tau\norm f_1=\norm f_1^2/K\le\varepsilon^2\norm f_1^2/400$.
In terms of the subset $H=\{i:p_i>\tau\}$ from \cref{sec:shared-seed}, the estimator sums the weights $\max\{f_i,\tau\}$ of the coordinates $i<x$ in $H$.
Next, we design a sketch that approximates the estimator $\widehat f_{[0,x)}$.

\paragraph{The priority sketch.}
Since the random variables $u_i$ are fixed before the stream, the priority vector $p$ is a linear function of the frequency vector $f$, and
an update $(i,\Delta)$ changes the priority $p_i$ by $\Delta/u_i$.  So we can
maintain CountSketches (\cref{def:countsketch}) of both the frequency
vector $f$ and the priority vector $p$. Recall from
\cref{sec:dyadic-upper} that a CountSketch with width $b$ and $r$ rows
estimates each coordinate by the median of its $r$ row estimates. We
use the following error bounds.
\begin{theorem}[CountSketch~\citep{CharikarCF04}]
\label{thm:countsketch}
Let $v\in\mathbb R^U$ be a vector and $b,r\ge1$.  Consider a CountSketch of $v$ with
width $b$ and $r$ rows (\cref{def:countsketch}) whose hashes and signs
are random.  Let $J\subseteq[U]$ be any set of at most $k$
coordinates such that the coordinates outside $J$ have $\ell_2$
norm at most $\gamma$.  Then for every $i\in[U]$ and $\theta>0$,
\begin{equation}
  \Prb\!\left[\left|\mathsf{CS}_{b,1}(v)_i-v_i\right|\ge\theta\right]
  \le\frac{k}{b}+\frac{\gamma^2}{b\theta^2}, \qquad
  \E\bigl|\mathsf{CS}_{b,1}(v)_i-v_i\bigr|
  \le\frac1b\sum_{j\in J}|v_j|+\frac{\gamma}{\sqrt b}.
  \label{eq:countsketch-tail}
\end{equation}
Moreover, if the first bound in \eqref{eq:countsketch-tail} is at
most $1/3$, then
$\Prb[|\mathsf{CS}_{b,r}(v)_i-v_i|\ge\theta]\le e^{-\Omega(r)}$.
\end{theorem}
We set the number of rows $m\coloneqq\ceil{C\log U}$. Besides the
exact $\ell_1$ norm $\norm f_1=\sum_if_i$, the sketch stores three
CountSketches of the priority vector $p$ and the frequency vector $f$,
with seeds independent of each other and of the random variables
$u_i$:
\[
  \widetilde p_i=\mathsf{CS}_{CK,m}(p)_i,\qquad
  \bar p_i=\mathsf{CS}_{CK/\varepsilon^2,1}(p)_i,\qquad
  \bar f_i=\mathsf{CS}_{CK^2,1}(f)_i.
\]
The \emph{coarse} estimate $\widetilde p$ has many rows, so that all
$U$ of its coordinates are accurate simultaneously. It plays the role of
the $\ell_2$-heavy-hitter sketch of \cref{sec:shared-seed}: it tells
whether $i\in H$ for every coordinate $i$ whose priority is not close
to $\tau$. The \emph{fine} estimate $\bar p$ has only one wide row, and
it decides whether $i\in H$ for the few remaining coordinates, whose
priorities are close to $\tau$. The \emph{fine} estimate $\bar f$ gives
the weights $\max\{f_i,\tau\}$ of the coordinates in $H$. This stores
$O(K\log U+K/\varepsilon^2+K^2)=O(\varepsilon^{-2}\log U)$ counters,
using the parameter $K=O(\varepsilon^{-2})$ and $\varepsilon^{-2}\le\log U$.

\paragraph{Answering queries.} At the end of the stream, the query
algorithm computes the threshold $\tau=\norm f_1/K$ from the stored
norm, and the estimates $\widetilde p_i$, $\bar p_i$, and $\bar f_i$
from the counters of the three CountSketches. In the estimator
$\widehat f_{[0,x)}$, it replaces the indicator $\ind\{p_i>\tau\}$ of
$i\in H$ by the indicator $\ind\{\widetilde p_i>3\tau/4,\ \bar p_i>\tau\}$,
built from the coarse and fine estimates of the priority, and the
weight $\max\{f_i,\tau\}$ by $\max\{\bar f_i,\tau\}$:
\begin{equation}
  \widetilde f_{[0,x)}\coloneqq\sum_{i<x}\max\{\bar f_i,\tau\}\,
    \ind\bigl\{\widetilde p_i>3\tau/4,\ \bar p_i>\tau\bigr\}.
  \label{eq:priority-estimator}
\end{equation}
For a rank query $x$, return $\widetilde f_{[0,x)}$. For a select query, run
a binary search on this cumulative estimate. Its summands are nonnegative, so $\widetilde f_{[0,y)}$ is
nondecreasing in $y$, and the search returns $\widehat x=y-1$ for the
least $y\in\{1,\ldots,U\}$ with $\widetilde f_{[0,y)}\ge r$, or
$\widehat x=U-1$ if none exists. For now, we allow both queries to scan
the coordinates of $[U]$ to evaluate \eqref{eq:priority-estimator}; we
replace the free randomness by a seed in \cref{app:priority-seeded}, and
avoid the scan in \cref{app:counter-implementations}.

\paragraph{Proof of correctness.}
We first show that, with good probability, only a few priorities are
large.  Since each priority satisfies $\Prb[p_i\ge t]\le f_i/t$, in expectation at most
$8\norm f_1/\tau=8K$ priorities are at least $\tau/8$, an eighth of the threshold $\tau$, and
$\E\sum_i\min\{p_i,\tau/8\}^2=\sum_i\int_0^{\tau/8}2t\,\Prb[p_i>t]\,dt\le\tau\norm f_1/4=\tau^2K/4$. Therefore, by Markov's inequality, with probability at least $0.99$, at most $O(K)$
priorities are at least $\tau/8$, and the other priorities have squared
sum $O(\tau^2K)$.

Let the \emph{good event} $\mathcal G$ be the event that the priorities
$p_i$ satisfy the above and all $U$ coordinates of the coarse estimate
$\widetilde p$ have additive error at most $\tau/8$, i.e.,
\[
  |\widetilde p_i - p_i| \le \tau/8 \quad \text{for all } i \in [U].
\]
Given the priorities, applying \cref{thm:countsketch} with the priority
vector $p$, the set $J=\{j:p_j\ge\tau/8\}$, the squared $\ell_2$ norm
$O(\tau^2K)$ outside $J$, width $b=CK$, and error threshold
$\theta=\tau/8$ bounds the failure probability of one row
$\mathsf{CS}_{b,1}(p)_i$ by $O(1/C)\le1/3$. So with $m\ge C\log U$ rows,
each coordinate of
$\widetilde p$ fails with probability at most $1/(200U)$, and a union
bound gives $\Prb[\mathcal G^c]\le0.01+0.005\le0.02$, where $\mathcal G^c$ denotes the complement of the good event $\mathcal G$. Note that $\mathcal G$ does not depend on the query $x$. 

Now, on the good event $\mathcal{G}$, we bound the difference between the estimate
$\widetilde f_{[0,x)}$ and the true rank $f_{[0,x)}$.  We use
the unbiased estimator $\widehat f_{[0,x)}$ from \eqref{eq:priority-rank-ideal} as a proxy, and
compare $\widetilde f_{[0,x)}$ with $\widehat f_{[0,x)}$ term by term.
Write $I_i\coloneqq\ind\{p_i>\tau\}$ for the indicator of $i\in H$ in $\widehat f_{[0,x)}$ and
$\widetilde I_i\coloneqq\ind\{\widetilde p_i>3\tau/4,\ \bar p_i>\tau\}$
for the replacement indicator in $\widetilde f_{[0,x)}$, built from the
coarse estimate $\widetilde p_i$ and the fine estimate $\bar p_i$.  Then
\[
  \widetilde f_{[0,x)}-f_{[0,x)}
  =\bigl(\widehat f_{[0,x)}-f_{[0,x)}\bigr)
   +\sum_{i<x}\max\{f_i,\tau\}\bigl(\widetilde I_i-I_i\bigr)
   +\sum_{i<x}\bigl(\max\{\bar f_i,\tau\}-\max\{f_i,\tau\}\bigr)\widetilde I_i.
\]
We bound the expectation of each of the three terms on $\mathcal G$,
that is, $\E[\,\cdot\,\ind_{\mathcal G}]$, where $\ind_{\mathcal G}$ is
the indicator of $\mathcal G$:
\begin{itemize}
\item \emph{Sampling error:} $|\widehat f_{[0,x)}-f_{[0,x)}|$.  Since
  the term is nonnegative, we may drop $\ind_{\mathcal G}$.  By
  \eqref{eq:priority-mean-var}, its expectation is at most
  $\sqrt{\Var \widehat f_{[0,x)}}\le\sqrt{\tau\norm f_1}\le\varepsilon\norm f_1/20$.
\item \emph{Decision error:}
  $\sum_{i<x}\max\{f_i,\tau\}\,\ind\{\widetilde I_i\ne I_i\}$.  This is the error from the coordinates $i$ for which we decide wrongly whether $i\in H$.  On
  the good event $\mathcal G$, $\widetilde I_i = 1$ and $I_i = 0$ imply that the fine estimate satisfies $\bar{p}_i > \tau > p_i > 3\tau /4 - \tau /8 = 5\tau/8 > \tau /2$. Conversely, $\widetilde I_i = 0$ and $I_i = 1$ imply $p_i > \tau$, hence the coarse estimate satisfies $\widetilde{p}_i > p_i - \tau /8 > 3\tau /4$, and therefore $\bar{p}_i \le \tau < p_i$. In either case, we have $p_i>\tau/2$ and $|\bar p_i-p_i|\ge|p_i-\tau|$.

  Next, we fix all randomness except the seed
  of the fine estimate $\bar p$, and assume the good event $\mathcal G$ occurs.  Then applying
  \cref{thm:countsketch} with the priority vector $p$, the set
  $J=\{j:p_j\ge\tau/8\}$, the squared $\ell_2$ norm $O(\tau^2K)$ outside
  $J$, width $b=CK/\varepsilon^2$, and error threshold $\theta=|p_i-\tau|$ bounds the probability that $\widetilde I_i\ne I_i$ by
  \[
    q_i\coloneqq\ind\{p_i>\tfrac\tau2\}\min\Bigl\{1,\
      O\Bigl(\frac{\varepsilon^2}C
      +\frac{\varepsilon^2\tau^2}{C(p_i-\tau)^2}\Bigr)\Bigr\} \geq 0.
  \]
  Therefore, taking the expectation over the seed of $\bar p$ alone, we have
  $\E_{\bar p}\bigl[\ind_{\mathcal G} \cdot \sum_{i<x}\max\{f_i,\tau\}\,\ind\{\widetilde I_i\ne I_i\}\bigr] \le \sum_{i<x}\max\{f_i,\tau\}\,q_i$, since the good event $\mathcal G$ does not depend on the seed of $\bar p$. Taking the expectation over the rest of the randomness, it suffices to bound $\E q_i$ for each $i$, where $q_i$ depends only on the priority $p_i=f_i/u_i$ with $u_i$ uniform on $(0,1]$. We prove that each $\max\{f_i,\tau\}\,\E q_i = O(\varepsilon f_i/\sqrt C)$ as follows:
  \begin{itemize}
  \item If $f_i\ge2\tau$, then $p_i = f_i / u_i \ge f_i$ gives $p_i-\tau\ge \tau$,
    so $q_i=O(\varepsilon^2/C)$ and the term is $O(\varepsilon^2f_i/C)$.
  \item If $f_i<2\tau$, then $\max\{f_i,\tau\}<2\tau$.  Since
    $\Prb[p_i>t]=\min\{1,f_i/t\}$, the density of $p_i$ on
    $(\tau/2,\infty)$ is at most $4f_i/\tau^2$, so
    $\Prb[p_i>\tau/2]\le2f_i/\tau$ and
    $\Prb[p_i>\tau/2,\ |p_i-\tau|<\theta]\le8f_i\theta/\tau^2$ for every
    $\theta>0$.  Since $\E Y=\int_0^1\Prb[Y>t]\,dt$ for $Y\in[0,1]$,
    \[
      \E\Bigl[\ind\{p_i>\tfrac\tau2\}
        \min\Bigl\{1,\frac{\varepsilon^2\tau^2}{C(p_i-\tau)^2}\Bigr\}\Bigr]
      =\int_0^1\Prb\Bigl[p_i>\tfrac\tau2,\
        |p_i-\tau|<\frac{\varepsilon\tau}{\sqrt{Ct}}\Bigr]\,dt
      \le\frac{16\varepsilon f_i}{\sqrt C\,\tau},
    \]
    and $\E[\ind\{p_i>\tfrac\tau2\}\,\varepsilon^2/C]\le2\varepsilon^2f_i/(C\tau) = O(\varepsilon^2 / C)$.
    So the term is $O(\varepsilon^2f_i/C+\varepsilon f_i/\sqrt C)$.
  \end{itemize}
  Summing over $i$, the expected decision error on $\mathcal G$ is
  $O(\varepsilon\norm f_1/\sqrt C)\le\varepsilon\norm f_1/100$ for a large constant $C$.
\item \emph{Weight error:}
  $\sum_{i<x}\bigl|\max\{\bar f_i,\tau\}-\max\{f_i,\tau\}\bigr|\,\widetilde I_i$.
  This is the error in the weights of the coordinates that we decide to be in $H$.
  On the good event $\mathcal G$, $\widetilde I_i=1$ implies $p_i>5\tau/8$, so the
  weight error times $\ind_{\mathcal G}$ is at most
  $\sum_i|\bar f_i-f_i|\,\ind\{p_i>\tau/2\}$, which no longer involves
  $\mathcal G$.  The seed of the frequency estimate
  $\bar f$ is independent of the priorities $p_i$.  At most $8K$ coordinates of
  $f$ are at least $\tau/8$, and the others have squared sum at most
  $\tau\norm f_1/8$.  So applying \cref{thm:countsketch} with the frequency
  vector $f$, the set $J=\{j:f_j\ge\tau/8\}$, the squared $\ell_2$ norm
  $\tau\norm f_1/8$ outside $J$, and width $b=CK^2$ gives
  $\E|\bar f_i-f_i|\le\norm f_1/(CK^2)+\sqrt{\tau\norm f_1/(8CK^2)}=O(\tau/\sqrt{CK})$.
  Since $\sum_i\Prb[p_i>\tau/2]\le2\norm f_1/\tau=2K$, the expected weight
  error is $O(\tau\sqrt{K/C})=O(\norm f_1/\sqrt{CK})\le\varepsilon\norm f_1/100$.
\end{itemize}
Hence
$\E\bigl[|\widetilde f_{[0,x)}-f_{[0,x)}|\,\ind_{\mathcal G}\bigr]\le7\varepsilon\norm f_1/100$,
so by Markov's inequality
\begin{equation}
  \Prb\bigl[|\widetilde f_{[0,x)}-f_{[0,x)}|>\varepsilon\norm f_1\bigr]
  \le\Prb[\mathcal G^c]+0.07
  \le0.09 .
  \label{eq:priority-pointwise-failure}
\end{equation}
Finally, we handle a fixed select query with target rank
$r\in[0,F]$, where $F=\norm f_1$. Let $a$ be the least
$x\in\{1,\ldots,U\}$ with $R_f(x)\ge r-\varepsilon F$, and let $c$ be the
largest $x\in\{1,\ldots,U\}$ with $R_f(x-1)\le r+\varepsilon F$; they
exist since $R_f(U)=F\ge r$ and $R_f(0)=0\le r$. Since $R_f$ is
nondecreasing, an answer $\widehat x$ is correct whenever
$a\le\widehat x+1\le c$. We show that a wrong answer forces an error of
more than $\varepsilon F$ in the rank estimate at $x=a-1$ or at $x=c$.
Recall that the answer is
$\widehat x=y-1$ for the least $y$ with $\widetilde f_{[0,y)}\ge r$, or
$\widehat x=U-1$ if no such $y$ exists, and that $\widetilde f_{[0,y)}$
is nondecreasing in $y$.
If $\widehat x+1<a$, then monotonicity and the minimality of $a$ give
$\widetilde f_{[0,a-1)}\ge\widetilde f_{[0,\widehat x+1)}\ge r>R_f(a-1)+\varepsilon F$.
If $\widehat x+1>c$, then the choice of $\widehat x$ and the maximality
of $c$ give $\widetilde f_{[0,c)}<r<R_f(c)-\varepsilon F$. Since $a-1$
and $c$ are fixed independently of the sketch, a union bound of
\eqref{eq:priority-pointwise-failure} over $x=a-1$, $x=c$, and the fixed
rank query shows that a fixed rank query and a fixed select query
are jointly correct with probability at least $1-3\cdot0.1>2/3$.

\subsubsection{Replacing the Free Randomness with a Seed}
\label{app:priority-seeded}
\label{app:priority-properties}
\label{app:priority-randomness}

We now replace the free randomness by a seed of $O(\log nU)$ bits. We
first list the properties of the randomness that the proof uses and
check that they suffice, and then construct a seed with these
properties.

\paragraph{Properties used by the proof.}
The proof in \cref{app:priority-free} uses the
randomness only through the following properties, which hold with
$\delta=0$ under free randomness.
\begin{enumerate}[label=(P\arabic*)]
\item For all $i$ and $t>0$, $\Prb[p_i\ge t]\le f_i/t$, and if
  $f_i\le t$, then $\Prb[p_i>t]\ge f_i/t-\delta$.  Moreover,
  $p_i\ge f_i$.  The $p_i$ are pairwise independent.
\item The $u_i$ and the seeds of the three CountSketch instances are
  mutually independent.
\item Every row of the three CountSketches satisfies \eqref{eq:countsketch-tail},
  and if each row errs by at least $\theta$ with probability at most $1/3$,
  then the median of $m$ rows does so with probability at most
  $1/(200U)$.
\end{enumerate}
Suppose (P1)--(P3) hold with some $\delta\le2/n$.  Each $i$ with
$f_i\le\tau$ has $\Prb[p_i>\tau]\in[f_i/\tau-\delta,f_i/\tau]$, each $i$
with $f_i>\tau$ has $p_i\ge f_i>\tau$, and at most $\norm f_1\le n$
coordinates have $f_i>0$.  The bound $\Prb[\mathcal G^c]\le0.02$ uses
only the upper tail in (P1), (P2), and (P3), so it is unchanged.  The three
error terms of \cref{app:priority-free} become:
\begin{itemize}
\item \emph{Sampling error:} $\widehat f_{[0,x)}$ now has bias at most
  $\tau n\delta\le2\norm f_1/K$, since only the nonzero coordinates with
  $f_i\le\tau$ are biased, each by at most $\tau\delta$.  Its variance is
  still at most $\tau\norm f_1$,
  since that bound needs only pairwise independence, $p_i\ge f_i$
  (so the coordinates with $f_i>\tau$ contribute no variance), and
  $\Prb[p_i>\tau]\le f_i/\tau$.  So the expectation is at most
  $\varepsilon\norm f_1/20+2\norm f_1/K\le3\varepsilon\norm f_1/50$.
\item \emph{Decision error:} the bound $q_i$ uses (P2) and (P3) for the
  fine estimate $\bar p$. Then the case $f_i\ge2\tau$ uses only
  $p_i\ge f_i$.  In the case $f_i<2\tau$, the bound on
  $\Prb[p_i>\tau/2,\ |p_i-\tau|<\theta]$ gains an additive $\delta$.
  This case involves at most $n$ nonzero coordinates, each with weight
  $\max\{f_i,\tau\}<2\tau$, so it adds at most
  $2\tau n\delta\le4\tau=4\norm f_1/K\le\varepsilon\norm f_1/800$ to the
  expected decision error.  For larger $C$, the decision error is still at
  most $\varepsilon\norm f_1/100$.
\item \emph{Weight error:} unchanged, at most $\varepsilon\norm f_1/100$,
  since it uses only the upper tail in (P1), the independence of the
  seed of $\bar f$ from the $u_i$ by (P2), and
  \eqref{eq:countsketch-tail} for $\bar f$ by (P3).
\end{itemize}
The expected total on $\mathcal G$ is then at most
$3\varepsilon\norm f_1/50+2\varepsilon\norm f_1/100=2\varepsilon\norm f_1/25$, and
\eqref{eq:priority-pointwise-failure} becomes
\begin{equation}
  \Prb\bigl[|\widetilde f_{[0,x)}-f_{[0,x)}|>\varepsilon\norm f_1\bigr]
  \le0.02+0.08=0.1,
  \label{eq:priority-seeded-failure}
\end{equation}
so the rest of the proof is unchanged. It remains to construct seeds of
$O(\log nU)$ bits in total that satisfy (P1)--(P3) with $\delta\le2/n$.
We draw all seeds below independently, which gives (P2).

\paragraph{Priorities.} Let $D_0$ be the least power of two with
$D_0\ge n^2$, and let $\mathbb F$ be the smallest binary field with
$|\mathbb F|\ge\max\{U,D_0,CK^2\}$, where $CK^2$ is the largest width
of the three CountSketches.  Embed $[U]$ in $\mathbb F$, evaluate a
uniformly random affine map $\mathbb F\to\mathbb F$ at $i \in [U]$, and let
$u_i\coloneqq j_i/D_0$, where $j_i$ is one plus the integer formed by
the top $\log_2D_0$ bits of the image of $i$ under the affine map. Then the $u_i$ are pairwise independent and uniform on
$\{1/D_0,\ldots,1\}$, and the seed has $O(\log nU)$ bits.  We round
the priorities to
\begin{equation}
  p_i\coloneqq\frac{a_if_i}{D_0}\quad\text{with }a_i\coloneqq\floor{D_0/u_i},
  \qquad\text{so that}\qquad
  \frac{f_i}{u_i}-\frac{f_i}{D_0}\le p_i\le\frac{f_i}{u_i}.
  \label{eq:priority-rounding-sandwich}
\end{equation}
We observe that the rounded priorities satisfy (P1) with
$\delta=2/D_0\le2/n^2$, as follows. By \eqref{eq:priority-rounding-sandwich},
$\Prb[p_i\ge t]\le\Prb[u_i\le f_i/t]\le f_i/t$. Conversely, if
$f_i\le t$, then $p_i>t$ whenever $u_i<c\coloneqq f_i/(t+f_i/D_0)$,
which has probability at least $c-1/D_0\ge f_i/t-2/D_0$, since
$f_i/t-c\le(f_i/t)^2/D_0\le1/D_0$. Moreover, $a_i\ge D_0$ as $u_i\le1$,
so $p_i\ge f_i$, and the $p_i$ are pairwise independent since the
$u_i$ are.

\paragraph{Rows of the CountSketch.} A row of width $b$, rounded up to a power of two, uses a
pairwise independent hash $h:[U]\to[b]$ and independent pairwise
independent signs $\sigma:[U]\to\{-1,1\}$, each a truncated affine map
over the above field $\mathbb F$: the hash $h(i)$ is formed by the top
$\log_2b$ bits of the image of $i$ under a uniformly random affine map
$\mathbb F\to\mathbb F$, and the sign $\sigma(i)$ by the top bit of the
image of $i$ under another. This needs $b\le|\mathbb F|$, which holds
since $|\mathbb F|$ is a power of two at least $CK^2$.  A row seed thus
consists of the four coefficients of the two affine maps, which is a
string of $s\coloneqq4\log_2|\mathbb F|=O(\log nU)$ bits, and pairwise
independence suffices for
\eqref{eq:countsketch-tail}~\citep{CharikarCF04}.

\paragraph{Error amplification.} Independent seeds for the $m=\ceil{C\log U}$ rows of the coarse sketch $\widetilde p$ would cost
$ms=O(\log U\log nU)$ bits. Instead, we seed these rows by the
vertices of a stationary random walk on the expander of
\cref{app:exact-fast}, now with vertex set $\{0,1\}^s$, the set of all
row seeds of $s$ bits.  The walk
takes $s+O(m)=O(\log nU)$ bits and each vertex is a uniform seed.  If one
row errs with probability at most $1/3$, then the median errs only if
at least half of the walk's rows err, which has probability
$e^{-\Omega(m)}\le1/(200U)$ for large $C$ by the expander Chernoff
bound~\citep{Gillman98}.  With the rows above, this gives (P3).

\subsubsection{Fast Implementation}
\label{app:counter-implementations}

Finally, we give the fast implementation of the priority sketch. Its
integer counters are stored as residues, as in \cref{app:exact-fast},
and its queries avoid scanning $[U]$ by a fast recovery of the
candidates for the subset $H$.

\paragraph{Counters as residues.}
\label{app:priority-counters}
The priority CountSketches of \cref{app:priority-randomness} sketch the
integer vector $D_0p$, to which an update $(i,\Delta)$ adds
$a_i\Delta$. Since $a_i\le D_0^2<4n^4$, their final counters lie in
$[-4n^5,4n^5]$, and the counters on $f$ lie in $[-n,n]$. Hence every
counter takes $O(\log n)$ bits as a residue. The whole sketch has
$O(K\log U+K/\varepsilon^2+K^2)=O(K\log U)$ counters, as
$K=O(\varepsilon^{-2})$ and $\varepsilon^{-2}\le\log U$, and an update touches one counter in each of its
$m+2=O(\log U)$ rows. Queries divide the priority counters by $D_0$
before taking medians.

\paragraph{Fast candidate recovery.}
\label{app:priority-fast}
The counters above give each summand of \eqref{eq:priority-estimator}
quickly, but the queries still scan $[U]$ to find the nonzero summands.
On the event $\mathcal G$, however, a nonzero summand requires
$\widetilde p_i>3\tau/4$, hence $p_i>3\tau/4-\tau/8=5\tau/8$, and at most
$O(K)$ priorities are at least $\tau/8$. So it suffices to recover a
list of $O(\poly(K))$ candidates that contains every coordinate with priority
at least $\tau/8$, and to evaluate the summands only at these
candidates. We do this with the following theorem, an adaptation of
the heavy-coordinate construction
of~\citep[Appendix~B.1, Lemma~14 and Remark~2]{LarsenNNT16}, with
implementation adjustments including pairwise independent signs, rows seeded by
one expander walk, and counters stored as residues.

\begin{theorem}[Candidate recovery]
\label{thm:priority-candidate-recovery}
Let $M\ge2$ and $1\le k=O(\log U)$. There is a randomized linear
sketch for turnstile streams with final vector $v\in\Z^U$ of $\ell_1$
norm at most $M$, together with a recovery algorithm that, given a
threshold $T>0$ at query time, reports at most $O(k^2)$ distinct
candidates, one at a time. If
\[
 S=\{i:|v_i|\ge T\},\qquad |S|\le k,\qquad
 \sum_{i\notin S}v_i^2\le kT^2,
\]
then the candidates include all of $S$ with probability at least
$0.999$. On a word RAM with $O(\log MU)$-bit words, the sketch uses
$O(k\log U\log M)$ bits throughout updates and recovery, including a
seed of $O(\log U)$ bits, with $O(\log U)$ update time and
$O(k^2\log^2U)$ recovery time.
\end{theorem}

\begin{proof}
We first hash the coordinates into buckets, so that each bucket
contains at most one coordinate of $S$, and then learn the coordinate
of $S$ in each bucket a few bits at a time.

Let $h:[U]\to[B]$ be a pairwise independent hash with
$B\coloneqq C^2k^2$. A union bound over the pairs in $S$ and Markov's
inequality show that, except with probability
$k^2/B+Ck^2/B\le0.0005$, the hash $h$ is injective on $S$ and
\begin{equation}
 \sum_{i\notin S:\ h(i)\in h(S)}v_i^2\le T^2/C.
 \label{eq:fast-good-hash}
\end{equation}
To learn the coordinate of $S$ in a bucket, we use the
PartitionCountSketch of~\citep[Appendix~B.1]{LarsenNNT16}. We write each
coordinate $i$ in binary with $\ceil{\log_2U}$ bits and split it into
$\ell\coloneqq\ceil{\ceil{\log_2U}/t}$ characters $c_j(i)$ of
$t\coloneqq\ceil{\log_2\log_2U}$ bits each, padding the last one. The
$j$-th label of $i$ is $P_j(i)\coloneqq(h(i),c_j(i))$, which takes
$N\coloneqq B2^t=O(\log^3U)$ values. For each $j\in [\ell]$, we keep the
PartitionCountSketch for the label map $P_j$ with
$R\coloneqq\Theta(\log(\ell N))$ rows and width $w\coloneqq\Theta(k)$:
\begin{equation}
 A_{j,r,q}\coloneqq\sum_{i:\ g_r(P_j(i))=q}\sigma_r(i)v_i,
 \qquad r\in[R],\ q\in[w].
 \label{eq:fast-counters}
\end{equation}
Here $g_r:[N]\to[w]$ and $\sigma_r:[U]\to\{-1,1\}$ are independent
pairwise independent hashes and signs. As in
\cref{app:priority-seeded}, their row seeds are the vertices of one
stationary expander walk with constant-time neighbor steps, independent
of $h$ and shared by all $j$.

To recover, we say that a label $\lambda$ \emph{passes} in block $j$ if
$\operatorname{median}_r|A_{j,r,g_r(\lambda)}|>T/2$. For each bucket
$b\in[B]$, if exactly one character $c$ with label $(b,c)$ passes in
every block, we concatenate these characters and output the result if it
lies in $[U]$ and hashes to $b$. This outputs at most $B$ distinct
candidates, one at a time.

Now suppose that $h$ is injective on $S$ and satisfies
\eqref{eq:fast-good-hash}. Then each label in a bucket of $h(S)$ is the
label of at most one coordinate of $S$, and the coordinates outside $S$
with this label have squared sum at most $T^2/C$. These are the
conditions of the single-row analysis in the proof
of~\citep[Lemma~14]{LarsenNNT16}, with the $\pm1$ signs
of~\citep[Remark~2]{LarsenNNT16},
so for each such label $\lambda$, one row satisfies
$|A_{j,r,g_r(\lambda)}|>T/2$ exactly when $\lambda$ is the label of a
coordinate of $S$, with probability at least $2/3$. The median is wrong
only if at least half of the rows are wrong, so the expander Chernoff
bound~\citep{Gillman98} and a union bound over the at most $\ell N$ such
labels show that all their tests are correct except with probability
$0.0005$. When all
these tests are correct, every $i\in S$ is output, since in the bucket
$h(i)$, exactly the character $c_j(i)$ passes in every block $j$. So
the recovery reports all of $S$ except with probability $0.001$.

Since the signs are $\pm1$, every counter has absolute value at most
$M$ at the end, so its residue takes $O(\log M)$ bits. The seed takes
$O(\log U+R)=O(\log U)$ bits. Since $\ell=O(\log U/\log\log U)$ and
$R=O(\log\log U)$, the number of counters, the number of counters that
an update touches, and the number of counter reads in the recovery are
respectively
\[
 \ell Rw=O(k\log U),\qquad \ell R=O(\log U),\qquad
 B\ell2^tR=O(k^2\log^2U).
\]
The recovery processes one bucket at a time, which needs only
$O(\log U+R\log M)$ additional bits.
\end{proof}

We apply \cref{thm:priority-candidate-recovery}, with a seed independent
of the priorities and of the three CountSketches, to $v=D_0p$ with
$M=4n^5$, $T=D_0\tau/8$, and $k=\ceil{CK}$. By the definition of
$\mathcal G$, on $\mathcal G$ we have
\begin{equation}
 S=\{i:p_i\ge\tau/8\},\qquad |S|\le k,\qquad
 \sum_{i\notin S}v_i^2\le kT^2,
 \label{eq:fast-target}
\end{equation}
and every nonzero summand of \eqref{eq:priority-estimator} belongs to
$S$, as observed above. As the recovery reports each candidate $i$, the
query algorithm evaluates the summand of $i$ in
\eqref{eq:priority-estimator} and keeps $i$ only if this summand is
nonzero. If more than $k$ candidates are kept, it discards all of them.
On $\mathcal G$, at most $|S|\le k$ candidates are kept. If, in
addition, the recovery reports all of $S$, then the kept candidates are
exactly the coordinates with nonzero summands. Like $\mathcal G$, this
event does not depend on the query $x$, so on it, for every $x$, the
sum of the summands of the kept candidates $i<x$ equals
$\widetilde f_{[0,x)}$. The rank and select queries use these sums in
place of $\widetilde f_{[0,x)}$, so on this event, they return the same
answers as in \cref{app:priority-free}. The recovery fails on
$\mathcal G$ with probability at most $0.001$, and by
\eqref{eq:priority-seeded-failure} and the end of
\cref{app:priority-free}, the answers of \cref{app:priority-free} are
jointly correct with probability at least $1-3\cdot0.1$. So a fixed rank
query and a fixed select query are jointly correct with probability
at least $1-3\cdot0.1-0.001>2/3$. Finally, evaluating the summands of the $O(k^2)$ candidates takes
$O(k^2\log U)$ time, sorting the kept candidates takes $O(k\log k)$
time, and storing them takes $O(k\log nU)$ bits, all within the bounds
of \cref{thm:priority-bits}. This completes the proof of
\cref{thm:priority-bits}. \qed

\end{document}